\documentclass[CRPHYS,Unicode,manuscript]{cedram}

\usepackage{ifthen}

\newboolean{ShowEdits}
\setboolean{ShowEdits}{true}

\usepackage{graphicx,subfigure}  % needed for figures
\usepackage{dcolumn}   % needed for some tables
\usepackage{bm}        % for bold math (bold \rho)
\usepackage{amssymb}   % for math
\usepackage{physics} %\usepackage{amsmath}
  \usepackage{epsfig}   % e.g. for boxes
\usepackage{verbatim}  % to use \begin{comment}
\usepackage{cases}
\usepackage{dsfont}
\usepackage{mathrsfs} % for the round capital S, \mathscr{S}
\usepackage{color}
\usepackage[normalem]{ulem}  %cross-out with \sout{}
\usepackage{cancel}   % for strike-out in math mode, with \cancel{}

\usepackage{xcolor}
\hypersetup{
    colorlinks,
    linkcolor={red!50!black},
    citecolor={blue!50!black},
    urlcolor={blue!80!black}
}

\definecolor{darkgreen}{rgb}{0.0, 0.5, 0.0}

\newcommand{\bea}{\begin{eqnarray}}
\newcommand{\eea}{\end{eqnarray}}
\newcommand{\be}{\begin{equation}}
\newcommand{\ee}{\end{equation}}
\newcommand{\bi}{\begin{itemize}}
\newcommand{\ei}{\end{itemize}}
\newcommand{\ds}{\displaystyle}
\newcommand{\rr}{\mathbf{r}}
\newcommand{\kk}{{\mathbf{k}}}
\newcommand{\KK}{{\mathbf{K}}}
\newcommand{\pp}{\mathbf{p}}
\renewcommand{\qq}{\mathbf{q}}

\newcommand{\RR}{\mathbf{R}}

\newcommand{\CC}{\mathbf{C}}

\newcommand{\PP}{\mathbf{P}}

\newcommand{\vn}{\mathbf{0}}
\newcommand{\ra}{\rangle}
\newcommand{\la}{\langle}

\newcommand{\UP}{\uparrow}
\newcommand{\down}{\downarrow}

\newcommand{\Ar}{\mathcal{A}}
\newcommand{\Br}{\mathcal{B}}
\newcommand{\Vr}{\mathcal{V}}

\newcommand{\Cr}{\mathcal{C}}
\newcommand{\Fr}{\mathcal{F}}

\newcommand{\Ir}{\mathcal{I}}
\newcommand{\Jr}{\mathcal{J}}

\newcommand{\Dr}{\mathcal{D}}

\renewcommand{\Pr}{\mathcal{P}}

\newcommand{\Nr}{\mathcal{N}}

\newcommand{\Mr}{\mathcal{M}}

\newcommand{\Oo}{{\bm \Omega}} % {\mathbf{\Omega}} % {\boldmath \Omega} 
\newcommand{\rrho}{\bm \rho} % {\mbox{\boldmath$\rho$}}
\newcommand{\md}{\mathsf{m}} % {{\rm m}}

\newcommand{\dhk}{d\hat{k}}
\newcommand{\dk}{d\hspace{-0.2mm}k}

\renewcommand{\j}{j}

\newcommand{\nn}{\nonumber}

\newcommand{\gt}{\tilde{g}}
\ifthenelse{\boolean{ShowEdits}}{}
  {\renewcommand{\sout}[1]{\unskip}
    
    \renewcommand{\comm}[1]{\unskip}
  }
  \let\oldtilde\tilde
\renewcommand*{\tilde}[1]{\mathchoice{\widetilde{#1}}{\widetilde{#1}}{\oldtilde{#1}}{\oldtilde{#1}}}
\let\oldhat\hat
\renewcommand*{\hat}[1]{\mathchoice{\widehat{#1}}{\widehat{#1}}{\oldhat{#1}}{\oldhat{#1}}}

\begin{abstract} 
A fundamental quantity characterizing Fermi gases with zero-range interactions is the three-body contact $C_3\,$, which determines several observables including the number of nearby triplets of fermions and the three-body loss rate in cold atom experiments, as shown in a companion article~\cite{C3_I}. Here, we compute $C_3$ to leading order in the non-degenerate limit for the homogeneous gas with negative or infinite scattering length $a$.\, At $a=\infty$\,, using a wavefunction approach, we obtain the analytical expression of $C_3$\,, which has a remarkably slow $1/T^{\,0.22728}$ dependence on the temperature $T$. In the Feynman diagram technique, which we use for $a<0$\,, the correct three-body short-distance correlations emerge only after a non-trivial cancellation between leading order two-body and three-body correlations, and the resulting power-law scaling comes from the large-wavevector tail of the 3-body T-matrix,  which we derive inspired by the analytical solution of the three-body problem at the unitary limit. 
\end{abstract}

\begin{document}
\title{Three-body contact for fermions. \\II. Non-degenerate limit}

%\alttitle{Contact à trois corps pour les fermions. \\II. Limite non dégénérée}

%% Authors, addresses and supports.
%% The optional argument is for shortened version appearing in the headings. Please
%% distinguish between first, middle and last names with the appropriate commands.

%\IsCorresp}
\author{\firstname{Xavier} \lastname{Leyronas} \CDRorcid{0000-0002-9499-6800}}
\address{Laboratoire de Physique de l'Ecole Normale Sup\'erieure, ENS - Universit\'e PSL, Sorbonne Universit\'e, Universit\'e Paris Cité, CNRS, 75005 Paris, France}
%\email{}
\author{\firstname{F\'elix} \lastname{Werner} \CDRorcid{0000-0002-5631-9024}}
\address{Laboratoire Kastler Brossel, Ecole Normale Sup\'erieure - Universit\'e PSL, CNRS, Coll\`ege de France, Sorbonne Universit\'e, 75005 Paris, France}
%\email{}

%% Keywords
%\keywords{Example, Optimization, Journal}

%\keywords{\kwd{Unitary gas} \kwd{fermions} \kwd{cold atoms} \kwd{virial expansion} \kwd{three-body problem}}
%\altkeywords{\kwd{Gaz unitaire} \kwd{fermions} \kwd{atomes froids} \kwd{développement du viriel} \kwd{problème à trois corps}}

%% Abstract should be placed before \maketitle (and, in fact, before
%% \begin{document is best)

\maketitle

%\newpage
\setcounter{tocdepth}{1}
\vspace*{-8pt}
\tableofcontents

%\vspace*{-18pt}

\section*{Introduction}

Over the last two decades, the BEC-BCS crossover has become  an archetypal example of quantum many-body physics. 
It was first studied theoretically ~\cite{Leggett_Chap_1980_book,Haussmann_Z_Phys,Haussmann_PRB} and then experimentally in ultracold gases \cite{RevueTrentoFermions,ZwergerBook2,ChapZwergerZwerger,ZwierleinChap2014,ThomasStronglyInteractingGaz,GrimmCrossover,JinPairCondensate,KetterlePairCondensate,SalomonCrossover,GrimmModes,ShinPhaseDiag,SylEOS,NirEOS,KuEOS,LENSRepulsivePolaron,ValeGoldstone,Boiling_MIT,ZwierleinSound,SagiPolaronMol,MoritzBragg3D,PanSpectralFct2024,ZwierleinSecondSound2024,EsslingerEntropyTransport2024}.
The system under consideration is a gas of spin $1/2$ fermions (experimentally, atoms in two internal states) interacting via a short-range potential of large scattering length~$a$.
When $1/a$ varies from $-\infty$ to $+\infty$, the interaction becomes increasingly attractive, and the BCS-BEC crossover takes place.
The limit of zero-range interaction can appear harmless at first sight. However, this has non-trivial consequences such as the existence of
Tan's two-body contact parameter $C_2$\,, which quantifies the bunching of two opposite-spin fermions
at short distance,
 and determines numerous observables~\cite{TanEnergetics,TanLargeMomentum,ChapLeggettBref,LeChapitreIn2,ChapBraatenBref,RanderiaRF_arxiv_B,ZwergerRFLong,RanderiaRF,BaymRF,BraatenC,TanViriel,FelixViriel,ZwergerRF,BraatenLong,WernerTarruellCastin,ZhangLeggettUniv,CombescotC,WernerCastinRelationsArxiv_B,HofmannAnomaly,TanTrapFunctional,Moelmer,WernerCastinRelationsFermions,HuLiuPhotoex_C_2021},
 which enabled its measurement by various probes~\cite{WernerTarruellCastin,HuletClosedChannel,AustraliensC,AustraliensT,ValeC_precise,JinUnivRel,NirEOS,Jin_C_homogeneous,SagiContactRF,ValeC_T_hom,ZwierleinC,LaurentC,Denschlag_C_2024,Thywissen_Contact_Projection}.
In a companion article~\cite{C3_I}, hereafter referred to as Article~I, we introduced the  three-body contact $C_3$\,, which quantifies the bunching of three particles, and
we
 showed that $C_3$ determines several observables including the three-body loss rate.
In short, $C_2$ (resp.~$C_3$) counts the number of pairs (resp.~triplets) with small interparticle distances, or equivalently, with large relative momenta.

In this article, we calculate the three-body contact $C_3$ in the non-degenerate regime, {\it i.e.} for $n\lambda_T^{\phantom{T}3}\ll1$, where $n$ is
the total density of fermions and
\be
\lambda_T \ = \ \sqrt{\frac{2\,\pi\,\hbar^2}{m\,k_B T}}
\nn\ee
is the thermal de Broglie wavelength (with $T$ the temperature and $m$ the particle mass). The calculation is a virial expansion at leading order.
 Accordingly, our result is exact to leading order in the non-degenerate limit.
For $a=\infty$\,, we compute $C_3$ analytically using a wavefunction approach similar to the one of~\cite{UnitaryBoseGas,PetrovWernerHetero},
whereas for $a<0$, based on the diagrammatic formalism of~\cite{Leyronas_b3},
we develop an original approach were we analyze the short-distance behavior of the three-body correlation function $g_3$ to {derive an expression of} $C_3$ in terms of
the three-body T-matrix,
which we evaluate numerically.
 
The article is organized as follows. In Section~\ref{secdefC3} we recall the definition of $C_3$ and the main observables determined by $C_3$\,. In Section~\ref{secmainres} we announce the main results of this work. In Section~\ref{waveful} we derive our analytical result for $C_3$ at the unitary limit.
In Section~\ref{secFeyndiag}
we compute the three-body contact for any negative value
of \,$\lambda_T/a$\,,
based on the Feynman diagrammatic virial-expansion technique of~\cite{Leyronas_b3}.
  Within this diagrammatic approach, we find that the correct 
  short-distance power law behavior of $g_3$ emerges as a result of a non-trivial cancellation between
 two-body and three-body leading contributions,
 and that this power-law scaling comes from a large-wavevector tail of the off-shell $3$-body T-matrix.  We derive this tail
 by drawing inspiration from the analytical solution of the three-body problem at the unitary limit 
    (as detailed in appendices,
     in particular App.~\ref{AppImt3largek}, \ref{app:eigUL} and \ref{Appeigenvec}).
 In Section~\ref{tail gamma}, we perform an independent calculation of the large momentum tail of  the center-of-mass momentum distribution of nearby fermion pairs, and using the general relation of Article I, we recover our result for $C_3$.
We conclude in Section~\ref{secconcl}.

We consider the homogeneous gas
in the thermodynamic limit at thermal equilibrium,
restricting to the unpolarized case.\footnote{The generalization to the polarized case is straightforward. In Eq.~(\ref{eq:C3_f}), the factor $n^3$ is replaced by \,$ 4 \,n_\uparrow^2 n_\downarrow$ 
to get the $\uparrow\uparrow\downarrow$ three-body contact density 
\,$\Cr_{2,1} = C_{2,1} / \, \Vr$, 
or by \,$4 \,n_\uparrow n_\downarrow^2$\, to get 
\,$\Cr_{1,2}= C_{1,2} / \, \Vr$\,; 
the function $f_3$ remains unchanged.}

\section{The three-body contact}\label{secdefC3}

In this section, we summarize the main general relations involving the three-body contact.\footnote{The derivations are given in Article~I,
 except for an additional relation
stated and derived in Appendix~\ref{app:N3K}.}
We write the relations in the slightly simplified form applicable to the unpolarized homogeneous gas at thermal equilibrium.
Accordingly, we consider the thermodynamic limit where the system volume $\Vr$ tends to infinity.
The three-body contact density 
\,$\Cr_3=C_3/\Vr$\, is finite,
{\it i.e.}, $C_3$ is extensive.

$\Cr_3$\, is defined as follows. If one measures the positions of all particles within a region of volume~$\Vr'$, 
then the number of triplets of particles with hyperradius $R<\epsilon$,
per unit volume, 
 is
\be
\frac{N_3(\epsilon)}{\Vr'}
\ \underset{\epsilon\to0}{\sim} \ \ 
           {\Cr}_3\ \, \epsilon^{\,2s\,+\,2}
\ \, = \ \, {\Cr}_3\ \, \epsilon^{5.54544853}.
\label{eq:N3}
\ee
Here, the hyperradius $R$ 
is defined as
\be
R \ = \ \sqrt{\,\frac{2}{3}
\left(r_{12}^{\phantom{aa}2} \, + \, r_{13}^{\phantom{aa}2} \, + \, r_{23}^{\phantom{aa}2}\,\right)}
\nonumber\ee
with $r_{ij} \, = \, \| \rr_i - \rr_j \|\,$,
and
\be
s=1.772724267\ldots
\label{eq:s_val}
\ee
is the lowest positive solution of
\be
  \frac{s \ \cos(s\pi/6) - \sqrt{3}\,\sin(s\pi/6)}{(1-s^2) \ \sin(s\pi/2)}
  \ = \ \frac{\sqrt{3}}{4}\,.
\label{eq:transc_s}
\ee
Accordingly, the triplet correlation function
\begin{align*}
g_3(\rr_1,\rr_2,\rr_3) \ &= \
\big\la \, \hat{\psi}^\dagger_\uparrow(\rr_1) \, \hat{\psi}^\dagger_\uparrow(\rr_2) \, \hat{\psi}^\dagger_\down(\rr_3) \, \hat{\psi}_\down(\rr_3) \, \hat{\psi}_\uparrow(\rr_2) \, \hat{\psi}_\uparrow(\rr_1) \, \big\ra
\\  \ &= \
\big\la \, \hat{\psi}^\dagger_\down(\rr_1) \, \hat{\psi}^\dagger_\down(\rr_2) \, \hat{\psi}^\dagger_\UP(\rr_3) \, \hat{\psi}_\UP(\rr_3) \, \hat{\psi}_\downarrow(\rr_2) \, \hat{\psi}_\downarrow(\rr_1)
\, \big\ra
\end{align*}
has the short-distance asymptotic behavior
\be
\int d^5\Omega\ \ g_3(\rr_1,\rr_2,\rr_3) \ \underset{R\to0}{\sim}
\ \ \frac{ 16(s+1)}{3\sqrt{3}}
\ \ \Cr_3\ R^{\,2s\,-\,4}\,.
\label{eq:g3_C3}
\ee
Thus $g_3  \, \underset{R\to0}{\propto} \, \Cr_3 \  R^{-0.45455146}$,
and accordingly, the many-body wavefunction \,$\psi
\underset{R\to0}{\propto}
R^{-0.227275733}$.
     This divergence of $g_3$ and $\psi$ for $R\to0$\, means that there is three-body bunching.
     In this sense, when three particles of spins $\UP\UP\down$ are nearby, the effect of the attractive zero-range interactions between $\UP$ and $\down$ particles dominates over the effect of the Pauli exclusion between the two $\UP$ particles. The three-body contact \,$\Cr_3$\, measures the magnitude of this bunching.
   
Here and in what follows, we use the notations
\bea
\rr \ &=& \ \rr_3 \ - \ \rr_1
\nonumber 
\\ \frac{\sqrt{3}}{2} \ \rrho \ &=& \ \rr_2  \ - \  \frac{\rr_1+ \, \rr_3}{2}
\label{eq:def_jaco}
\eea
for the Jacobi coordinates, and
\be
\CC \ = \ \frac{\rr_1  \ + \  \rr_2 + \rr_3}{3}
\label{eq:def_C}
\ee
for the center-of-mass.
The relative positions of
the three particles
are then fixed by the six-dimensional vector
\be
\RR \ = \ (\rr, \rrho)
\nonumber\ee
whose norm $R=\sqrt{r^2+\rho^2}$\, is the hyperradius,
and whose direction
\be
\Oo \ \equiv \ \RR \, / R
\nonumber\ee
can be parameterized by five hyperangles.
In Eq.~(\ref{eq:g3_C3}),
the change of coordinates from $(\rr_1, \rr_2, \rr_3)$
to $(\CC, R, \Oo)$ is implied,
and $d^5\Omega$ is such that $d^6\!R = d^5\Omega\ R^5\,dR$.

The exponent $s$
follows from the analytical solution of the unitary three-body problem~\cite{Efimov}, and it
is related to the ground state energy of three particles in a harmonic trap of frequency $\omega$ by $E = (s+5/2)\,\hbar\omega$\,~\cite{TanScalingREVTEX,Werner3corpsPRL}.
In the non-relativistic field theory formalism, $s+5/2$ is the scaling dimension of a local three-fermion operator~\cite{son_cft}.

The asymptotic behavior of the three-particle momentum distribution at large relative momenta is a power law whose prefactor is determined by \,$\Cr_3$\,, see Eq.~(\ref{eq:N3K}) in Appendix~\ref{app:N3K}.

$\Cr_3$\,
also determines the large-momentum tail of the center-of-mass momentum distribution
of nearby fermion pairs $n_P(K\,)$, defined as follows.
Let $N_2(\epsilon,K\,)$ be the probability distribution that a pair of opposite-spin particles have center-of-mass momentum ${\bf K}$ and are separated by a distance smaller than $\epsilon$,
with the normalization condition that
$\int N_2(\epsilon,K\,)\ d^3K/(2\pi)^3$
equals the total number of opposite-spin pairs of separation $<\epsilon$.
We can then define $n_P(K\,)$ by
\be
\frac{N_2(\epsilon, K\,)}{\Vr} \ \, \underset{\epsilon\to0}{\sim}\ \, n_P(K\,)\ \, \frac{\epsilon}{4\pi}\,. \nn
\ee
We have
\be
    n_P(K\,) \ \underset{K\to\infty}{\sim} \ \ \Mr_P\ \, \frac{\Cr_3}{K^{\,2s+4}}
\label{eq:NP_tail}
  \ee
with the prefactor
\be
 \Mr_P \ = \ 32\, \pi^3\,\frac{4^s}{3^{s+1/2}}\,(s+1)\ \Gamma(s+2)^2\,\sin^2(s\pi)\ \Nr^2\,.
\label{eq:Mr}
\ee
The constant $\Nr$
(the normalization constant of the unitary hyperspherical wavefunction)
has the numerical value~\cite{SonRangeCorrections}\footnote{We give here the value from Ref.~\cite{SonRangeCorrections}, which has one more significant digit that the value \,$\Nr = 0.31149$\, from Ref.~\cite{C3_I}.}$^,$\footnote{The relation between $\Nr$ and
the constant $f_s^1$ from Ref.~\cite{SonRangeCorrections}
is \,$\Nr = \sqrt{32 \ \pi \, / \, (3 f_s^1)}$\,
(taking into account the convention for the hyperangular integration measure and the non-standard normalization of spherical harmonics in Ref.~\cite{SonRangeCorrections}\,).}$^,$\footnote{In Eq.~(\ref{eq:NP_tail}), the large exponent $2s+4 =7.54544853\ldots$ is unfavorable for the observability of the tail, but the large value of the prefactor $\Mr_P$ is favorable. These two effects mostly compensate each other, in the sense that
\,$\Mr_P/K^{2s+4} \,=\, (\Mr_P'\,/\,K\,)^{2s+4}$ where $\Mr_P' = 2.785\ldots$ is of order unity.}
\be
\Nr \ = \ 0.311489
\label{eq:Nr_val}
\ee
which yields
\be
\Mr_P \ = \ 2272.3
\label{eq:Mr_val}
\ee

Beyond the zero-range model,
\,$\Cr_3$\, also appears in two additional relations.
  \bi
  \item
    For finite-range interactions supporting deeply bound dimers,
the rate $\Gamma_3$ of inelastic three-body recombination events is
\be
\frac{\Gamma_3}{\Vr} \ = \ -\,8 \,s \, (s+1) \ \frac{\hbar}{m} \ \Cr_3\ \Im \, \bar{a}_3
\label{eq:Gamma3}
\ee
where $\bar{a}_3$\, is the mean three-body parameter.
 \item If a short-range three-body interaction potential is added to the zero-range two-body interactions, then the energy is shifted by
\be
\frac{\delta \!E_3}{\Vr} \ = \ 4 s \, (s+1) \ \frac{\hbar^2}{m} \ \Cr_3 \ \bar{a}_3\,.
\label{eq:dE3}
\ee
This holds for each eigenstate (and thus also at thermal equilibrium for fixed entropy).
Equation~(\ref{eq:dE3}) generalizes
   to finite-range two-body interactions,
by considering the energy difference \,$\delta\!E$\, between finite-range and zero-range models,
and defining $\delta \!E_3$ as the contribution to \,$\delta\!E$ coming from configurations where three particles are nearby.

\ei
Equations~(\ref{eq:Gamma3}) and~(\ref{eq:dE3}) hold to leading order
in the zero-range limit,
and
\,$\bar{a}_3$\, is a small parameter proportional to the range to the power $2s$.
Accordingly, \,$\Cr_3$\, can be evaluated within the zero-range model
(whereas
%and the only ingredient beyond the zero-range model is the parameter
$\bar{a}_3$
%\,, which
depends on the finite-range three-body problem).
%(more specifically, on the finite-range correction to the three-body zero-energy wavefunction at interparticle distances much larger than the range and much smaller than $|a|$\,).

\section{Main results}\label{secmainres}

\begin{figure}[h]
\begin{center}
  \includegraphics[width=0.8\linewidth
    ,trim={0.5cm 0.2cm 0.5cm 0.5cm}, clip
  ]{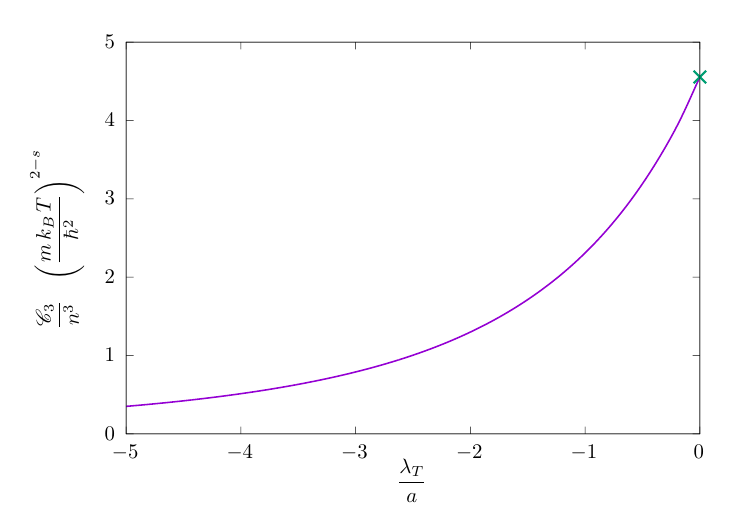}
\caption{Adimensionalized
  three-body contact
  \,$f_3\!\left( \frac{\lambda_T}{a} \right) \,=\, \frac{\mathcal{C}_3}{n^3} \times\left(\frac{m\,k_B\,T}{\hbar^2}  \right)^{2-s}$\,
  as a function of
\vskip .05cm
  the ratio
  between thermal wavelength $\lambda_T$ and scattering length $a$\,,
  for the homogeneous gas in the non-degenerate limit.
  The cross at \,$a=\infty$\,
  is the  analytical result from~Eq.~(\ref{eq:f(0)}).}
\label{figC3dimensionless}
\end{center}
\end{figure}

In the non-degenerate limit,
{\it i.e.} for $n\,\lambda_T^{\phantom{T}3}\to0$
  with fixed $T$ and $a$\,,
we find that
the three-body contact has the asymptotic expression
\be
\boxed{
\Cr_3(n , T , a) \ \, \simeq \ \, n^3 \  \left(\frac{\hbar^2}{m k_B T} \right)^{2-s}
\ f_3\!\left(\frac{\lambda_T}{a}\right)
\label{eq:C3_f}
}
\ee
with $f_3$\, a dimensionless function. At the unitary limit $a=\infty$\,,
we find the analytical result
\be
\boxed{
  f_3(0) \ = \ \frac{
    3^{5/2}\ \pi^3}{2^{2s+1}\,\Gamma(s+2)}
\ = \ 4.5552892 \ldots}
\label{eq:f(0)}
\ee
while the result of our numerical computation of \,$f_3$\, for negative $a$\, is plotted in Fig.~\ref{figC3dimensionless}.

In the grand-canonical ensemble,
\be
\Cr_3(\mu , T , a) \ \,
\simeq \ \,
e^{\,3\beta\mu} \ \ \frac{(m k_B T)^{s+5/2}\ f_3(\lambda_T/a)}{2^{3/2}\,\pi^{9/2} \ \hbar^{2s+5}}
\label{eq:C3_f_gcan}
\ee
in the limit $e^{\,\beta\mu} \to 0$ with fixed $T$ and $a$\,.
Here $\mu$ is the chemical potential, and $\beta=1/(k_B T)$.
Equation~(\ref{eq:C3_f}) follows from Eq.~(\ref{eq:C3_f_gcan}), because to leading-order in the non-degenerate limit, we can use the equation of state of the Boltzmann gas for each spin state,
  \be
  \frac{n}{2}\ \lambda_T^3 \ = \ e^{\,\beta\mu}\,.
  \label{eq:boltz_eos}
\ee
The expansion of \,$\Cr_3$ in Eq.~(\ref{eq:C3_f}) [resp.~Eq.~(\ref{eq:C3_f_gcan})] starts at order $n^3$ \big[resp.~$(e^{\beta\mu})^3$\,\big],
because $\Cr_3$ is a $3$-body observable.

For $a=\infty$\,, the three-body contact has a remarkably slow temperature dependence, $\Cr_3 \propto 1/T^{2-s} = 1/T^{0.227275733}$\,. This comes from the fact that the value of \,$s$\, for fermions at the unitary limit happens to be quite close to 2, which is the value corresponding to distinguishable non-interacting particles  ({\it i.e.}, roughly speaking, the bunching effect due to interactions is mostly compensated by the anti-bunching due to Pauli exclusion between the two same-spin fermions within the triplet).

\section{Wavefunction approach at the unitary limit}\label{waveful}

We set $\hbar=k_B=1$ from now on.
In this subsection, we consider the unitary limit $a = \infty$\,, and derive the analytical expression for the three-body contact in the non-degenerate regime given by Eqs.~(\ref{eq:C3_f}) and~(\ref{eq:f(0)}).
We first compute the three-body contact for a three-body scattering state (Sec.~\ref{sec:3b_scatt}),
and deduce the final result by thermally averaging over the incoming wavevectors (Sec.~\ref{sec:therm_avg}).\footnote{This approach is similar to the one used in Refs.~\cite{UnitaryBoseGas,PetrovWernerHetero} to compute the three-body loss rate $\Gamma_3$ for Bose gases or mixtures subject to the Efimov effect
(and for bosons, the calculation of \,$\Gamma_3$~\cite{UnitaryBoseGas} also yields $C_3$~\cite{BraatenC3unitary}
{\it via} the relation~\cite{WernerCastinRelationsBosons} between $\Gamma_3$ and~$C_3$\,).}

\subsection{Three-body scattering state} \label{sec:3b_scatt}

A three-body scattering state with zero center-of-mass momentum,
$\psi(\rr_1,\rr_2,\rr_3) = \psi(\RR)$\,, 
is defined by the following conditions.
\bi
\item It is an eigenstate of the three-body problem in free space,
{\it i.e.}\,, it satisfies
the Schr\"odinger equation
\be
-\,\frac{1}{m} \ \Delta_{\RR} \,\psi(\RR) \ =\  E \ \psi(\RR)
\label{eq:schro_3b}
\ee
together with the following contact condition:
There exists $\Ar$ such that
\be
\psi(\RR) \, \underset{r\to0}{=} \ \left( \frac{1}{r} - \frac{1}{a} \right)\ \Ar(\rrho) \, + \, O(r)\,.
\label{eq:cc_3b}
\ee
\item
  It is antisymmetric with respect to the same-spin fermionic particles 1 and 2,
\be
\psi(\rr_1,\rr_2,\rr_3) = - \psi(\rr_2,\rr_1,\rr_3).
\label{eq:antisym}
\ee

\item Its large-distance asymptotic behavior is
\be
\psi(\RR) \, \underset{R\to\infty}{\simeq} \ \psi^{(0)}(\RR) \, + \, \psi_{\rm sc}(\RR)
\label{eq:scatt_asym}
\ee
where
 $\psi_{\rm sc}(\RR)$ is a purely outgoing scattered wave,
and 
\be
\psi^{(0)}(\RR) \ = \ \frac{1-\hat{P}}{\sqrt{2}}\ e^{\,i \,\kk \cdot \RR} 
\label{eq:sym_eikR}
\ee
is a plane wave, anti-symmetrized thanks to the operator $\hat{P}$ that exchanges particles $1$ and $2$.
The relation between $k$ and the collision energy is
$E=k^2/m$.
\ei
The goal is to determine the $R\to0$ behavior of $\psi$.
The key step is to  expand the scattering state as
\be
\psi(\RR) \ = \ \sum_\nu \ \frac{F_\nu(R)}{R^2} \ \phi_\nu(\Oo)
\label{eq:sum_s}
\ee
where $\phi_\nu(\Oo)$ are the unitary hyperspherical wavefunctions,
{\it i.e.}\,, the hyperangular parts of the eigenstates of the unitary three-body problem,
which form an orthonormal basis for the scalar product $(f|g) \equiv \int d^5\Omega\ f^*(\Oo)\,g(\Oo)$\,,
see Appendix~A of Article~I (see also Refs.~\cite{Werner3corpsPRL,WernerThese}).
Hence
$F_\nu(R) / R^2 = (\phi_\nu | \psi) = \int d^5\Omega\ \phi_\nu^*(\Oo)\,\psi(\RR)$.

Each hyperradial wavefunction $F_\nu(R)$ solves the hyperradial Schr\"odinger equation
\be
\left( - \, \frac{d^2}{dR^2} - \frac{1}{R} \frac{d}{dR} + \frac{s^2}{R^2} \right)
\ F(R) \ = \ m\,
E\
F(R)
\label{eq:schro_R}
\ee
since we consider $a=\infty$.~\footnote{For finite scattering length, even though the expansion over the unitary hyperangular wavefunctions Eq.~(\ref{eq:sum_s}) formally still holds, the Hamiltonian does not act as the Laplacian on each term of the sum over $\nu$, since these terms do not belong to the domain of the Hamiltonian, defined by the finite-$a$ contact condition (see, {\it e.g.}, Ref.~\cite{LeChapitreIn2}).}
The three-body scattering problem is thereby reduced to a set of decoupled
effective one-body problems\footnote{The corresponding fictitious particle lives in 2D,
due to the term $R^{-1}\,(d/dR)$ in Eq.~(\ref{eq:schro_R}).
This results from the conventional factor $1/R^2$ in Eq.~(\ref{eq:sum_s}).
  Alternatively, one can consider $\tilde{F}_\nu(R) := F_\nu(R)/R^2$,
  which corresponds to a fictitious particle in 6D, in an effective potential $(s_\nu^2-4)/R^2$\,, which is attractive for $s_\nu=s=1.77\ldots$,
  directly explaining the bunching
  ($\psi \propto \tilde{F}_\nu(R)\to\infty$).}
in an effective potential $s_\nu^2/R^2$.

In the zero-range model, $F_{\nu}(R)$ is bounded for $R\to0$ (this corresponds to having no three-body resonance, see~\cite{LeChapitreIn2} and Refs. therein).
Hence $F_{\nu}(R)$ is
the Bessel function of the first kind
$J_{s_{\nu}}(k R)$,
up to a normalization factor. 
In the short-distance limit $R\to 0$,
$F_\nu(R)\propto R^{s_\nu}$.
The leading-order behavior is thus set by
the smallest $s_\nu$\,, which is $s_{l=1,n=0}$
in the present case of equal-mass fermions\footnote{As in Ref.~\cite{C3_I},
for each $l$, we denote by $\{s_{l,n}; n\in\mathbb{N} \}$ the allowed values of $s$, in increasing order ($s_{l,0} < s_{l,1} < \ldots$). 
In particular, the $s_{l=1,n}$ are the positive solutions
of Eq.~(\ref{eq:transc_s}).}.
We use the shorthand notations
$s:=s_{l=1,n=0}$\,,
$\phi_\md(\Oo) := \phi_{(l=1,m,n=0)}(\Oo)$\,,
and $\nu_\md := (l{=}1,\md,n{=}0)$\,.
The hyperradial wavefunction $F_{\nu_\md}(R)$ has the large-distance behavior
\be
F_{\nu_\md}(R) \ \underset{R\to\infty}{\simeq} \ 
\left(
A^{\rm in}_\md\ e^{- i k R}
+
A^{\rm out}_\md \ e^{\,i k R}
\,\right)
\ R^{-1/2}\,.
\label{eq:in_and_out_amps}
\ee
The coefficient $A^{\rm in}_\md$ of the incoming wave is determined by the incoming three-body plane-wave $\psi^{(0)}$ projected onto the unitary hyperspherical 
wavefunction $\phi_{\md}(\Oo)$.
Indeed, the overlap $\left(\phi_{\md} \, \big| \, \psi^{(0)} \,\right)$ behaves at large $R$ as
\be
\left( A^{{\rm in}\, (0)}_\md \, e^{-i k R} + A^{{\rm out}\, (0)}_\md \, e^{\,i k R} \,\right)
 \ R^{-5/2}
\nonumber\ee
with $A^{{\rm in}\, (0)}_\md = 8\, (\pi/k)^{5/2}\, e^{\,i\,5\pi/4}\ \phi_{\md}^*\!(-\kk/k)$,
as follows from the stationary phase method;
and $A^{\rm in}_\md$ has to be equal to $A^{{\rm in}\, (0)}_\md$,
in order for $\big(\phi_{\md}\,\big|\, \psi - \psi^{(0)}\,\big)$ to have a purely outgoing-wave large-distance behavior, as imposed by Eq.~(\ref{eq:scatt_asym}).

By matching this with the large-argument asymptotic behavior of the Bessel-$J$ function,
the normalization factor $F_{\nu_\md}(R) / J_s(k R)$ is determined to be
$A_\md^{{\rm in}\,(0)}\,\sqrt{2 k \pi}\,e^{-i(s\pi/2+\pi/4)}$. Using the small-argument asymptotic behavior of the Bessel-$J$ function then yields
\be
\psi(\RR) \ \underset{R\to0}{\sim} \ R^{\,s\,-2}\,\sum_{\md=-1}^1 \phi_\md(\Oo)\,B_\md(\,\kk\,)
\label{eq:psi_k_R0}
\ee
where
\be
B_\md(\,\kk\,) \ = \ - \, \frac{2^{7/2-s} \pi^3}{\Gamma(s+1)}\ e^{-i\pi s/2}\ k^{\,s\,-2}\ \phi_\md^*\big(-\hat{k}\ \big)
\label{eq:Bmk}
\ee
with \,$\hat{k} = \kk/k$.

\subsection{Thermal averaging} \label{sec:therm_avg}

We turn to the homogeneous gas. The physical intuition is that in the non-degenerate limit, 
the short-distance three-body correlations are determined by the three-body problem, the effect of additional
many-body correlations being negligible to leading order. To formalize this, let us introduce a cubic box $\Br$ of linear size $L$, large compared to the thermal wavelength but small compared to the typical interparticle distance:
\be
\lambda_T \ \ll \ L \ \ll \ n^{-1/3}
\nonumber\ee
which is possible in the non-degenerate regime $n \lambda_T^3 \,{\ll}\, 1$.
Let $N_{2,1}(\epsilon)$ stand for the number of triplets of particles with spins $\uparrow\uparrow\down$ and hyperradius $R<\epsilon$,
in a measurement of the positions of all the particles in this box $\Br$.
Using Eq.~(\ref{eq:N3}) with $\Vr'=L^3$,
and with $N_{2,1}(\epsilon) = N_3(\epsilon) / 2$
since we consider the unpolarized gas,
we get
\be
N_{2,1}(\epsilon) \ \underset{\epsilon\to0}{\sim} \ 
\Cr_3
\ \epsilon^{\,2s\,+\,2}\ \frac{L^3}{2}\,.
\label{eq:N21_C3}
\ee
We can thus determine $\Cr_3$
by computing $N_{2,1}(\epsilon)$
in the  $\epsilon\to0$ limit ({\it i.e.}\,, for $\epsilon \ll \lambda_T$).
We use the grand-canonical ensemble to describe the open system delimited by the box $\Br$.
Thanks to the condition $L \gg \lambda$, we can use periodic boundary conditions in the box $\Br$.
Denoting by $|i\,\ra$ and $E_i$ the 
eigenstates and energies of the unitary three-body problem in the box with two $\uparrow$ and one $\down$  particles, we have
\be
N_{2,1}(\epsilon) \ \simeq \ \frac{1}{Z} \ \sum_i e^{-\beta \,(E_i {\,-\, 3 \mu})}\ \la \, i \, | \, \hat{N}_{2,1}(\epsilon)\, | \, i \, \ra
\nonumber
\ee
where $Z$ is the grand partition function for the box $\Br$ and 
\begin{equation}
\hat{N}_{2,1}(\epsilon) \ := \ \int_{\Br,\ R < \epsilon} d^3\!r_1\,d^3\!r_2\,d^3\!r_3\ \, \hat{\psi}^\dagger_\uparrow(\rr_1)\,\hat{\psi}^\dagger_\uparrow(\rr_2) \, \hat{\psi}^\dagger_\down(\rr_3)\,
\hat{\psi}_\downarrow(\rr_3) \, \hat{\psi}_\uparrow(\rr_2) \, \hat{\psi}_\uparrow(\rr_1) 
\nonumber
\end{equation}
is the operator counting the number of triplets of particles  of hyperradius smaller than $\epsilon$ (and with spins $\uparrow\uparrow\down$).
We restricted the sum to states $|i\,\ra$ with
two $\uparrow$ and one $\down$  particles,
because states with less particles do not contribute to $N_{2,1}(\epsilon)$, while states with more particles have a negligible probability due to the condition $L \ll n^{-1/3}$
[their contribution is of higher order in fugacity, $O(e^{4\beta\mu})$\,].
The condition \,$L \ll n^{-1/3}$\, also implies that $Z$ is dominated by the zero-particle-number contribution, $Z\simeq1$.

Next, based on the condition $\lambda_T \ll L$, we replace the eigenstates in the box $\la \rr_1,\rr_2,\rr_3|\,i\,\ra$ by $D \ e^{\,i\, \KK \cdot \CC}\ \psi_{\kk}(\RR)$,
where $\psi_{\kk}$ is the three-body scattering state introduced in Sec.~\ref{sec:3b_scatt} (making explicit the dependence on the incoming wavevector $\kk$) and $D $ is a normalization factor.
The index $i$ and the wavevectors $(\,\KK,\kk)$ are linked as follows:
Identifying $i$ with a triplet of single-particle wavevectors $(\,\kk_1,\kk_2,\kk_3)$,
we have $\kk_1\cdot\rr_1 + \kk_2\cdot \rr_2 + \kk_3 \cdot \rr_3 = \KK\cdot \CC + \kk \cdot \RR$ [which is consistent with Eqs.~(\ref{eq:def_KK}) and~(\ref{eq:def_kk})].
We have the restriction $\kk_1 \neq \kk_2$ in order for
the incoming plane wave $\psi^{(0)}$ defined by (\ref{eq:sym_eikR}) to be non-zero.
One has $|D|^2  \simeq L^{-9}$ in the limit $\lambda_T \ll L$,
because the normalization integral $\int d^3r_1\,d^3r_2\,d^3r_3 \ \big|\la \rr_1,\rr_2,\rr_3|\,i\,\ra\big|^2$ is dominated by the contribution of $\psi^{(0)}$,
and we get
$\int d^3r_1\,d^3r_2\,d^3r_3 \ | \psi^{(0)} |^2 = L^9$
using the fact that $\kk_1 \neq \kk_2$.
Moreover, using again the condition $\lambda_T \ll L$,
we can replace the energy $E_i$ by the non-interacting energy $\big(k_1^2+k_2^2+k_3^2\big)/(2m)$.
This yields
\be
N_{2,1}(\epsilon)
\ \simeq \
\frac{1}{2}\,\frac{e^{3\beta\mu}}{L^9}
\ \sum_{\kk_1 \neq \kk_2} \ \sum_{\kk_3} \
e^{- \frac{\beta}{2m}\,(k_1^2+k_2^2+k_3^2)}
\ \la\psi_\kk | \hat{N}_{2,1}(\epsilon) | \psi_\kk \ra
\nonumber
\ee
where the factor $1/2$ comes from the fact that exchanging the wavevectors $\kk_1$ and $\kk_2$ of the indistinguishable same-spin particles leads to the same physical state (the same wavefunction $|i\,\ra$ up to a global sign).
Then, since $L \gg \lambda_T$, we can replace the sums by integrals,
$L^{-3} \sum_{\kk_i} \ \longrightarrow \  (2\pi)^{-3}\int d^3\!k_i$\,.
Also using the Boltzmann-gas equation of state Eq.~(\ref{eq:boltz_eos}), {\it i.e.} $e^{\,\beta\mu} = \frac{n}{2} \ / \int \frac{d^3 k}{(2\pi)^3}\ e^{-\frac{\beta}{2m}k^2}$\,,
we obtain
  \be
N_{2,1}(\epsilon) \ \simeq\ 
\frac{n^3}{16}\ \ 
\frac{\int d^3\!k_1\, d^3\!k_2\, d^3\!k_3\ \ e^{-\frac{\beta}{2m}\left(k_1^2+ k_2^2+k_3^2\right)}\ \la\psi_\kk | \hat{N}_{2,1}(\epsilon) | \psi_\kk \ra}{\int d^3\!k_1\, d^3\!k_2\, d^3\!k_3\ \ e^{-\frac{\beta}{2m}\left(k_1^2+ k_2^2+k_3^2\right)}}\,.
  \nn
  \ee
Changing variables $(\,\kk_1,\kk_2,\kk_3) \longrightarrow (\,\KK,\kk)$,
{and using the expression of the energy in terms of the new variables
$\frac{1}{2m}(k_1^2+ k_2^2+k_3^2) = \frac{K^2}{6m} + \frac{k^2}{m}$
\big[which follows from the identity $\frac{1}{2m}(\Delta_{\rr_1} + \Delta_{\rr_2} + \Delta_{\rr_3}) = \frac{\Delta_{\CC}}{6m} + \frac{\Delta_{\RR}}{m}$\,\big]\,,
we get}
\be
N_{2,1}(\epsilon) \ \simeq\
\frac{n^3}{16}\ \ 
\frac{\int d^6\!k\ e^{-\beta k^2/m}\ \la\psi_\kk | \hat{N}_{2,1}(\epsilon) | \psi_\kk \ra}{\int d^6\!k\ e^{-\beta k^2/m}}\,.
\label{eq:N3_int_k}
\ee
At small $\epsilon$, the integrand behaves as
\be
\la\psi_\kk | \hat{N}_{2,1}(\epsilon) | \psi_\kk \ra
\ \underset{\epsilon\to0}{\sim} \ C_{2,1}(\,\kk\,)\ \epsilon^{\,2s\,+\,2}
\label{eq:integrand_eps0}
\ee
with
\be
C_{2,1}(\,\kk\,) \ := \
L^3 \ \frac{
3^{3/2}
}{16\,(s+1)}
\ \sum_{\md=-1}^1 |B_\md(\,\kk\,)|^2,
\nonumber
\ee
as follows from the short-distance behavior Eq.~(\ref{eq:psi_k_R0}) of $\psi=\psi_\kk$\,.
Here, the factor 
$L^3$ comes from
the integral of the center-of-mass $\CC$ over the box, the Jacobian contributes a factor $|\partial(\rr_1,\rr_2,\rr_3)/\partial(\rr,\rrho,\CC)| = 3^{3/2}\, /\, 8$\,,
and we used $(\phi_{\md}|\phi_{\md'}) = \delta_{\md,\md'}$.
Then, injecting Eq.~(\ref{eq:integrand_eps0}) into Eq.~(\ref{eq:N3_int_k}) yields
\be
N_{2,1}(\epsilon) \underset{\epsilon\to0}{\sim} C_{2,1}(T)\ \epsilon^{\,2s\,+\,2},
\label{eq:N21_C21}
\ee
with $C_{2,1}(T)$ proportional to the thermal average of $C_{2,1}(\,\kk\,)$,
\be
C_{2,1}(T) \ = \ 
\frac{n^3}{16}\ \,
 \frac{\int d^6\!k\ e^{-\beta k^2/m}\ \ C_{2,1}(\,\kk\,)}{\int d^6\!k\ e^{-\beta k^2/m}}\,.
\nonumber\ee
Substituting the expression of $B_\md(\,\kk\,)$ given in Eq.~(\ref{eq:Bmk}),
the numerator can be computed analytically:
Using Eq.~(\ref{eq:def_dhk}),
we get the hyperangular integral $\int \dhk\ \left|\phi_\md\left(-\hat{k}\ \right)\right|^2=1$,
leading to
\be
C_{2,1}(T) \ = \ L^3\ 
n^3\ \frac{
3^{5/2}
  \ \pi^3}{2^{2s+2}\,\Gamma(s+2)}\ (m\,k_B\,T)^{s\,-\,2}.
\nonumber\ee
We have \,$\Cr_3 = 2\,C_{2,1}(T) / L^3$
from Eqs.~(\ref{eq:N21_C3},\ref{eq:N21_C21}),
which yields the final analytical result (\ref{eq:C3_f},\ref{eq:f(0)}) for $\Cr_3$\,,
after restoring the factors $\hbar$ and $k_B$ by dimensional analysis.
 The same approach is applicable to compute the $l=0$ three-body contact, with the result given in Appendix~\ref{app:C3_l0}.

\section{Diagrammatic approach}\label{secFeyndiag}

In this section, we 
show how the three-body contact appears within the Feynman diagram technique,
and extend to arbitrary negative scattering lengths
the calculation of $\Cr_3$ in the non-degenerate limit.
The main message of this section is that we recover the
power law scaling
of $g_3$
with the correct exponent, see Eq.~(\ref{eq:g3_C3}),
and therefore we can calculate
the prefactor
$\mathcal{C}_3$\,.
Moreover,  we recover numerically the analytic result
Eq.~(\ref{eq:f(0)})
for $\mathcal{C}_3$ at unitarity. 
A key ingredient of our analysis is the high-momentum power-law scaling of the $3$-body T-matrix, given in Eq.~(\ref{eqimt31}).\footnote{For bosons, a diagrammatic computation of the third virial coefficient, and of $C_3$ by numerical differentiation
w.r.t. the three-body parameter, was
performed in Ref.~\cite{BarthHofmannEfimov}. This approach is not directly applicable in the case of mass-balanced fermions considered here, because
a simple zero-range model with two parameters (scattering length and three-body parameter) is not well defined~\cite{C3_I,PricoupenkoNbodyRes}.}

\subsection{Three-body correlator and three-body propagator}

\subsubsection{Basic definitions and identities}

The three-body propagator in position imaginary-time representation is defined by
\begin{equation}
  \mathcal{G}_3({\bf r}_1,{\bf r}_2,{\bf r}_3;\tau)
  \ = \ - \, \bigg\langle T_{\tau}\big[
\,\hat{\psi}_\down(\rr_3,\tau)\,\hat{\psi}_\UP(\rr_2,\tau)\,\hat{\psi}_\UP(\rr_1,\tau)
\ \hat{\psi}_\UP^\dagger(\rr_1,0)\,\hat{\psi}_\UP^\dagger(\rr_2,0)\,\hat{\psi}_\down^\dagger(\rr_3,0)
\,\big]\ \bigg\rangle\nn
\end{equation}
{for $\tau \in ]-\beta , \beta [$\,.}
  Here, 
  $T_{\tau}$ is the imaginary-time-ordering operator,
  and
  \,$\hat{\psi}_\sigma(\rr,\tau) = e^{\tau H}\ \hat{\psi}_\sigma(\rr)\ e^{-\tau H}$\,
  with $H$ the grand-canonical hamiltonian.
  The three-body correlation function is given by the three-body propagator at $\tau=0^-$,
\be
g_{3}({\bf r}_1,{\bf r}_2,{\bf r}_3) \ = \ \mathcal{G}_3({\bf r}_1,{\bf r}_2,{\bf r}_3;0^{-})
\nn\ee
Equivalently, by the anti-periodicity property
\,$\mathcal{G}_3(\tau) \, = \, - \,\mathcal{G}_3(\beta+\tau)$\,, we have
\be
g_{3}({\bf r}_1,{\bf r}_2,{\bf r}_3) \ = \ -\mathcal{G}_3({\bf r}_1,{\bf r}_2,{\bf r}_3;\beta^{-})
\nn\ee
In momentum imaginary-time representation,
we define the three-body propagator
\begin{equation}
\mathcal{G}_3(\,{\bf k}_1,{\bf k}_2,{\bf k}_3\,;\,{\bf k}'_1,{\bf k}'_2,{\bf k}'_3\,;\,\tau)
 \ = \ - \,
\Big\langle T_{\tau}\big[
\,  \hat{c}^{}_{{\bf k}_3\downarrow}(\tau)\
  \hat{c}^{}_{{\bf k}_2\uparrow}(\tau)\
  \hat{c}^{}_{{\bf k}_1\uparrow}(\tau)\
  \hat{c}^{\,\dagger}_{{\bf k}^{'}_1\uparrow}(0) \
  \hat{c}^{\,\dagger}_{{\bf k}^{'}_2\uparrow}(0)\
  \hat{c}^{\,\dagger}_{{\bf k}^{'}_3\downarrow}(0)
\, \big]\ \Big\rangle\nonumber
\end{equation}
From
the relation
\,$\hat{\psi}_{\sigma}({\bf r})=\sum_{{\bf k}}\ e^{\,i\,{\bf k}\cdot{\bf r}}\ \hat{c}_{{\bf k}\sigma} / \sqrt{\Vr}$\,
between annihilation operators in position and momentum representation,
we have
\bea
g_{3}({\bf r}_1,{\bf r}_2,{\bf r}_3)&=&
-
\,\frac{1}{\Vr^3}
\sum_{\{{\bf k}_j,{\bf k}'_j\}}
\ \exp\left(
i \sum\limits_{j=1}^3 (\,{\bf k}_j-{\bf k}'_j\,)\cdot {\bf r}_j
\right)
\ \, \mathcal{G}_3({\bf K},{\bf K}';\beta)
\label{eqg3G3k}
\eea
Here and in what follows, we use the notations ${\bf K}=\big({\bf k}_1,{\bf k}_2\,,{\bf k}_3\big)$ and 
${\bf K}'=\big({\bf k}'_1,{\bf k}'_2,{\bf k}'_3\big)$\,,
as well as the convention
$\mathcal{G}_3({\bf K},{\bf K}';\beta) \, := \, \mathcal{G}_3({\bf K},{\bf K}';\beta^-)$\,.
    By conservation of the total momentum,
in order for $\mathcal{G}_3({\bf K},{\bf K}';\beta)$ to be non-zero, it is necessary to have
    ${\bf k}_1 + {\bf k}_2 + {\bf k}_3 = {\bf k}'_1 + {\bf k}'_2 + {\bf k}'_3$\,,
    and we will implicitly assume that this condition holds
    in what follows.
    This leaves us with five independent wavevectors: ${\bf k}_1$\,, ${\bf k}_2$\,, ${\bf k}'_1$\,, ${\bf k}'_2$\,, and the total momentum
\be
   {\bf P} \ = \ {\bf k}_1 + {\bf k}_2 + {\bf k}_3  \ = \  {\bf k}'_1 + {\bf k}'_2 + {\bf k}'_3
   \label{eq:defP}
\ee
We can thus rewrite Eq.~(\ref{eqg3G3k}) as
\bea
g_{3}({\bf r}_1,{\bf r}_2,{\bf r}_3) \ &=& \ 
-\,
\frac{1}{\Vr^3}
\ \sum_{{\bf k}_1, {\bf k}_2 \, , \, {\bf k}'_1, \, {\bf k}'_2 \, , \, {\bf P}}
\ \, 
e^{\,i
\phi(\,{\bf k}_1, \, {\bf k}_2\,, \, {\bf k}'_1, \, {\bf k}'_2 \,)}
\ \, \mathcal{G}_3({\bf K},{\bf K}';\beta)
\label{eqg3G3k2}
\eea
with
\be
\phi({\bf k}_1, {\bf k}_2, {\bf k}'_1, {\bf k}'_2) \ = \ ({\bf k}_1- {\bf k}'_1)\cdot({\bf r}_1-{\bf r}_3)
\ + \ ({\bf k}_2-{\bf k}'_2)\cdot({\bf r}_2-{\bf r}_3)
\nn\ee

It will prove convenient to work with the relative momenta
\be
    {\bf q} \  = \ {\bf k}_2 \, - \, \frac{{\bf P}}{3}
    \ ,
    \ \ \ \ \ \ 
        {\bf p} \  =  \ \frac{{\bf k}_3 \, - \,  {\bf k}_1}{2}
        \label{eqdefpq}
\ee
{\it i.e.}\,, ${\bf q}$ is the momentum of particle $2$ in the center of mass reference frame of the $3$ particles, and ${\bf p}$ is the momentum of the relative motion between particles $1$ and $3$.
We use similar definitions for the primed wavevectors,
\bea
    {\bf q}' \  = \ {\bf k}'_2 \ - \ \frac{{\bf P}}{3}
    \ ,
    \ \ \ \ \ \ 
        {\bf p}' \  =  \ \frac{{\bf k}'_3 \, - \,  {\bf k}'_1}{2}
\label{eqdefpq'}
\eea
The phase $\phi$ can be expressed in terms of these relative momenta and of the Jacobi coordinates $(\rr, \boldsymbol{\rho})$ defined in Eq.~(\ref{eq:def_jaco}),
 \be
 \phi(\,{\bf k}_1, {\bf k}_2, {\bf k}'_1, {\bf k}'_2 \, ) \ = \ \varphi({\bf p},{\bf q})
 \ - \ \varphi({\bf p}',{\bf q}')
\ \ \ \ {\rm  with} \ \ \ \
\varphi({\bf p},{\bf q}) \ = \ {\bf p}\cdot {\bf r} \ + \ \tfrac{\sqrt{3}}{2} \ {\bf q}\cdot \boldsymbol{ \rho} 
 \nn
 \ee

\subsubsection{Permutations}\label{subsectionpermutation}

This technical subsection will be useful to avoid computing separately the contributions of diagrams differing by permutations of external $\UP$ propagator lines.
Let $({\bf r}' , \boldsymbol{\rho}'\,)$ be
the Jacobi coordinates obtained from the original Jacobi coordinates $({\bf r},\boldsymbol{\rho})$ by exchanging the positions ${\bf r}_1$ and ${\bf r}_2$\, of the $\uparrow$ fermions, {\it i.e.},
\bea
{\bf r}' \ &=& \ \tfrac{1}{2}\ {\bf r} \ - \ \tfrac{\sqrt{3}}{2}\ \boldsymbol{\rho}\nn\\
\boldsymbol{\rho}' \ &=&  \ - \, \tfrac{\sqrt{3}}{2}\ {\bf r} \ - \ \tfrac{1}{2}\ \boldsymbol{\rho}
\nn\eea
To determine the effect of permutation of momenta of the $\uparrow$ fermions, we consider the operator $\,\Pr$ which exchanges ${\bf k}_1$ and ${\bf k}_2$\,,
and $\,\Pr'$ which exchanges ${\bf k}'_1$ and ${\bf k}'_2$\,.
In other words,
for any function $f$ of $\,{\bf k}_1 \,,\,{\bf k}_2 \,,\,{\bf k}'_1$ and ${\bf k}'_2$\,, 
\bea
\Pr f({\bf k}_1,\,{\bf k}_2,\,{\bf k}'_1,\,{\bf k}'_2)& \ := \ &f({\bf k}_2,\,{\bf k}_1,\,{\bf k}'_1,\,{\bf k}'_2)\nn\\
\Pr' f({\bf k}_1,\,{\bf k}_2,\,{\bf k}'_1,\,{\bf k}'_2)& \ := \ &f({\bf k}_1,\,{\bf k}_2,\,{\bf k}'_2,\,{\bf k}'_1)
\nn\eea
Due to the anticommutation properties of the fermionic creation and annihilation operators, we have
$\Pr \, \mathcal{G}_3({\bf K},{\bf K}';\beta) =-\mathcal{G}_3({\bf K},{\bf K}';\beta)$ and 
$\Pr' \, \mathcal{G}_3({\bf K},{\bf K}';\beta)=-\mathcal{G}_3({\bf K},{\bf K}';\beta)$.
This yields the identity
\be
\mathcal{G}_3({\bf K},{\bf K}';\beta) \ =  \ \frac{1}{4}\ \big(1-\Pr-\Pr'+\Pr\Pr'\,\big)\ \mathcal{G}_3({\bf K},{\bf K}';\beta)\,.
\nn\ee
Let us inject this identity into Eq.~(\ref{eqg3G3k2}),
and make changes of variables such that the permutations now act on the phases. For instance, the permutation $\Pr$, which exchanges ${\bf k}_1$ and ${\bf k}_2$\,, changes the phase $\varphi({\bf p},{\bf q})$ into
\be
   {\varphi}'({\bf p},{\bf q}) \ = \ {\bf p}\cdot {\bf r}' \ + \ \tfrac{\sqrt{3}}{2} \ {\bf q}\cdot \boldsymbol{ \rho}'
   \nn
   \ee
which amounts to exchange the positions ${\bf r}_1$ and ${\bf r}_2$\,, as it should be.
In this way, we obtain
\begin{multline}
  g_{3}({\bf r}_1,{\bf r}_2,{\bf r}_3) \ = \ - \, \dfrac{1}{4\,\Vr^3}
\ \   \ \sum_{{\bf k}_1, \, {\bf k}_{2}, \, {\bf k}'_1, \, {\bf k}'_2, \, {\bf P}}
\ \ \ \left(
 e^{\,i \varphi({\bf p},{\bf q})}
 -
  e^{\,i \varphi'({\bf p},{\bf q})}
\right)
\
\mathcal{G}_3({\bf K},{\bf K}';\beta)
\ 
\left(
e^{-i\varphi({\bf p}',{\bf q}')}
-e^{-i \varphi'({\bf p}',{\bf q}')}
\right)
\label{eqg3symm}
\end{multline}
This is the expression we shall use in our high-temperature expansion. 
We can already note that this expression only depends on
the Jacobi coordinates ${\bf r}$ and $\boldsymbol{ \rho}$\,,
and not on the center of mass coordinate ${\bf C}$\,,
in agreement with translational invariance.
The reason why Eq.~(\ref{eqg3symm}) will be useful is the following property.

\vskip.2cm
\noindent {\bf Property 1.}
{\it Let
\,$\mathcal{G}_3^{(a)}$ , 
$\mathcal{G}_3^{(b)}$ , 
$\mathcal{G}_3^{(c)}$ , 
$\mathcal{G}_3^{(d)}$
be four contributions to $\,\mathcal{G}_3$ such that
\be
\mathcal{G}_3^{(b)} \ = \ -\Pr\,\mathcal{G}_3^{(a)}\,, \
\ \mathcal{G}_3^{(c)} \ = \ -\Pr'\,\mathcal{G}_3^{(a)}\,, \
\ \mathcal{G}_3^{(d)} \ = \ \Pr\,\Pr'\,\mathcal{G}_3^{(a)}\,.
\nn\ee
Let
$g_3^{(a)}$ be the contribution to $g_3$ obtained by replacing $\,\mathcal{G}_3$ by $\,\mathcal{G}_3^{(a)}$ in Eq.~(\ref{eqg3symm}).
Similarly, let 
$g_3^{(b)}$\,,
$g_3^{(c)}$ and
$\,g_3^{(d)}$ be the respective contributions to $g_3$ obtained by replacing 
$\,\mathcal{G}_3$ by $\,\mathcal{G}_3^{(b)}$ , 
$\mathcal{G}_3^{(c)}$ and 
$\,\mathcal{G}_3^{(d)}$
in Eq.~(\ref{eqg3symm}).
Then we have $g_3^{(a)} = g_3^{(b)} = g_3^{(c)} = g_3^{(d)}$.}

\vskip.1cm

To derive this property,
we define
$A({\bf k}_1,{\bf k}_2;{\bf P})=e^{\,i \varphi({\bf p},{\bf q})}-e^{\,i\varphi'({\bf p},{\bf q})}$,
and $A'$ the corresponding phase factor for primed variables in Eq.~(\ref{eqg3symm}).
We can write schematically
$g_3^{(b)} = \sum
A()\mathcal{G}_3^{(b)}()A'()=\sum A()(-\Pr\,\mathcal{G}_3^{(a)}())A'()$.
Using again the change of variables which permutes ${\bf k}_1$ and ${\bf k}_2$\,, this is equal to
  $\sum(-\Pr\,A())\mathcal{G}_3^{(b)}()A'()$,
and thus to $\sum A()\mathcal{G}_3^{(b)}()A'()$
since
$A$ is antisymmetric ($\Pr\,A=-A$).
Hence $g_3^{(b)} = g_3^{(a)}$.
The effect of the action of $-\Pr'$ is derived in the same way,
and the effect of the action of $\,\Pr\,\Pr'$ immediately follows.

\subsection{Virial expansion of $\ \mathcal{G}_3$}

\begin{figure}
\includegraphics[width=0.9\columnwidth]{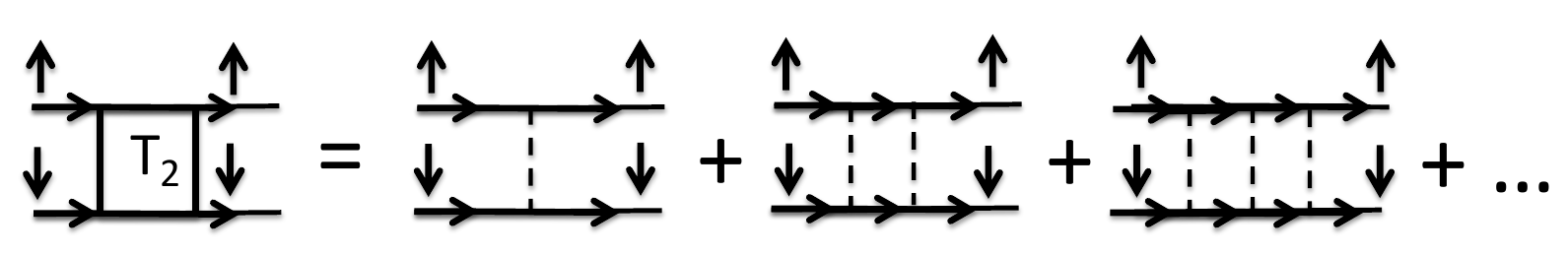}
\caption{The two-body T-matrix $T_2$ is given by a sum of ladder diagrams built with the bare interaction vertex.
  \label{fig:T2_serie}}
\end{figure}

\begin{figure}
\includegraphics[width=0.8\columnwidth, trim={0 1.5cm 0 1cm}, clip]{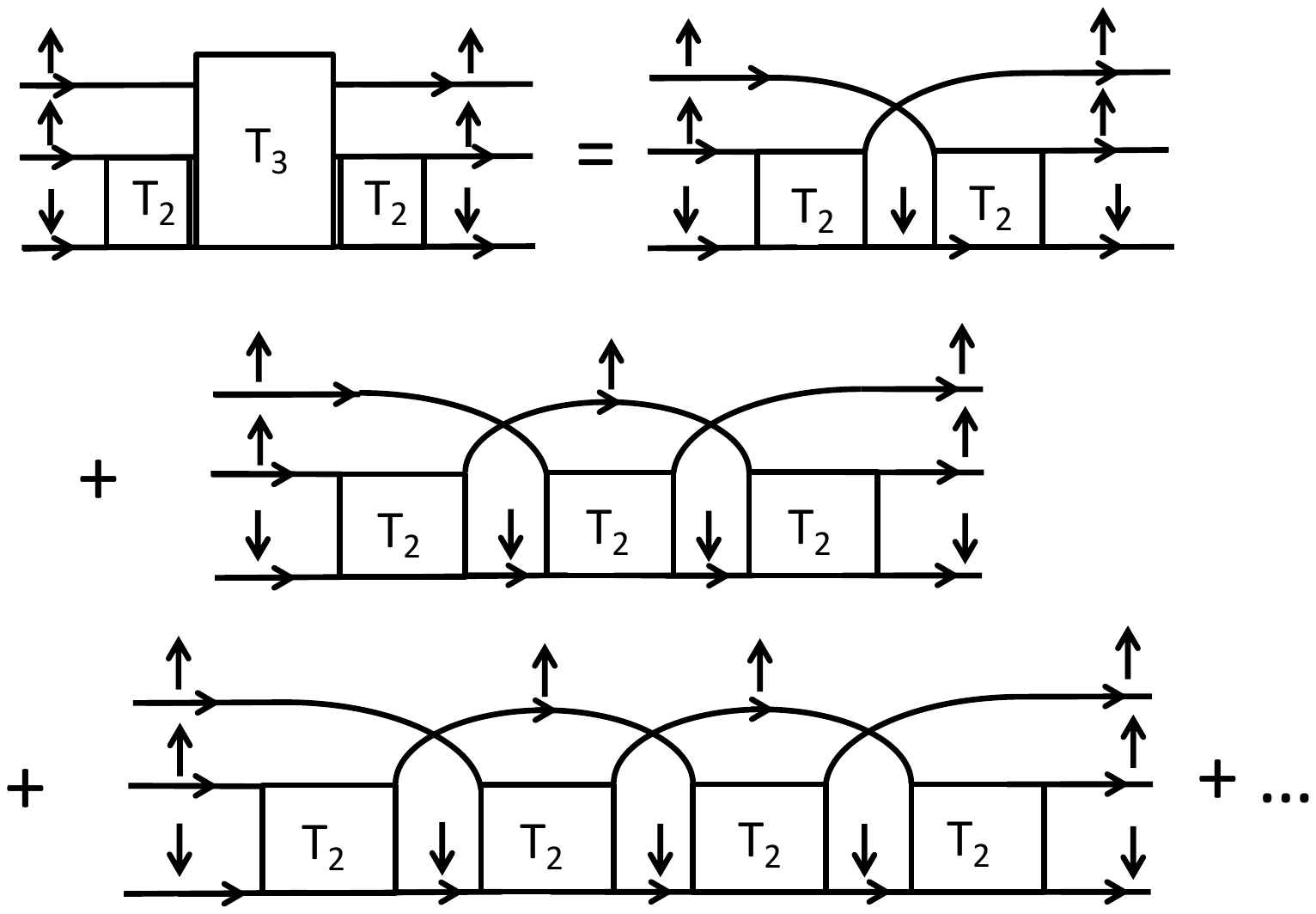}
  \caption{The three-body T-matrix $T_3$ is given by an infinite sum of diagrams built with the two-body T-matrix $T_2$.
  \label{fig:T3_serie}}
\end{figure}

We follow the method of Refs.\cite{Leyronas_b3,SunVirial3}  (see also \cite{BarthHofmannEfimov,SunCuiEfimovCorrelPRL,SunCuiEfimovCorrelPRA} for extensions to bosons) in order to obtain the high temperature expansion of \,$\mathcal{G}_3({\bf K},{\bf K}';\beta)$ in powers of the fugacity
\be
z \ = \ e^{\,\beta\mu}
\nn
\ee
In this approach, one uses the Feynman diagram technique and expands the
free particle propagators
in powers of $z$. At lowest order in $z$, diagrams contain only vacuum propagators, which are retarded: Their origin is to the left of their destination, with the convention that the imaginary time increases from left to right. They physically represent the imaginary time propagation of a particle in vacuum. $\mathcal{G}_3({\bf K},{\bf K}';\beta)$ is the propagator of two spin~$\uparrow$ and one spin~$\downarrow$ particles in vacuum, from time $\tau=0$ to time \,$\tau=\beta$. Each fermionic line is proportional to the fugacity $z$.
To leading order,
$\mathcal{G}_3({\bf K},{\bf K}';\beta)$
  is of order $z^3$ 
  and is a sum of Feynman diagrams which we classify into three groups:
\begin{itemize}
\item the diagrams without any interaction (section~\ref{AppG31})
\item the diagrams with one $T_2$  (section~\ref{AppG32})
\item the diagrams with one $T_3$
  (section~\ref{AppG33}).
\end{itemize}
Here, $T_2$ is the two-body T-matrix, see Fig.~\ref{fig:T2_serie},
while $T_3$ is the three-body T-matrix,
see Fig.~\ref{fig:T3_serie}.\footnote{Since we consider contact interactions, $T_2$ can be viewed as a line, connecting the common destination of two ingoing single-particle propagator lines  with the common origin of two outgoing single-particle propagator lines. Accordingly, $T_3$ is a four-point object, with two external single-particle lines (one ingoing and one outgoing), and two external $T_2$ lines (one ingoing and one outgoing).}
 The three-body and two-body T-matrices in the center-of-mass frame, $t_3$ and $t_2$\,, are related through the
  Skorniakov--Ter-Martirosian (STM) equation, see Appendix~\ref{app:STM}.
Note that the diagrams with one $T_3$ are also the diagrams with two or more $T_2$'s, so that our classification is also a classification according to the number of $T_2$'s.\footnote{Diagrams containing slashed
(with the notation of~\cite{Leyronas_b3})
single-particle lines propagating backwards in time are of order $z^4$ or higher, so that they are negligible in the non-degenerate limit.
Furthermore we expect their contribution to $g_3$ to be negligible even if the limit $R\to0$ is taken before the non-degenerate limit\,;
for example the diagrams with one four-body T-matrix $T_4$ closed by one slashed single-particle line should give rise to a
small-$R$ singular term
$\propto \! R^{\,2s^{(4)}-4}$,
which is negligible compared to the leading three-body term $\propto \! R^{\,2s-4}$,
since we have~\cite{DailyBlume_4body_spectrum} $s^{(4)}=2.509(1)>s$.}

\subsubsection{No interaction: \,$\mathcal{G}_3^{(1a)}$,\,$\mathcal{G}_3^{(1b)}$}\label{AppG31}
\begin{figure}[h]
\subfigure[\label{fig1a}]{\includegraphics[width=0.49\linewidth]{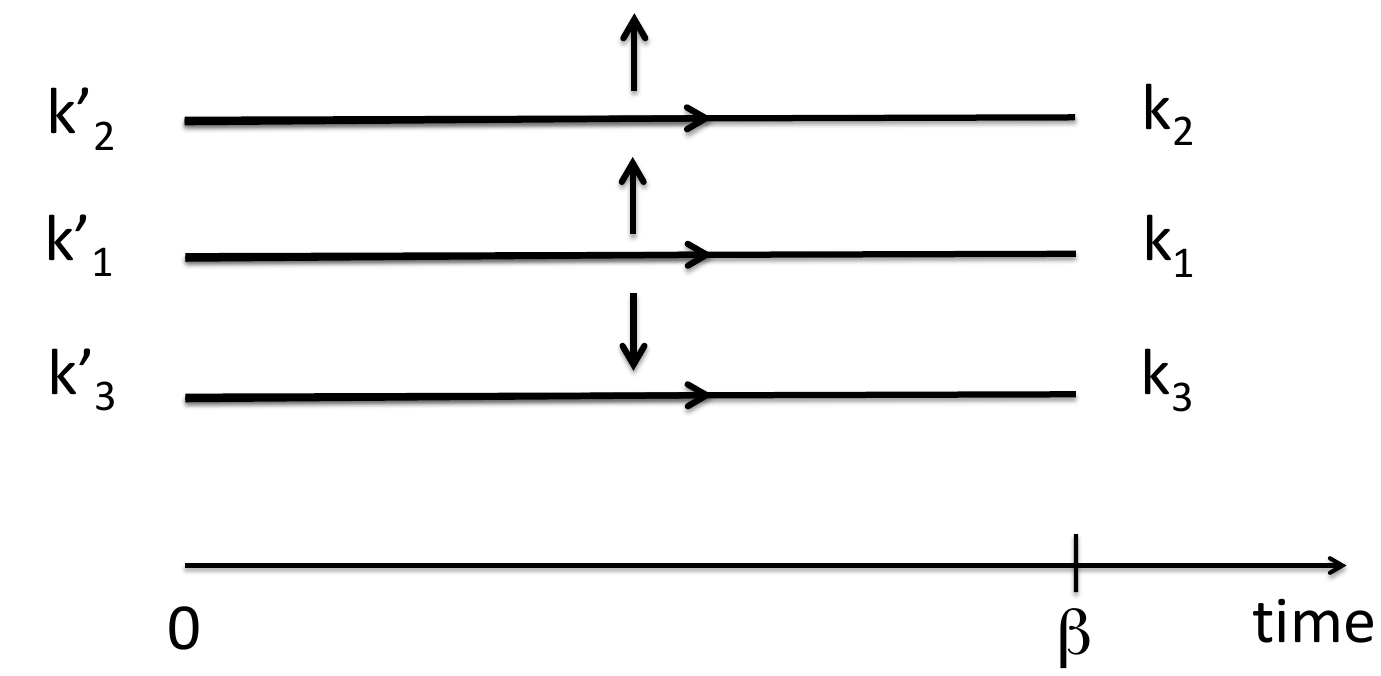}}
\subfigure[\label{fig1b}]{\includegraphics[width=0.49\linewidth]{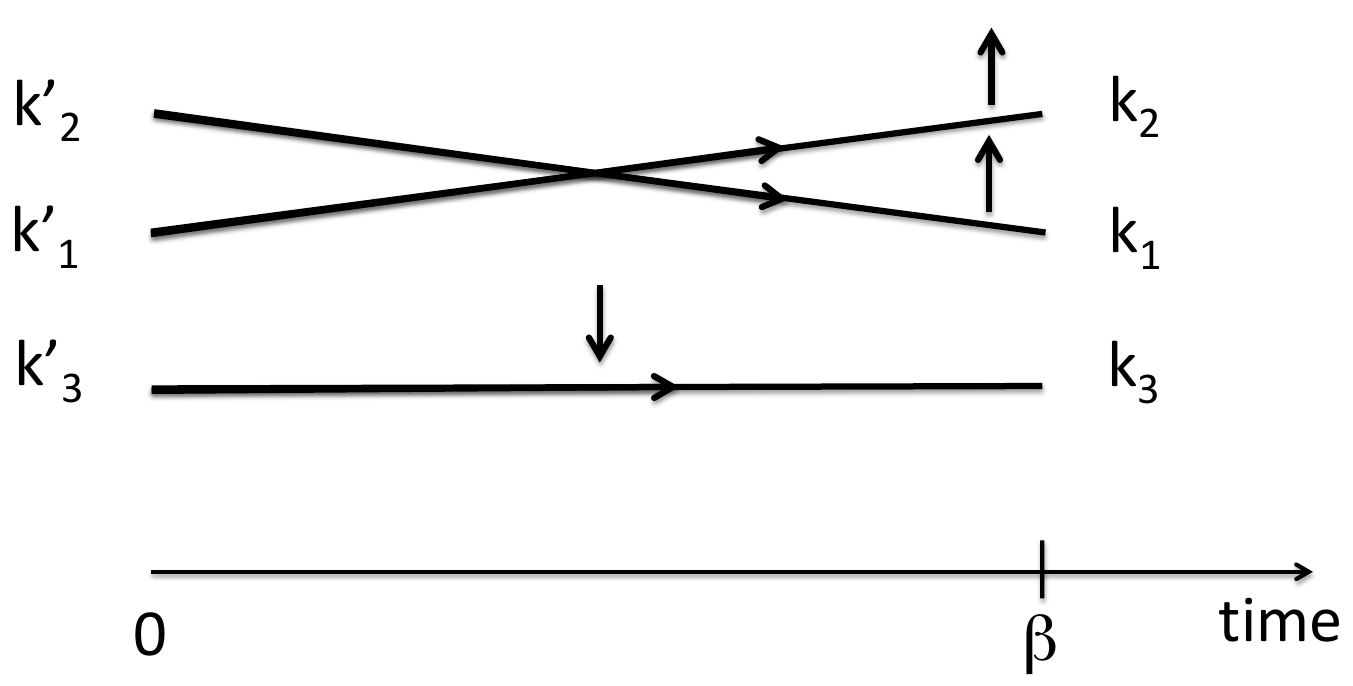}}
\caption{The two diagrams which do not contain any interaction vertex, $\mathcal{G}_3^{(1a)}$ and $\mathcal{G}_3^{(1b)}$\,.
\label{fig:G31}}
\end{figure}

The two diagrams which do not contain any interaction are shown in Fig.~\ref{fig1a} and \ref{fig1b}.
Their contributions to $\mathcal{G}_{3}$ are denoted by
\,$\mathcal{G}_3^{(1a)}$\, and \,$\mathcal{G}_3^{(1b)}$\, respectively.

$\mathcal{G}_3^{(1b)}$ is obtained from $\mathcal{G}_3^{(1a)}$ by permutation of the variables ${\bf k}_1$ and ${\bf k}_2$\,.
    Due to the fermionic character of particles, a permutation is accompanied by a minus sign. Therefore we have
$\mathcal{G}_3^{(1b)} = -\Pr\,\mathcal{G}_3^{(1a)}$.
Using Property~1, $-\Pr\,\mathcal{G}_3^{(1a)}$ and $\,\mathcal{G}_3^{(1a)}$ yield
  the same contribution to $g_{3}$\,  in Eq.~(\ref{eqg3symm}), so that we only need to consider the contribution of $\,\mathcal{G}_3^{(1a)}$. 
The expression of $\,\mathcal{G}_3^{(1a)}$ is
\be
\mathcal{G}_3^{(1a)}({\bf K},{\bf K}';\beta)
\ = \ 
 - \,
z^3\ \delta_{{\bf k}_1,{\bf k}'_1}\,\delta_{{\bf k}_2,{\bf k}'_2}\
e^{-\beta\ \left(
\frac{{P}^2}{6 m}+\frac{3 q^2}{4 m}+\frac{p^2}{m}
\right)}\label{eqG31a}
\ee
where we used the rewriting of the total energy
\,$\frac{{k}_1^2}{2 m}+\frac{{k}_2^2}{2 m}+\frac{{k_3^2}}{2 m}
\, = \,
\frac{P^2}{6 m}+\frac{3q^2}{4 m}+\frac{p^2}{m}$\,.

\subsubsection{One $T_2$ vertex: \,$\mathcal{G}_3^{(2a)}$,\,$\mathcal{G}_3^{(2b)}$,\,$\mathcal{G}_3^{(2c)}$ and \,$\mathcal{G}_3^{(2d)}$}\label{AppG32}

\begin{figure}[h]
  \vskip0.4cm
  \subfigure[\label{fig2a}]{\includegraphics[width=0.45\linewidth]{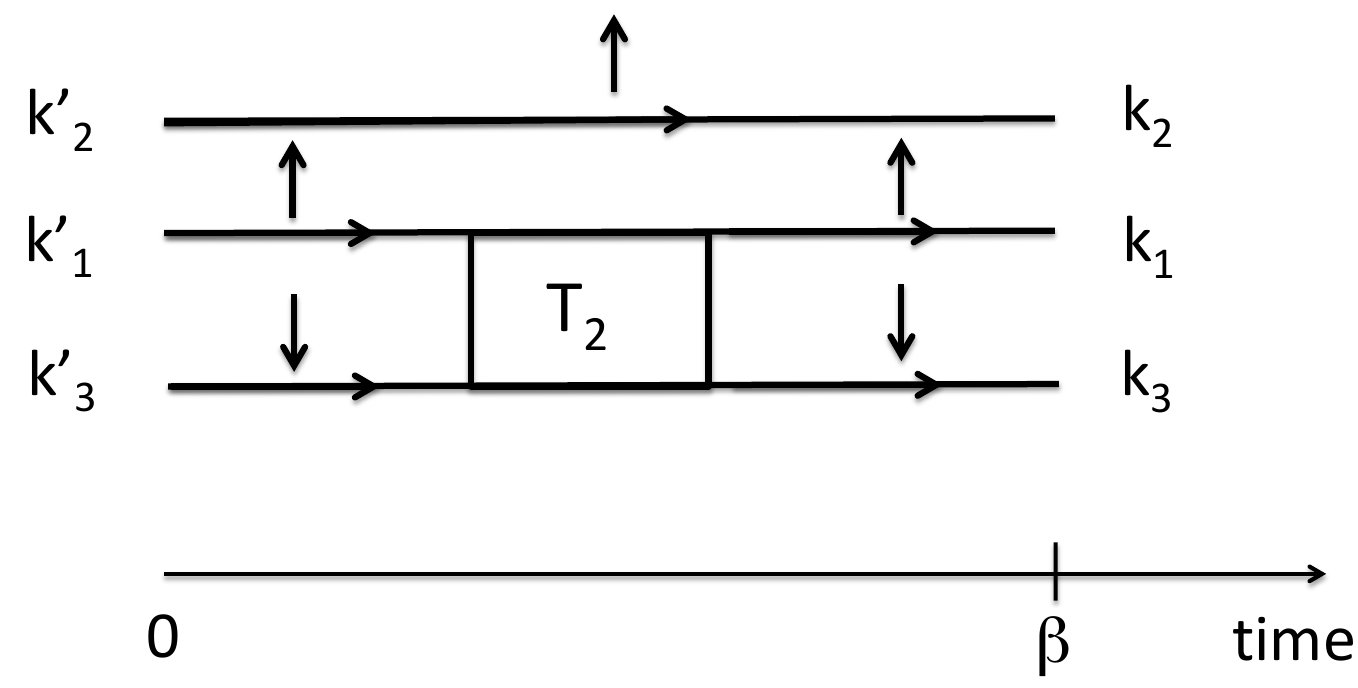}}
  \hfill
\subfigure[\label{fig2b}]{\includegraphics[width=0.45\linewidth]{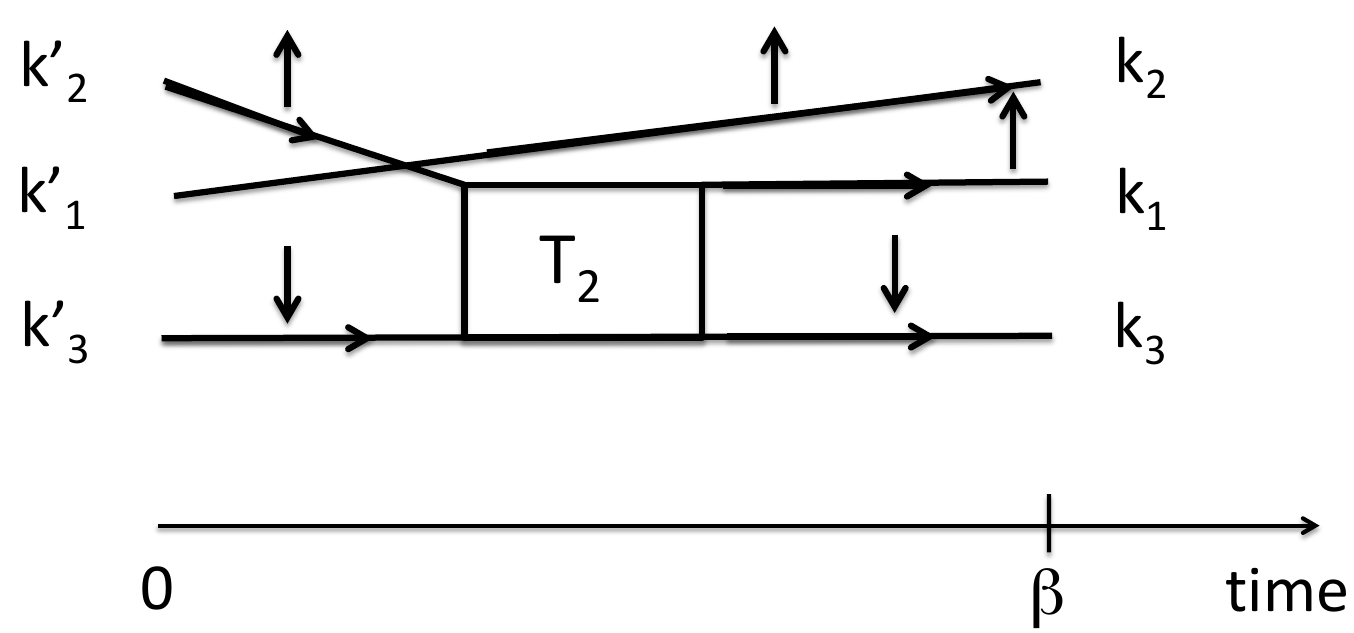}}
  \vskip1cm
  \subfigure[\label{fig2c}]{\includegraphics[width=0.45\linewidth]{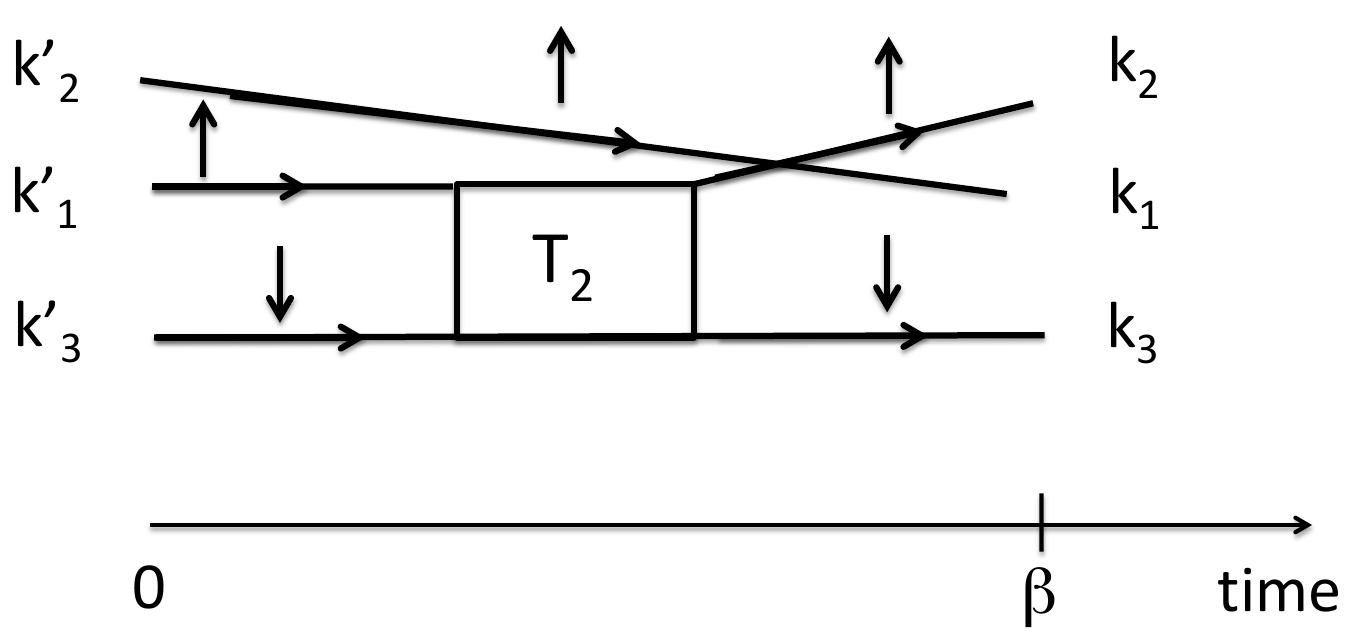}}
  \hfill
\subfigure[\label{fig2d}]{\includegraphics[width=0.45\linewidth]{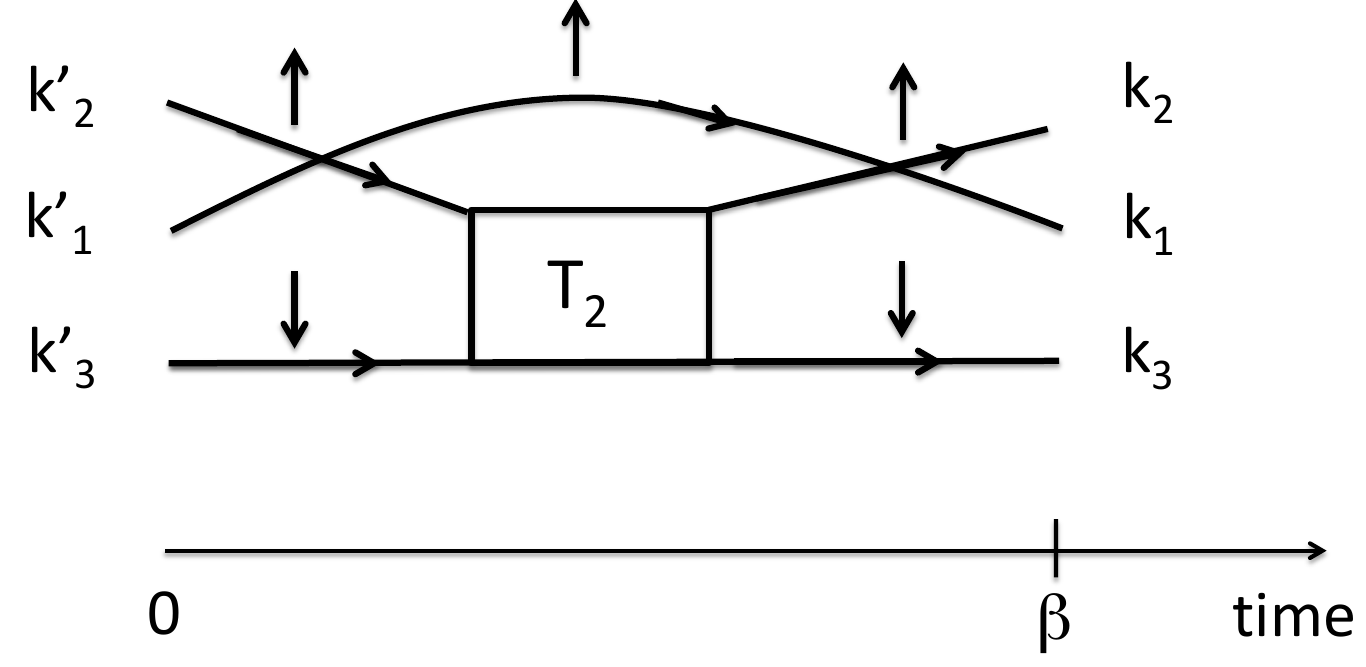}}

\caption{
  The four diagrams
  containing a single $T_2$\, vertex,
  $\mathcal{G}_3^{(2a)}$,\,$\mathcal{G}_3^{(2b)}$,\,$\mathcal{G}_3^{(2c)}$ and \,$\mathcal{G}_3^{(2d)}$\,.
\label{fig:G32}}
\end{figure}

\begin{figure}
\begin{center}
\includegraphics[width=0.5\linewidth,trim={0 1cm 0 0},clip]{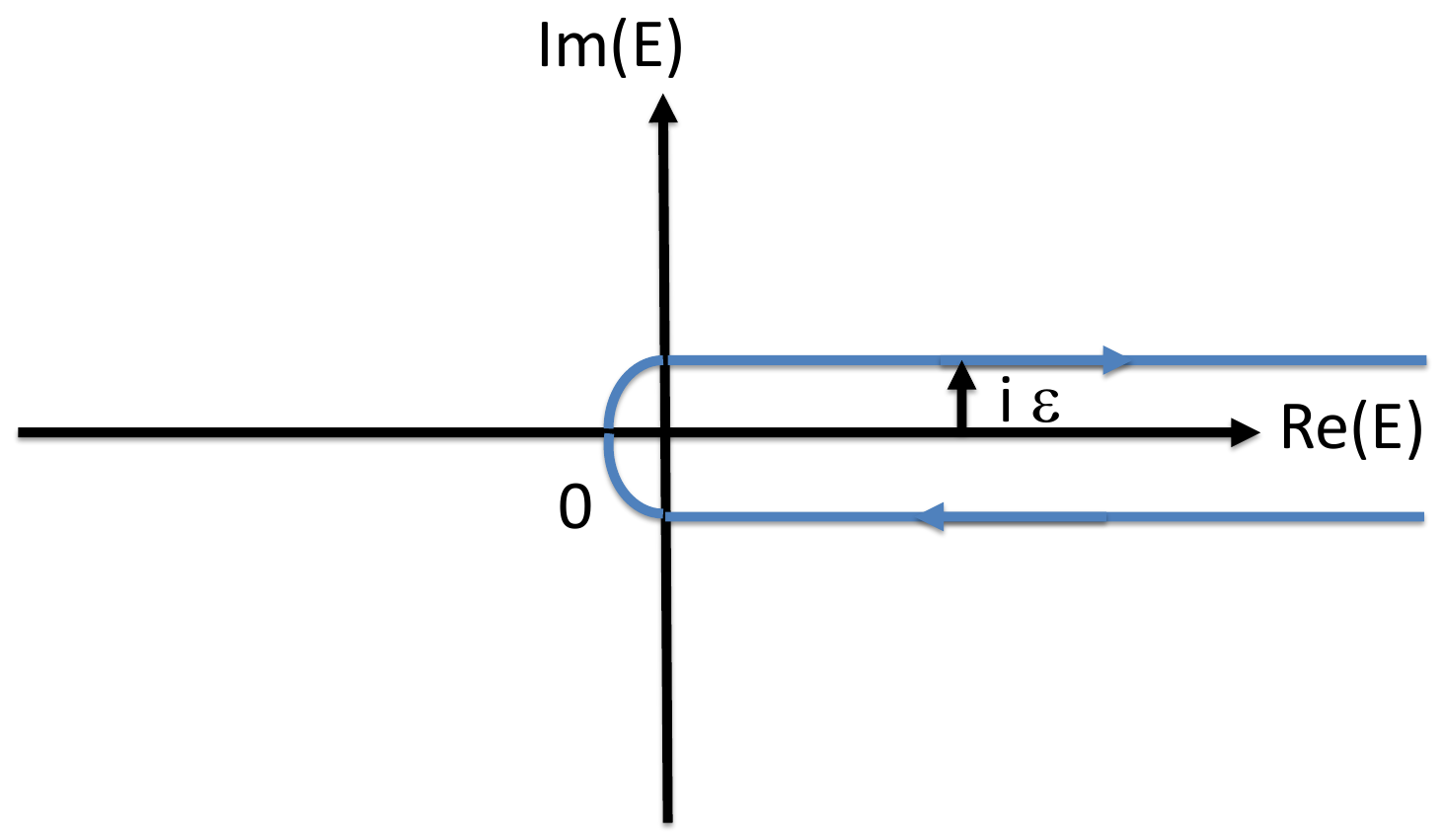}
\caption{Integration contour $\Cr_+$ surrounding the positive real axis (the limit $\epsilon \to 0^+$ is implied).
}
\label{fig:C+}
\end{center}
\end{figure}

The four diagrams contributing to $\mathcal{G}_3$ at order $z^3$
  containing a single $T_2$ are shown in Figs.~\ref{fig2a}, \ref{fig2b}, \ref{fig2c} and \ref{fig2d}.
Their contributions to $\mathcal{G}_{3}$ are denoted by \,$\mathcal{G}_3^{(2a)}$,\,$\mathcal{G}_3^{(2b)}$,\,$\mathcal{G}_3^{(2c)}$ and $\mathcal{G}_3^{(2d)}$ respectively.
 We have
 \be
 \mathcal{G}_3^{(2b)} \ = \ -\Pr'\,\mathcal{G}_3^{(2a)}\,,
\ \  \ \ \mathcal{G}_3^{(2c)} \ = \ -\Pr\,\mathcal{G}_3^{(2a)}\,,
\ \  \ \ \mathcal{G}_3^{(2d)} \ = \ \Pr\Pr'\,\mathcal{G}_3^{(2a)}
 \nn
 \ee
and thus by Property~1,
the contribution to $g^{(3)}$\, in Eq.~(\ref{eqg3symm}) coming from
$\,\mathcal{G}_3^{(2a)}$,\,$\mathcal{G}_3^{(2b)}$,\,$\mathcal{G}_3^{(2c)}$\, and $\,\mathcal{G}_3^{(2d)}$
are identical.
Hence we will only need to compute the contribution of
$\,\mathcal{G}_3^{(2a)}\,$.
As shown in Appendix~\ref{app:G3},
 \begin{equation}
  \mathcal{G}_3^{(2a)}({\bf K},{\bf K}';\beta)=z^3\,\Vr^{-1}\,
  {\delta_{{\bf k}_2,{\bf k}'_2}} \,
e^{-\beta\frac{{\bf P}^2}{6m}}
\int_{\mathcal{C}_+}\, \frac{dE}{2\pi i}\ e^{-\beta\,E}\ 
\frac{t_2\left(E-
  \frac{3{q}^2}{4m}\right)}
     {\left(E-\frac{3{q}'^{\phantom{.}2}}{4m}-\frac{{p}'^{\phantom{.}2}}{m}\right)\left(E-\frac{3{q}^2}{4m}-\frac{{p}^2}{m}\right)}
     \label{eqG32a}
 \end{equation}
where $t_2$ is the two-body T-matrix in the center-of-mass frame whose expression is given by Eq.~(\ref{eq:t2}),
     and the contour $\Cr_+$ surrounds the real positive axis in the clockwise direction, see Fig.~\ref{fig:C+}.

 \subsubsection{One $T_3$ vertex: \,$\mathcal{G}_3^{(3a)}$,\,$\mathcal{G}_3^{(3b)}$,\,$\mathcal{G}_3^{(3c)}$,\,$\mathcal{G}_3^{(3d)}$}\label{AppG33}
\begin{figure}[h]
   \vskip0.6cm
   \subfigure[\label{fig3a}]{\includegraphics[width=0.45\linewidth]{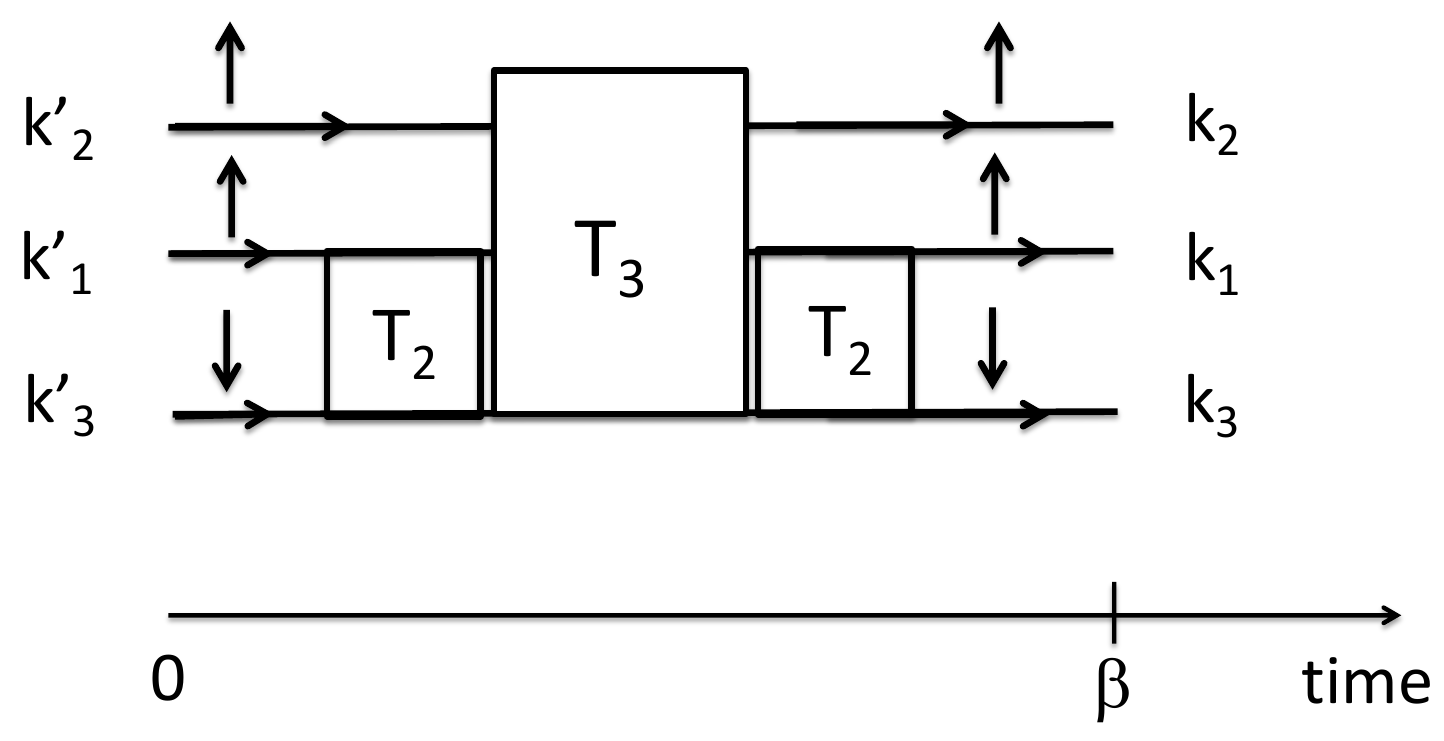}}
\hfill\subfigure[\label{fig3b}]{\includegraphics[width=0.45\linewidth]{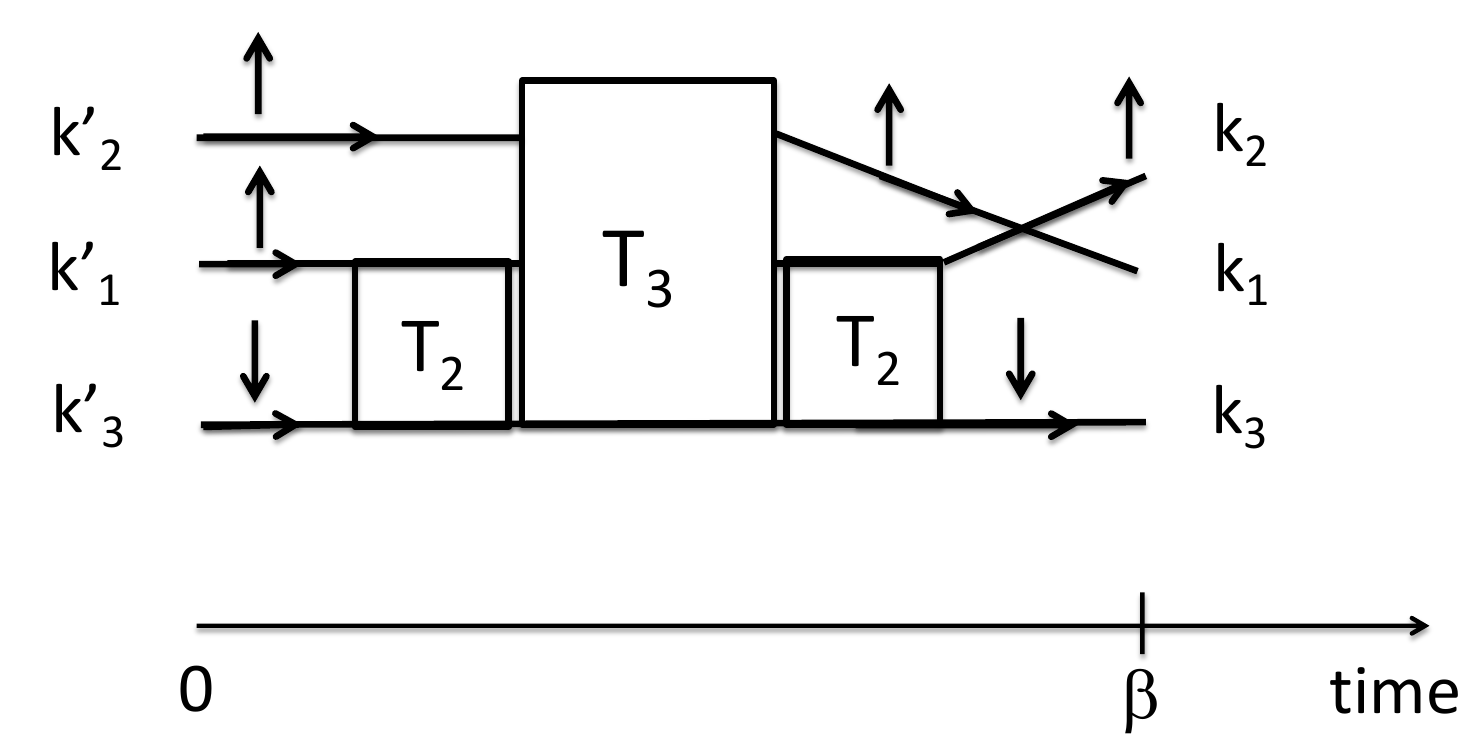}}
\vskip1cm
\subfigure[\label{fig3c}]{\includegraphics[width=0.45\linewidth]{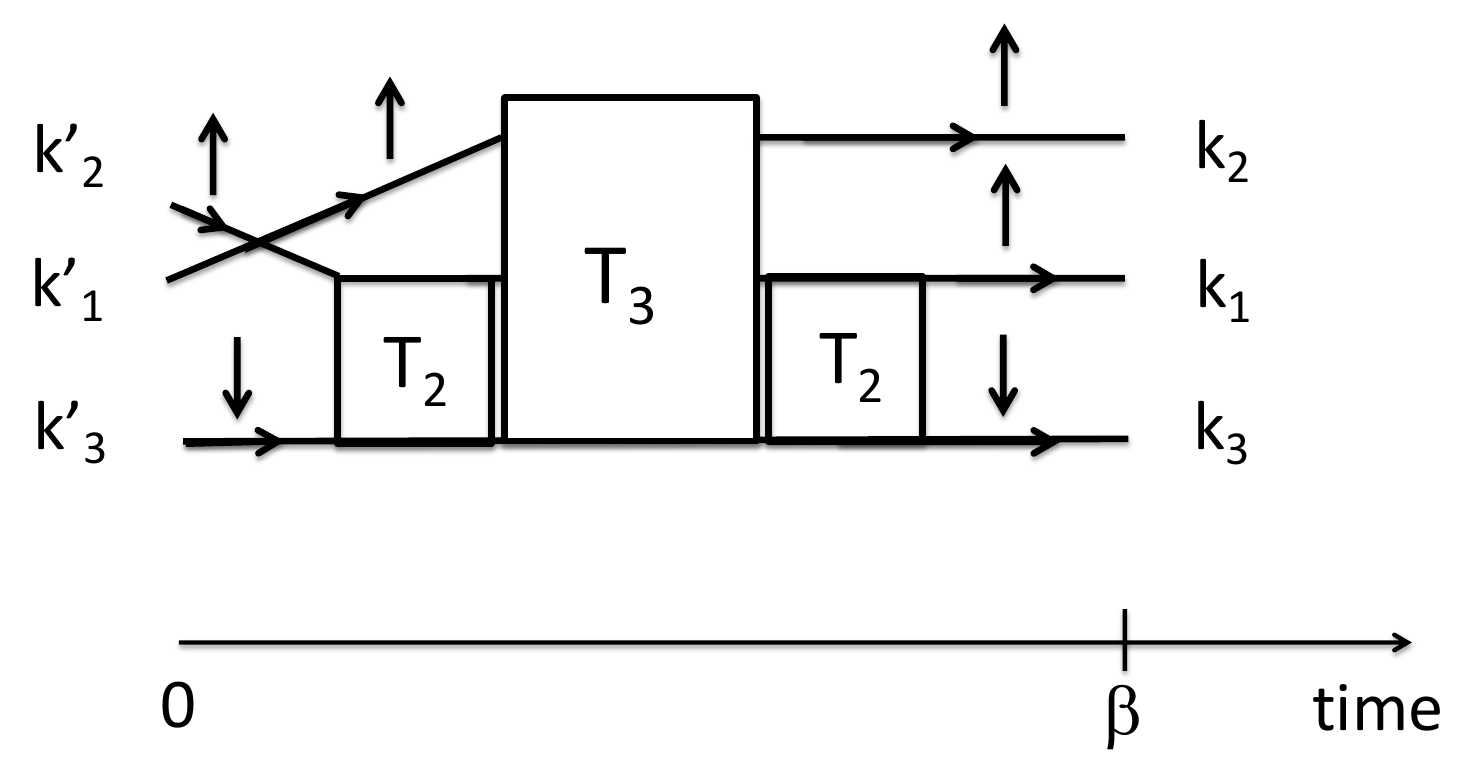}}
\hfill
\subfigure[\label{fig3d}]{\includegraphics[width=0.45\linewidth]{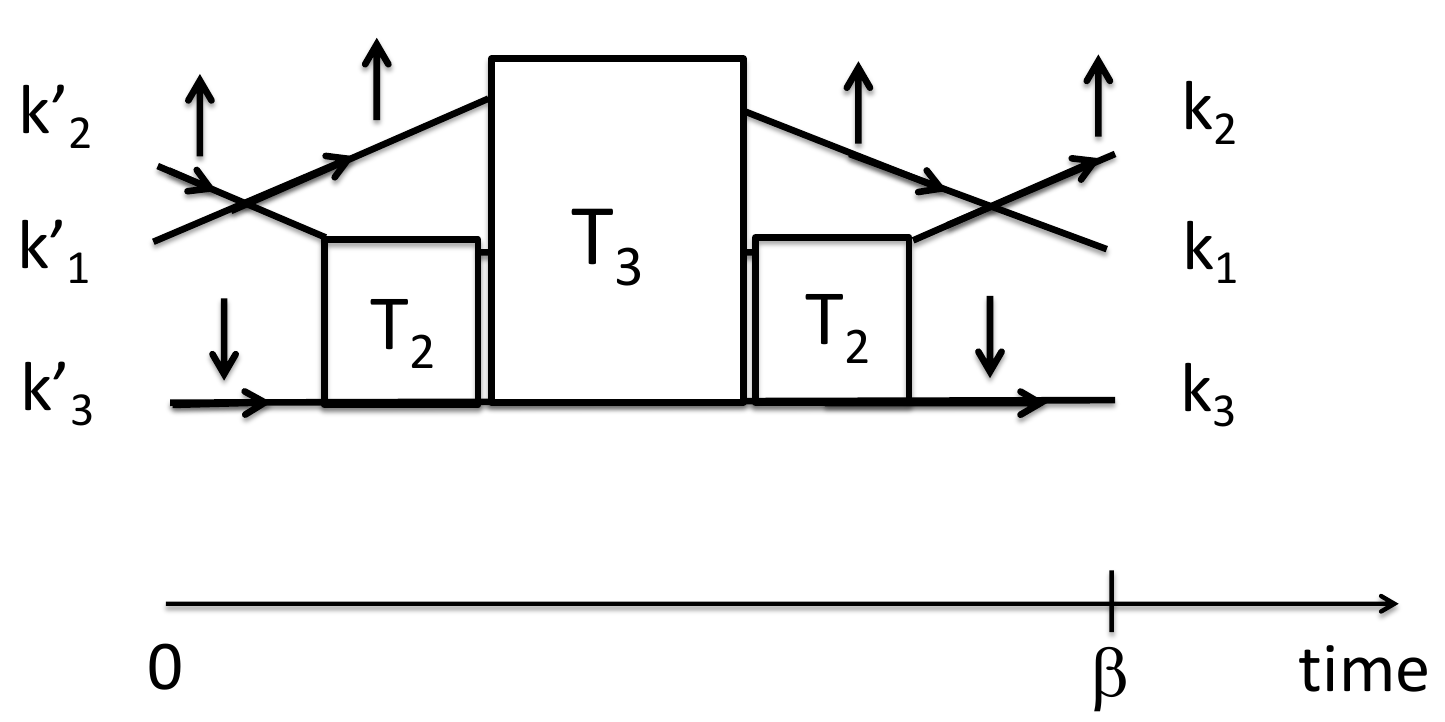}}

\caption{The four diagrams containing at least two $T_2$ vertices,
$\mathcal{G}_3^{(3a)}$,\,$\mathcal{G}_3^{(3b)}$,\,$\mathcal{G}_3^{(3c)}$ \,and~\,$\mathcal{G}_3^{(3d)}$\,.
\label{fig:G33}}
\end{figure}

 The four diagrams involving more than two $T_2$'s are shown in Figs.~\ref{fig3a}, \ref{fig3b}, \ref{fig3c} and~\ref{fig3d}.
 {Their contributions to $\mathcal{G}_{3}$ are denoted by \,$\mathcal{G}_3^{(3a)}$,\,$\mathcal{G}_3^{(3b)}$,\,$\mathcal{G}_3^{(3c)}$ and $\mathcal{G}_3^{(3d)}$ respectively.}
 We have
 \be
 \mathcal{G}_3^{(3b)}\ =\ -\Pr\,\mathcal{G}_3^{(3a)}\,,
\ \ \mathcal{G}_3^{(3c)}\ =\ -\Pr'\,\mathcal{G}_3^{(3a)}\,,
\ \ \mathcal{G}_3^{(3d)}\ =\ \Pr\,\Pr'\,\mathcal{G}_3^{(3a)}
\nn
\ee
and thus by Property~1, the
contribution to $g^{(3)}$\, in Eq.~(\ref{eqg3symm}) coming from $\,\mathcal{G}_3^{(3a)}$,\,$\mathcal{G}_3^{(3b)}$,\,$\mathcal{G}_3^{(3c)}$\, and $\,\mathcal{G}_3^{(3d)}$
are identical.
Hence we only need to compute the contribution of
$\,\mathcal{G}_3^{(3a)}\,$.
\,As shown in Appendix~\ref{app:G3},
 \begin{multline}
 \mathcal{G}_3^{(3a)}({\bf K},{\bf K}';\beta)=
 z^3\,\Vr^{-2}\,
 e^{-\beta\frac{{P}^2}{6 m}}\ 
 \int_{\mathcal{C}_+} \, \frac{dE}{2\pi i}\ e^{-\beta\,E}\ 
  \frac{t_2\big(E-\frac{3{q}^2}{4m}\big)}{E-\frac{3{q}^2}{4m}-\frac{{p}^2}{m}}
 \
 t_3({\bf q},{\bf q}';E\,)
 \
  \frac{t_2\big(E-\frac{3{q}'^{\phantom{.}2}}{4m}\big)}{E-\frac{3{q}'^{\phantom{.}2}}{4m}-\frac{{p}'^{\phantom{.}2}}{m}}
 \label{eqmG33a}
 \end{multline}
 where $t_3$ is the three-body T-matrix in the center-of-mass frame,
which solves the STM equation Eq.~(\ref{eq:STMt3}).

\subsection{Virial expansion of \,$g_3$}
In Eq.~(\ref{eqg3symm}),
  we need to evaluate inverse Fourier transforms of the type
\be
\sum_{{\bf k}_1,{\bf k}_{2},{\bf k}'_1,{\bf k}'_2,{\bf P}}
\ \ \ \ 
e^{\,i \big({\bf p}\,\cdot\,{\bf r}_a+
\frac{\sqrt{3}}{2}\,{\bf q}\,\cdot \boldsymbol{ \rho}_a
\,-\,{\bf p}'\cdot{\bf r}_b
\,-\,
\frac{\sqrt{3}}{2}\,{\bf q}'\cdot \boldsymbol{ \rho}_b
\big)}
\ \ \mathcal{G}_3^{(\dots)}({\bf K},{\bf K}';\beta)\nn
\ee
for each of the three contributions to $\mathcal{G}_3$ given by Eqs. (\ref{eqG31a}), (\ref{eqG32a}) and~(\ref{eqmG33a}),
with
  $({\bf r}_a \, , \boldsymbol{ \rho}_a)$ equal to
  $({\bf r} \, , \boldsymbol{ \rho})$ or $({\bf r}' , \boldsymbol{ \rho}')$\,,
and similarly for
$({\bf r}_b \, , \boldsymbol{ \rho}_b)$\,.
The sum over the center of mass momentum ${\bf P}$
factorizes, and yields
$\int
e^{-\beta\frac{{P}^2}{6m}}
\,d^3\!P/(2\pi)^3 \, = \, \lambda_T^{-3}\,3^{3/2}$
in the considered thermodynamic limit $\Vr\to\infty$.

\subsubsection{No interaction: $g_{3}^{(1)}$} \label{subsec:g31}

Inserting the expression~(\ref{eqG31a})
of \, $\mathcal{G}_3^{(1a)}$
into Eq.~(\ref{eqg3symm}),
and multiplying by 2 to account for the contributions of the other diagram $\mathcal{G}_3^{(1b)}$,
we obtain
\be
g_{3}^{(1)}({\bf r},\boldsymbol{ \rho}) \ = \  z^3
\, \lambda_T^{-3} \
  \frac{3^{3/2}}{2} \ \int \frac{d^3\!p\,d^3\!q}{(2\pi)^6}\ 
e^{-\beta\big(\frac{3{q}^2}{4m}+\frac{{p}^2}{m}
\big)}
\ \big|
e^{\,i
  \varphi({\bf p},{\bf q})
}
-
e^{\,i
  \varphi'({\bf p},{\bf q})
}
\big|^2
\nn\ee
where we performed the change of variables
     $({\bf k}_1 , {\bf k}_2) \rightarrow ({\bf p} , {\bf q})$
     at fixed ${\bf P}$, whose Jacobian equals unity.
Evaluating the gaussian integrals, 
and using the identity
$\|\boldsymbol{ \rho}-\boldsymbol{ \rho}'\|^2+\|{\bf r}-{\bf r}'\|^2=4\,\|{\bf r}_2-{\bf r}_1\|^2$ 
as well as the leading-order equation of state
Eq.~(\ref{eq:boltz_eos}),
we get
\be
g_{3}^{(1)}({\bf r},\boldsymbol{ \rho}) \ \simeq \ \frac{n^3}{8}\ \left(1 \, - \, e^{-\frac{m}{\beta}\|{\bf r}_2-{\bf r}_1\|^2}\,\right)
\label{eq:g3_1}
\ee
which can also be obtained by simply applying Wick's theorem for the ideal Fermi gas.
This expression does not diverge when 
${\bf r}_1$\,, ${\bf r}_2$ and ${\bf r}_3$ tend to the same limit, so that it does not contribute to the three-body contact \,$\mathcal{C}_3$\,,
 and its contribution to \,$g_{3}({\bf r},\boldsymbol{ \rho})$\, is negligible in the $R\to0$ limit.

\subsubsection{One $T_2$ vertex: $g_{3}^{(2)}$}
Injecting the expression~(\ref{eqG32a}) {of \, $\mathcal{G}_3^{(2a)}$}
into Eq.~(\ref{eqg3symm}),
{and multiplying by 4 to account for the contributions of the three other diagrams $\mathcal{G}_3^{(2b)}, \mathcal{G}_3^{(2c)}$ and $\mathcal{G}_3^{(2d)}$,}
we find an expression for $g_3^{(2)}$.
The integrations over ${\bf p}$ and ${\bf p}'$ can be done analytically using
\be
\int \frac{d^3\! p}{(2\pi)^3}\ \frac{e^{\,i\, {\bf p}\cdot{\bf x}}}
{Z-\frac{{p}^2}{m}}
\ = \ 
-\,\frac{m}{4\pi}\ 
\frac{e^{-\sqrt{-m\,Z}\,x}}{x}
\label{eqintonp}
\ee
which holds for any complex $Z$ outside the real positive axis.
Using
Eq.~(\ref{eqintonp})
for ${\bf x} =\rr$ or $\rr'$
and $Z= {E}- 3 q^2/(4m)$\,,
we obtain
\be
g_3^{(2)}({\bf r}, \boldsymbol{ \rho}) \ = \ - c \ 
\int_{\mathcal{C}_+}\frac{d E}{2\pi i} \ e^{-\beta\,E}\,
\int
\frac{d^3\! q}{(2\pi)^3}\ \,
\Psi_{{\bf q}}({\bf r}, \boldsymbol{ \rho})
\ t_2\left(E-
\tfrac{3 q^2}{4m}\right)
\ \Psi_{-{\bf q}}({\bf r}, \boldsymbol{ \rho})
\nn
\ee
where
$\Psi_{{\bf q}}({\bf r}, \boldsymbol{ \rho}) \ = \ 
e^{\frac{\sqrt{3}}{2}\left(i \, {\bf q}\cdot\boldsymbol{\rho}-\widetilde{q}\,r\right)}/r
 \ - \ 
e^{\frac{\sqrt{3}}{2}\left(i \, {\bf q}\cdot\boldsymbol{\rho}'-\widetilde{q}\,r'\right)}/r'$
\,with
\be
\tilde{q} \ = \ \sqrt{-\tfrac{4}{3}m E+q^2}
\label{eq:def_qtilde}
\ee
and
\be
  c \ = \ z^3 \ \lambda_T^{-3} \ \, 3^{3/2} \ \Bigg(\frac{m}{4\pi}\Bigg)^2
\nn\ee
Integrating over angles yields
    \begin{multline}
  g_3^{(2)}({\bf r}, \boldsymbol{ \rho}) \ = \
  -\,
c \ 
\int_{\mathcal{C}_+}
\, \frac{dE}{2\pi i} \ e^{-\beta\,E} \ 
\int_0^{\infty}
\dfrac{d q}{2\pi^2}\ q^2\ 
t_2\left(E-\frac{3q^2}{4m}\right)
\\
\times
\left[
\frac{e^{-\sqrt{3}\,\widetilde{q}\, r}}{r^2}
\, + \, \frac{e^{-\sqrt{3}\,\widetilde{q}\, r'}}{r'^{\phantom{.}2}}
\ -\ \frac{2 \ {\rm sinc}\!\left(\tfrac{\sqrt{3}}{2}q\, \|\boldsymbol{\rho}-\boldsymbol{\rho}'\|\right)\ e^{-\frac{\sqrt{3}}{2}\widetilde{q}\,(r+r')}}{r\,r'}
\ \right]
\label{eqg32}
    \end{multline}
  where
${\rm sinc}(Z) \, := \, \sin(Z)/Z$.

\subsubsection{One $T_3$ vertex: $g_{3}^{(3)}$}

We use the angular-momentum decomposition of the $3$-body T-matrix, see Eq.~(\ref{eqt3Pl}) in App.~\ref{app:STM}.
Using the addition theorem for spherical harmonics,
$\sum_{\md=-l}^{l}Y_l^{\md}\!\left(\hat{q}\,\right) \, Y_l^{\md}\!\left(\hat{q}\,'\,\right)^* \ = \ \frac{2l+1}{4\,\pi}\ P_l\!\left(\,\hat{q}\cdot\hat{q}\,'\,\right)$\,,
Eq.~(\ref{eqt3Pl}) becomes
\be
t_3({\bf q},{\bf q}';E\,)
\ =\ \sum_{l=0}^\infty \ \, \frac{4\,\pi}{2l+1} \
\ \sum_{\md=-l}^{l} \ \, Y_l^{\md}\!\left(\hat{q}\,\right)
\ t_{3,l}(q,q';E\,)
\ Y_l^{\md}\!\left(\hat{q}\,'\,\right)^*
\nn\ee
{Inserting the expression~(\ref{eqmG33a}) of  $\,\mathcal{G}_3^{(3a)}$
into Eq.~(\ref{eqg3symm}),
and multiplying by 4 to account for the contributions of the three other diagrams $\mathcal{G}_3^{(3b)}, \mathcal{G}_3^{(3c)}$ and $\mathcal{G}_3^{(3d)}$,}
we find an expression for $g_3^{(3)}$.
We can again integrate over ${\bf p}$ and ${\bf p}'$ using Eq.~(\ref{eqintonp}). We can also integrate over the angles of  ${\bf q}$ and ${\bf q}'$, using the identity
\be
\int d\hat{q} \ \, e^{\,i A\,\hat{q}\cdot\hat{u}}\ Y_l^{\md}(\hat{q}\,) \ = \ 
4\pi\,i^l \ j_l(A)\ Y_l^{\md}(\hat{u}\,)
\label{eq:plane-wave_Ylm}
\ee
    with $j_l$\, the spherical Bessel function of order $l$,
$\hat{u}$\, a unit vector, and $A$ real.
This yields
\begin{multline}
  g_3^{(3)}({\bf r}, \boldsymbol{ \rho}) \ = \ -
  c
    \ \sum_{l=0}^\infty\ 
  \ \sum_{\md=-l}^{l} \ \ \frac{4\,\pi}{2l+1}
\ \  \int_{\mathcal{C}_+}\,\frac{d E}{2\pi i}
  \ e^{-\beta\,E}
  \ \int_0^{\infty}\frac{dq\,q^2}{2\,\pi^2}
  \ \int_0^{\infty}\frac{dq'\,q'^{\phantom{.}2}}{2\,\pi^2}
  \\
  \times \, \Phi_{q} \ \ t_2\left(E-\frac{3q^2}{4m}
\right)
\ t_{3,l}( q, q';E)
\ t_2\left(E-\frac{3q'^{\phantom{.}2}}{4m}
\right)
\ \ \Phi'_{q'}
\label{eqg33-2}
\end{multline}
where
\begin{align}
  \Phi_{q}
  & \ = \ Y_l^{\md}\!\left(\hat{\rho}\,\right)\ \ \dfrac{e^{-
     \frac{\sqrt{3}}{2}\,\widetilde{q}
      \, r}}{r}
\ \ j_l\left(\tfrac{\sqrt{3}}{2}\, q\,\rho\right)
\ - \
\big[ \, (r, \rrho) \ \longrightarrow \ (r', \rrho') \ \big]
\label{eq:def_Phi_q}
\\
\Phi'_{q'}& \ = \ Y_l^{\md}\!\left(\hat{\rho}\,\right)^*\ \ \dfrac{e^{-
    \frac{\sqrt{3}}{2}\,\widetilde{q}\,'
    \, r}}{r} \ \ j_l\left(\tfrac{\sqrt{3}}{2}\, q' \rho\right)
    \ - \
    \big[ \, (r, \rrho) \ \longrightarrow \ (r', \rrho') \ \big]
\label{eq:def_Phi'_q'}
  \end{align}
with $\tilde{q}$ defined in Eq.~(\ref{eq:def_qtilde}), and similarly $\tilde{q}\,'=\sqrt{q'^{\phantom{.}2} -\frac{4}{3}m E}$\,.

We will use the fact that in both Equations (\ref{eqg32}) and (\ref{eqg33-2}),
due to the factor $e^{-\beta E}$, the integral over $E$ is dominated by
the range
\be
E \ \ \lesssim \ \ \frac{1}{\beta}
\label{eq:E_range}
\ee

\subsection{Short-distance limit}

 We turn to the evaluation of $g_3$ in the \,$R\to0$\, limit.
We have seen that
\be
g_3 \ = \ g_3^{(1)} \ + \ g_3^{(2)} \ + \ g_3^{(3)}
\nn\ee
Note that \,$g_3^{(1)}$\, can be viewed as the one-body contribution, in the sense that the corresponding diagrams do not contain any interaction vertices (see Fig.~\ref{fig:G31});
\,$g_3^{(2)}$\, can be viewed as the two-body contribution, in the sense that in the corresponding diagrams, two particles interact while the third particle is a spectator
(see Fig.~\ref{fig:G32});
and
\,$g_3^{(3)}$\, can be viewed as the three-body contribution, in the sense that in the corresponding diagrams, all three particles interact 
(see Fig.~\ref{fig:G33}).

By Eq.~(\ref{eq:g3_1}), the one-body contribution $g_3^{(1)}$ is bounded and thus negligible for $R\to0$. Thus
\,$ g_3 \ \simeq \ g_3^{(2)} \ + \ g_3^{(3)}$\,,
where the two-body contribution $g_3^{(2)}$ is given by Eq.~(\ref{eqg32}),
and the three-body contribution $g_3^{(3)}$ is given by Eq.~(\ref{eqg33-2}).
Further decomposing $g_3^{(3)}$ into the sum of
  \bi
\item $g_3^{(3,l=0)}$, the $l=0$ contribution to (\ref{eqg33-2}),
  \vskip .1cm
  \item $\tilde{g}_3^{(3)}$, the $l\geq1$ contribution to (\ref{eqg33-2}),
    \ei
we get
\be
g_3 \ \simeq \ g_3^{(2)} \ + \ g_3^{(3,l=0)} \ + \ \tilde{g}_3^{(3)}
\nn
\ee

Let us announce the main conclusions.
The $R\to0$ behavior of $g_3$ is determined by $\tilde{g}_3^{(3)}$, which has the power-law divergence $R^{2s-4} = R^{-0.45455146}$
expected from Eq.~(\ref{eq:g3_C3})
(more specifically, this comes from the $l=1$ contribution, while the $l>1$ contributions are negligible). The terms $g_3^{(2)}$ and $g_3^{(3,l=0)}$ diverge more strongly, like $R^{-2}$, but they cancel each other to leading order,
and their sum 
\,$g_3^{(2)} \, + \, g_3^{(3,l=0)}$\, is a negligible contribution to $g_3$ in the $R\to0$ limit.

\vskip .2cm

\noindent {\bf Remark.}
This means that the ``two-body'' diagrams of Fig.~\ref{fig:G32} yield a contribution to $g_3$ which diverges as
%$R^{-2}$ for $R\to0$,
\be
g_3^{(2)} \ \underset{R\to0}{\propto} \ R^{-2}\,,
  \label{eq:g32_R-2}
\ee
an unphysical divergence that is exactly canceled out by the ``three-body'' diagrams of Fig.~\ref{fig:G33}.
The sum of all diagrams yields the correct scaling \,$g_3 \propto R^{2s-4}$.
We also note that,
using the decomposition of Fig.~\ref{fig:T3_serie},
the diagrams of Fig.~\ref{fig:G33} can be expressed in terms of $T_2$ as an infinite sum of diagrams.
In each of these diagrams, the $\down$ particle interacts successively with the first and the second $\UP$ particle
(up to four successive interactions are shown in Fig.~\ref{fig:T3_serie}).
The prefactor of the unphysical $R^{-2}$ divergence vanishes provided one sums up this series of diagrams to infinite order, {\it i.e.}, provided one allows for an arbitrary number of successive interactions of the $\down$ particle with each of the two $\UP$ particles.

\subsubsection{Three-body \,$l\geq1$ contribution}
\label{subsec:gtilde33}

We begin by evaluating  $\tilde{g}_3^{(3)}$, {\it i.e.}, the $l\geq1$ contribution to $g_3^{(3)}$.
We introduce a cutoff $\Lambda$,
which is much larger than any characteristic wavevector of the problem,
and much smaller than $1/R$\,,
\be
\max\left(\,|a|^{-1} \, , \,\lambda_T^{-1}\,\right)
\ \ll \
\Lambda
\ \ll \
R^{-1}
\label{eq:<<Lambda<<}
\ee
    More formally, we take the double limit
     \be
     \Lambda
     \to\infty \ \ \
    {\rm and} \ \ \
    \Lambda R \to 0.
    \label{eq:Lambda_double_lim}
     \ee
     Also, recall that $e^{\beta\mu}\to0$, at fixed $T$ and $a$.
   
Since the expression~(\ref{eqg33-2}) contains an integral over two momenta ($q$ and $q'$), we define the high-wavevector (resp. low-wavevector) contribution where both momenta are larger (resp. smaller) than $\Lambda$, and the mixed contribution where one of the momenta is $>\Lambda$ and the other momentum is $<\Lambda$\,:
\be
\gt_3^{(3)} \ = \ \gt_3^{(3)}\big|_{\rm high}
\ +\  \gt_3^{(3)}\big|_{\rm mixed}
\ + \  \gt_3^{(3)}\big|_{\rm low}
\label{eq:decomp_gt}
\ee
where
\bi
\item $\gt_3^{(3)}\big|_{\rm high}$ is the contribution
to Eq.~(\ref{eqg33-2})
  from $\{ q>\Lambda \, , q' > \Lambda \}$
\item
  $\gt_3^{(3)}\big|_{\rm mixed}$ is the contribution
to Eq.~(\ref{eqg33-2})
  from $\{ q>\Lambda \, , q' < \Lambda \}$
  or $\{ q<\Lambda \, , q' > \Lambda \}$
  \vskip .05cm
\item $\gt_3^{(3)}\big|_{\rm low}$ is the contribution
to Eq.~(\ref{eqg33-2})
  from $\{ q<\Lambda \, , q' < \Lambda \}$
  \ei
      and in all cases, the sum over $l$ in Eq.~(\ref{eqg33-2}) is restricted to $l\geq1$.
The corresponding explicit expressions are
\begin{multline}
  \gt_3^{(3)}\big|_{\rm high}({\bf r}, \boldsymbol{ \rho}) \ = \ -
  c
    \ \sum_{l=1}^\infty\ 
  \ \sum_{\md=-l}^{l} \ \ \frac{4\,\pi}{2l+1}
\ \  \int_{\mathcal{C}_+}\,\frac{d E}{2\pi i}
  \ e^{-\beta\,E}
  \ \int_\Lambda^{\infty}\frac{dq\,q^2}{2\,\pi^2}
  \ \int_\Lambda^{\infty}\frac{dq'\,q'^{\phantom{.}2}}{2\,\pi^2}
  \\
  \times \, \Phi_{q} \ \ t_2\left(E-\frac{3q^2}{4m}
\right)
\ t_{3,l}( q, q';E)
\ t_2\left(E-\frac{3q'^{\phantom{.}2}}{4m}
\right)
\ \ \Phi'_{q'} \ ,
\end{multline}
\begin{multline}
  \gt_3^{(3)}\big|_{\rm mixed}({\bf r}, \boldsymbol{ \rho}) \ = \ -
  c
    \ \sum_{l=1}^\infty\ 
  \ \sum_{\md=-l}^{l} \ \ \frac{4\,\pi}{2l+1}
\ \  \int_{\mathcal{C}_+}\,\frac{d E}{2\pi i}
  \ e^{-\beta\,E}
  \ \int_0^\infty\frac{dq\,q^2}{2\,\pi^2}
  \ \int_0^{\infty}\frac{dq'\,q'^{\phantom{.}2}}{2\,\pi^2}
  \\
  \times \,
    \ \left(
    1_{\{q>\Lambda \, , \, q'<\Lambda\}}
   \, + \,
      1_{\{q<\Lambda \, , \, q'>\Lambda\}} \,
  \right) \ 
\Phi_{q} \ \ t_2\left(E-\frac{3q^2}{4m}
\right)
\ t_{3,l}( q, q';E)
\ t_2\left(E-\frac{3q'^{\phantom{.}2}}{4m}
\right)
\ \ \Phi'_{q'} \ ,
\label{eq:gt3mixed}
\end{multline}
     \begin{multline}
  \gt_3^{(3)}\big|_{\rm low}({\bf r}, \boldsymbol{ \rho}) \ = \ -
  c
    \ \sum_{l=1}^\infty\ 
  \ \sum_{\md=-l}^{l} \ \ \frac{4\,\pi}{2l+1}
\ \  \int_{\mathcal{C}_+}\,\frac{d E}{2\pi i}
  \ e^{-\beta\,E}
  \ \int_0^\Lambda\frac{dq\,q^2}{2\,\pi^2}
  \ \int_0^\Lambda\frac{dq'\,q'^{\phantom{.}2}}{2\,\pi^2}
  \\
  \times \, \Phi_{q} \ \ t_2\left(E-\frac{3q^2}{4m}
\right)
\ t_{3,l}( q, q';E)
\ t_2\left(E-\frac{3q'^{\phantom{.}2}}{4m}
\right)
\ \ \Phi'_{q'} \ .
\label{eq:gt3low}
     \end{multline}

Let us first consider $\gt_3^{(3)}\big|_{\rm high}$\,. 
As we will see, this yields the dominant contribution to $\gt_3^{(3)}$.
     We can restrict to the range Eq.~(\ref{eq:E_range}). 
     Hence $q^2 \gg m E$, so that $q^2- \frac{4}{3}m E$ is positive.
     Thus the function
     $E \mapsto \Phi_q\ t_2\left(E-\frac{3q^2}{4m}\right)$
     is continuous when $E$ crosses the real positive axis.
     We have the same simplification for the $q'$-dependent terms.
     Also using Eq.~(\ref{eq:t3l_im}),
we obtain
            \begin{multline}
        \gt_3^{(3)}\big|_{\rm high} \ = \
        -c\
            \ \sum_{ l=1}^\infty\ 
  \ \sum_{\md=-l}^{l} \ \ \frac{4\,\pi}{2l+1}
\ \
  \int_0^\infty\,\frac{d E}{\pi}
  \ e^{-\beta\,E}
  \ \int_\Lambda^{\infty}\frac{dq\,q^2}{2\,\pi^2}
  \ \int_\Lambda^{\infty}\frac{dq'\,q'^{\phantom{.}2}}{2\,\pi^2}
  \\
  \times
\, \Phi_{q}
\ \ t_2\left(E-\frac{3q^2}{4m}
\right)
\ \ \Phi'_{q'}
\ t_2\left(E-
\frac{3q'^{\phantom{.}2}}{4m}
\right)
\ \Im t_{3,l}( q, q';E+ i\,0^+ )
\label{eq:g3high_intermediaire}
\end{multline}

        The imaginary part of $t_{3,1}( q, q';E\,)$ has a power-law asymptotic behavior, in the limit of large $q$ and $q'$, at fixed real positive $E$:
\be
\boxed{
\Im\, t_{3,1}(q,q'; E+i\,0^+)
\ \sim \ m\,\frac{\mathcal{B}_1(m\,E\,;a^{-1})}{(q\,q')^{s+1}} \,,
\,\, \textrm{for}\  q,q'\to\infty\ \, \textrm{and}\ E \in \mathbb{R}^+\ \, \textrm{fixed}}
\label{eqimt31}
\ee
see Appendix \ref{AppImt3largek} for the derivation. 
This is the key ingredient for all the power laws scalings.
We also show in App.~\ref{AppImt3largek} that for
arbitrary $l$,
\be
{\rm Im} \, t_{3,l}(q,q'; E+i\,0^+)
\ \sim \ m\,\frac{\mathcal{B}_l(m\,E\,;a^{-1})}{(q\,q')^{s_l+1}} \,,
\,\, \textrm{for}\  q,q'\to\infty\ \, \textrm{and}\ E \in \mathbb{R}^+\ \, \textrm{fixed}
\label{eqimt3l}
\ee
where $s_l$ is the smallest scaling exponent of the unitary three-body problem
with angular momentum $l$~\cite{Werner3corpsPRL,WernerThese,C3_I}.
As a consequence, for $l\neq 1$, we have $s_l>s_1=s$, so that Eq.~(\ref{eqimt3l}) is subdominant compared to Eq.~(\ref{eqimt31}).

In Eq.~(\ref{eq:g3high_intermediaire}), $q$ and $q'$
  are larger than $\Lambda$ which tends to $\infty$,
  and $E$ is essentially bounded by Eq.~(\ref{eq:E_range}),
    so that we can use the asymptotic behavior Eq.~(\ref{eqimt31}).
      For similar reasons,
  we have
  ${t_2\big(E-3q^2/(4m)\big)} \simeq - 8\pi/(\sqrt{3} m q)$.
  Finally,
  we have
  $|\tilde{q}r - q r| \ \lesssim \ m E\, r / \Lambda \ \lesssim \ m R / (\Lambda \beta) \, \sim \,  (R/\lambda_T) / (\Lambda \,\lambda_T) \, \ll \, 1$,
and thus
  \be
  |\tilde{q}r - q r| \ \ll \ 1 \ \ \ \ {\rm for} \ \ \ q>\Lambda
  \label{eq:qtilde->q}
\ee
so that 
we can approximate
$\tilde{q}$\, by $q$
in the expression~(\ref{eq:def_Phi_q}) of $\Phi_q$\,.
The $q'$-dependent factors can be approximated in the same way.
As a result, the integrals over
    $E$, $q$ and $q'$ factorize.
Moreover, $\Phi'_{q'} = \Phi_{q'}^*$ so that the integrals over $q$ and $q'$ are complex conjugate. Also using 
      the identity $\hat{P}f(\rr,\rrho) = f(\rr',\rrho')$\,,
          we obtain
            \be
        \gt_3^{(3)}\big|_{\rm high} \ \simeq \
        - \, z^3\ \lambda_T^{-3}\ \, \frac{4\sqrt{3}}{\pi^3} \,
            \ \sum_{ l=1}^\infty\ \,
            \frac{D_l}{2l+1}
            \ \sum_{\md=-l}^{l} \ \
            \left| \left(1 - \hat{P}\ \right)\ \frac{Y_l^\md\!\left(\hat{\rho}\,\right)}{r}\ \,
\Ir_l\ 
\right|^2
\label{eq:g3high_Il}
            \ee
with
             \be
             D_l\left(\frac{\beta}{m};a^{-1}\right)
             \ = \ m\,\int_{0}^{\infty}\ \dfrac{d E}{\pi}\ e^{-\beta\,E}\ \,\mathcal{B}_l\left(m\,E\,;a^{-1}\right)
             \label{eqdefDl}
             \ee
                and
                \be
                \Ir_l \ = \ \int_\Lambda^{\infty}\ dq\ q^{- s_l}\ e^{-\frac{\sqrt{3}}{2} q r}\ j_l\left(\tfrac{\sqrt{3}}{2} q \rho\right)
                \nn
                \ee
                To evaluate $\Ir_l$\,, we introduce the hyperangle
$\alpha \, = \, {\rm arctan}(r/\rho)$,
                and we perform the change of variable $t=\frac{\sqrt{3}}{2}q\,R$\,,
                which yields
                \be
\Ir_l \ = \ \left(\frac{\sqrt{3}\ R}{2}\right)^{s_l-1}\ \int_{\frac{\sqrt{3}}{2} \Lambda R}^{\infty}\ dt\ t^{-s_l}\ e^{-t\, \sin \alpha}\ j_l\left(t\, \cos\alpha\right)
                \nn
                \ee
                The lower bound of this integral is $\ll 1$.
                For $t\to0$ the integrand behaves as $t^{l-s_l}$.
                The exponent $l-s_l$ can be smaller or larger than $-1$
                (we have $s_l>l+1$ for $l$ even, and $s_l<l+1$ for $l$ odd~\cite{Werner3corpsPRL,WernerThese}), and
                we must distinguish between these two cases.
                \bi
              \item If $s_l < l+1$,
                we can replace the lower bound of the integral by 0,
                so that $\Ir_l$ equals $R^{s_l-1}$ times a dimensionless function of $\alpha$.
                The resulting contribution to $g_3$ scales as $R^{\,2s_l-4}$.
                In the considered limit $R\to0$,
                the dominant contribution comes from $l=1$, since $s_1 \equiv s$\, is the smallest of the exponents~$s_l$\,.
Evaluating the integral over $t$ for $l=1$ yields
                \be
                \Ir_1 \ \simeq \ - \, \left(\frac{\sqrt{3}\ R}{2}\right)^{s\,-1}\
                \frac{\pi}{\sin(\pi s)\ \Gamma(s+2)}\ \times \  \frac{\varphi(\alpha)}{\cos\alpha}
                \label{eq:I1}
                 \ee
                 where
                 \be
\varphi(\alpha)
\ = \ -s\,\cos\bigg[s\,\bigg(\frac{\pi}{2}-\alpha\bigg)\bigg]
\ + \
\sin\bigg[s\,\bigg(\frac{\pi}{2}-\alpha\bigg)\bigg]
\ \tan\alpha
\nn\ee
is the function entering the expression of the $l=1$ unitary hyperspherical wavefunction,
\,$\phi_\md(\Oo) \, = \, \Nr\, \tilde{\phi}_\md(\Oo)$\, with
\be
\tilde{\phi}_{\md}(\Oo)\ = \ \left(
1-
\hat{P}\
\right)
\, \left(\frac{\varphi(\alpha)}{\sin(2\,\alpha)} \ Y_{1}^{\md}\!\left(\hat{\rho}\,\right)\,\right)
\label{eq:phi_tilde_m=}
\ee
              \item
                If $s_l > l+1$,
                the integral over $t$ scales like $(\Lambda\,R)^{l+1-s_l}$ and thus 
                $\Ir_l/\Ir_1\propto R^{l+1-s}\Lambda^{l+1-s_l}$. Since $l\geq2$, we have $l+1-s>0$, so that $\Ir_l/\Ir_1$ is a product of two
                factors which both tend to zero
                  for $R\to0$ and $\Lambda\to\infty$.
                \ei
                
                \noindent    Hence the dominant contribution in Eq.~(\ref{eq:g3high_Il}) comes from $l=1$, and scales as $R^{\,2s-4}$.
By injecting Eq.~(\ref{eq:I1}) into Eq.~(\ref{eq:g3high_Il}), using Eq.~(\ref{eq:phi_tilde_m=}), and omitting the subleading $l\neq1$ terms,
we get
 \begin{equation}
   \gt_3^{(3)}\big|_{\rm high}({\bf r}, \boldsymbol{ \rho})  \
\simeq
   \,  \ - \ 
 \frac{4^{3-s}\ 3^{\,s\,-\,3/2}}{\pi\ \sin^2(\pi s)\,\Gamma(s+2)^2}
   \ \, D_1
   \ z^3\,\lambda_T^{-3}
\,\ R^{\,2\,s-4} \ 
 \sum_{\md=-1}^{1}\,\big|\tilde{\phi}_{\md}(\Oo)\big|^2
 \label{eq:g3t_asympt}
 \end{equation}

 In Appendix~\ref{app:g3tilde_low+mixed}, we show that the
       low-wavevector and mixed contributions 
       $\gt_3^{(3)}\big|_{\rm low}$ and $\gt_3^{(3)}\big|_{\rm mixed}$
       are negligible compared to the $\gt_3^{(3)}\big|_{\rm high}\propto R^{2s-4}$ contribution. Hence, in Eq.~(\ref{eq:decomp_gt}), the second and third term are negligible, and the asymptotic behavior of $\gt_3^{(3)}({\bf r}, \boldsymbol{ \rho})$ for $R\to0$ is given by the RHS of Eq.~(\ref{eq:g3t_asympt}).

       \subsubsection{Cancellation between two-body and $l=0$ three-body contributions} \label{subsec:g33_l=0+g32}

       We turn to the evaluation of \,$g_3^{(2)} + g_3^{(3,l=0)}$\,.
       Since $j_0={\rm sinc}$, Eqs.~(\ref{eq:def_Phi_q}) and~(\ref{eq:def_Phi'_q'}) reduce for $l=0$ to
\be
  \Phi_{q}
   \ = \
  \dfrac{e^{-
     \frac{\sqrt{3}}{2}\,\widetilde{q}
      \, r}}{\sqrt{4\pi} \, r}
\ \ {\rm sinc}\left(\tfrac{\sqrt{3}}{2}\, q\,\rho\right)
\ - \
\big[ \, (r, \rho) \ \longrightarrow \ (r', \rho') \, \big]
\label{eq:Phi_q_l0}
\ee
and $\Phi'_{q'}=\Phi_{q'}$.
Summing up Eq.~(\ref{eqg32}) and the $l=0$ contribution to Eq.~(\ref{eqg33-2}) yields
\begin{multline}
  g_3^{(2)}({\bf r}, \boldsymbol{ \rho}) \ + \ g_3^{(3,l=0)}({\bf r}, \boldsymbol{ \rho}) \ = \ -
  c\,
\ \  \int_{\mathcal{C}_+}\,\frac{d E}{2\pi i}
  \ e^{-\beta\,E}
  \ \int_0^{\infty}\frac{dq\,q^2}{2\,\pi^2} \ \ t_2\left(E-\frac{3q^2}{4m}\right)
  \\
\times \left[ \, \chi_q  \ + \ 4\,\pi\ \Phi_{q}
  \ \int_0^{\infty}\frac{dq'\,q'^{\phantom{.}2}}{2\,\pi^2}
\ t_{3,0}( q, q';E)
\ t_2\left(E-\frac{3q'^{\phantom{.}2}}{4m}
\right)
\ \ \Phi_{q'} \, \right]
\label{eqg32-g33l=0}
\end{multline}
with
\be
\chi_q \ := \ \frac{e^{-\sqrt{3}\,\widetilde{q}\, r}}{r^2}
\, + \, \frac{e^{-\sqrt{3}\,\widetilde{q}\, r'}}{r'^{\phantom{.}2}}
\ -\ \frac{2 \ {\rm sinc}\!\left(\tfrac{\sqrt{3}}{2}q\, \|\boldsymbol{\rho}-\boldsymbol{\rho}'\|\right)\ e^{-\frac{\sqrt{3}}{2}\widetilde{q}\,(r+r')}}{r\,r'}
\label{eq:chi_q_l0}
\ee
In the limit where $R$ (and thus also $r$ and $\rho$) tend to 0, Eqs.~(\ref{eq:Phi_q_l0}) and (\ref{eq:chi_q_l0}) reduce to
  \be
  \Phi_q \ \, \underset{R\to0}{\sim} \ \, \frac{1}{\sqrt{4\,\pi}} \ \left(\frac{1}{r} \ - \ \frac{1}{r'} \,\right)
\ \ \ {\rm and} \ \ \
       \chi_q \ \underset{R\to0}{\sim} \ \left(\frac{1}{r} \ - \ \frac{1}{r'} \, \right)^2
\label{eq:chiq_phiq_sim}
\ee
Thus, if we simply take the limit $R\to0$ under the integral in
Eq.~(\ref{eqg32-g33l=0}), we find that
  \,$g_3^{(2)}({\bf r}, \boldsymbol{ \rho}) \, + \, g_3^{(3,l=0)}({\bf r}, \boldsymbol{ \rho})$\, behaves for $R\to0$ like
\begin{multline}
-c \ \left(\frac{1}{r} \ - \ \frac{1}{r'} \, \right)^2 \
   \int_{\mathcal{C}_+}\,\frac{d E}{2\pi i}
  \ e^{-\beta\,E}
  \ \int_0^{\infty}\frac{dq\,q^2}{2\,\pi^2} \ \ t_2\left(E-\frac{3q^2}{4m}\right)
 \\ \times \left[ \ 1 \ + \ 
  \int_0^{\infty}\frac{dq'\,q'^{\phantom{.}2}}{2\,\pi^2}
\ t_{3,0}( q, q';E)
\ t_2\left(E-\frac{3q'^{\phantom{.}2}}{4m}
\right) \
\right]
\label{eqg32-g33l=0bis}
\end{multline}
 This is the sum of two terms proportional
     to $\left(1/r-1/r'\right)^2 \, \propto \, R^{-2}$\,,
which cancel out due to the sum rule
\be
\boxed{
 \int_0^{\infty}\dfrac{d q'}{2\pi^2} \ q'^{\phantom{.}2}\ t_{3,0}(q,q';E\,)
 \ 
  t_2\left(E-\frac{3q'^{\phantom{.}2}}{4m}\right)
\  = \  -1}
\label{sumrulel=0}
\ee
valid for any complex value of the energy $E$ and any wavevector $q'$,
see Appendix \ref{appsumrule} for the derivation.
Hence, to evaluate
Eq.~(\ref{eqg32-g33l=0})
in the $R\to0$ limit, we need to go beyond the leading-order behavior given in Eq.~(\ref{eq:chiq_phiq_sim}).
This is done in Appendix~\ref{app:g32+g330}, where we show that
\,$g_3^{(2)} \, + \, g_3^{(3,l=0)}$\, is 
bounded, hence negligible compared to the leading $\propto R^{2s_1-4}$ term.

\subsection{Three-body contact}

We have shown in the previous section that the leading-order $R\to0$ behavior of \,$g_3(\rr,\rrho)$\, is given by the RHS of Eq.~(\ref{eq:g3t_asympt}).
The three-body contact is obtained by integrating \,$g_3$\, over the hyperangles,
see Eq.~(\ref{eq:g3_C3}).
This yields
\be
\mathcal{C}_3 \ \, \simeq \ \,
-\ \frac{3^{s+1}\ 4^{1-s}\ \pi^2}{(s+1)\ \sin^2(\pi\,s)\ \Gamma(s+2)^2\ \Nr^2}
\ \ D_1\ \, \left(\frac{\beta}{m}\right)^3\ n^3
\label{eqC3}
\ee
where we used $({\phi}_{\md}\,|\,{\phi}_{\md}\,) = \Nr^2\ \left(\tilde{\phi}_{\md}\,\big|\,\tilde{\phi}_{\md}\right) = 1$
as well as Eq.~(\ref{eq:boltz_eos}).

It remains to compute $D_1$, which is given by Eq.~(\ref{eqdefDl}),
with $\Br_1$ given by Eq.~(\ref{eqimt31}).
Accordingly, we first compute $t_3$ by numerically solving the STM equations on the real axis,
we then determine $\mathcal{B}_1$ by fitting the large-momentum tail of \,${\rm Im}\,t_3$, and we deduce $D_1$ by numerical integration.
More specifically,
we numerically determine the solution $t_{3,l=1}(q,q';E + i\,0^+)$ of the STM equation for positive real $E$
by solving Eqs.~(\ref{eq:STM_x}) and~(\ref{eq:STM_y}) (using linear grids for the variables $\theta$ and $u$\,).
We then extract numerically the prefactor $\Br_1$ appearing in the large-momentum behavior of \,${\rm Im}\,t_{3,1}$\,, see Eq.~(\ref{eqimt31}),
  and we deduce $D_1$ by numerical integration of Eq.~(\ref{eqdefDl}).

By dimensional analysis, $D_1\!\left(\beta/m \, ; \, a^{-1}\right)$ equals $(m/\beta)^{s+1}$ times a dimensionless function of 
$\lambda_T/a$. 
Hence, by Eq.~(\ref{eqC3}), $\Cr_3$ is of the form Eq.~(\ref{eq:C3_f}), where $f_3(\lambda_T/a)$ is a dimensionless function proportional to $D_1$.
The numerical procedure to compute $f_3$ is naturally carried out using dimensionless variables.
 Setting $k_0=\sqrt{m\,E}$\,,
 we tabulate the dimensionless function
 \be
 F\left(\frac{1}{k_0\,|a|}\right) \ \, = \ \, - \, \mathcal{B}_1(m\,E\,;a^{-1}) \ k_0^{-2\,s}
 \label{eq:def_F}
 \ee
 by solving the STM equation for different values of the single dimensionless parameter $\tfrac{1}{k_0\,|a|}$\,.
We then evaluate the function $f_3$ using
\begin{equation}
f_3\!\left(\frac{\lambda_T}{a}\right)
  \ = \
  \frac{3^{s+1} \, 2^{3-2s} \, \pi}{(s+1)  \, \sin^2(\pi s) \, \Gamma(s+2)^2 \, \Nr^2}\
  \int_0^{\infty} dx \ x^{-(2s+3)}\, \exp(-\frac{1}{x^2}) \ \, F\left(x \ \, \frac{\lambda_T}{\sqrt{2\pi}\ |a|}\right)
  \label{eq:f_vs_F}
\end{equation}
We find numerically that $F$ is a decreasing function, so that the integral is convergent.\footnote{More specifically, our data is compatible with the power law $F(y)\propto 1/y^{6-2s}$ for $y\to\infty$. This corresponds to $\Br_1 \propto E^3$\, for $E\to0$, which may be expected given that for the non-interacting problem, the scaling exponent $s$ is replaced with $s_{\rm free}=3$~\cite{SonPrivateF,SonHammer_production_3neutrons}.}
The obtained function $f_3$ is plotted in Fig.~\ref{figC3dimensionless}.
We have estimated all sources of numerical errors, and the final error is smaller than the linewidth in Fig.~\ref{figC3dimensionless}.

For $a=\infty$, Eq.~(\ref{eq:f_vs_F}) reduces to
\be  
f_3(0) \ = \ 
\frac{3^{s+1}\ 4^{1-s}\ \pi}{(s+1)^2\ \sin^2(\pi\,s)\ \Gamma(s+2)\ \Nr^2}
\ \ F(0)
\label{eqC3nUL}
\ee
Our numerical procedure yields $F(0)=0.29256(21)$,
and thus $f_3(0)=4.5566(33)$, in agreement with the analytical result Eq.~(\ref{eq:f(0)}).
We also computed $F(0)$ using a different approach and obtained a consistent result, as described in Appendix~\ref{AppImt3largek}, Section~\ref{AppasympUL}.

\section{Large momentum tail of the center-of-mass momentum distribution of nearby fermion pairs}\label{tail gamma}
In this section, we investigate the virial expansion of the center-of-mass momentum distribution of nearby fermion pairs $n_P(K\,)$.
We will check that we recover the large momentum asymptotic behavior Eq.~(\ref{eq:NP_tail}), with the {same expression of} the three-body contact $\mathcal{C}_3$ 
than in the previous section.
$n_P(K\,)$ is given by the pair propagator $\Gamma$
  at imaginary time $0^-$,
 \be
 n_P(K\,) \ = \ - \, m^2\ \Gamma(K,0^-)
 \label{eqnpGamma}
 \ee
[see, {\it e.g.}, Ref.~\cite{RossiContact} for the definition of $\Gamma$,
and footnote 8 of Article I for a derivation of Eq.~(\ref{eqnpGamma})].
By $\beta$-periodicity,
$\Gamma(K,0^{-})=\Gamma(K,\beta^{-})$.
As shown in Ref.~ \cite{SunVirial3}, the high temperature expansion of $\Gamma(K,\beta^-)$ is given by
  \,$\Gamma(K,\beta^-) = \Gamma^{(2)}(K,\beta^-)
  + \Gamma^{(3)}(K,\beta^-) + \ldots$
  where the order $z^2$ term reads
 \be
 \Gamma^{(2)}(K,\beta^-)
 \ = \
 z^2\,e^{-\beta\frac{K^2}{4m}}
 \ \int_{\mathcal{C}_{\gamma}}\, \frac{dE}{2\pi i}\ e^{-\beta E}\,t_2(E\,)
 \label{eq:Gamma(2)}
 \ee
while the order $z^3$ term reads\footnote{The expression~(F1) of Ref.~\cite{SunVirial3} corresponds to a single diagram; we thus multiply it by 2 to obtain the total $\Gamma^{(3)}$.}
\bea
\Gamma^{(3)}(K,\beta^{-})
\ &=&\ 
\frac{9 m\,z^3}{4 \pi^2 \beta K}
\ \int_0^{\infty}
dq\  q\ 
\ \left(e^{-\frac{3\,\beta(K-q)^2}{8 m}}    -e^{-\frac{3\,\beta(K+q)^2}{8 m}}\right)\nn\\
&&\times\int_{\mathcal{C}_+}
\, \frac{dE}{2\pi i}\ e^{-\beta\,E}
\ t_2\!\left(E-
\frac{3q^2}{4m}\right)
^2
\ \ \sum_{l=0}^\infty\ t_{3,l}(q,q;E\,)
\label{Gamma3}
\eea
Here we deformed the integration contour from $\Cr_\gamma$ to $\Cr_+$\,,
  which is justified since
  $e^{-\beta\,E}\to 0$ for $\Re E \to\infty$ and $t_2\!\left(E-
\frac{3q^2}{4m}\right)$ and $ t_{3,l}(q,q;E\,)$ are analytic in the whole complex  $E$ plane outside of the positive real axis.
From Eq.~(\ref{eq:Gamma(2)}), we see that $\Gamma^{(2)}(K,\beta^-)$ is a gaussian function of $K$ and therefore does not
give rise to a power-law large-$K$ tail. In contrast, $\Gamma^{(3)}$ does give rise to a power-law tail, as we will see.
We consider the limit $K\to\infty$,
{\it i.e.}\,, $K \gg {\rm Max}\left(\,|a|^{-1} \, , \, \lambda_T^{-1}\,\right)$.
The term $e^{-\frac{3\,\beta(K+q)^2}{8 m}} \leq e^{-\frac{3\,\beta K^2}{8 m}}$
gives a contribution to $\Gamma^{(3)}(K,\beta^{-})$ which is
exponentially small, and can thus be neglected compared to the
leading power-law behavior derived in the sequel.
We are left with the term $e^{-3\,\beta(K-q)^2/(8 m)}$\,,
which is a gaussian function of $q$, centered at $q=K$, of width $\sim \lambda_T^{-1} \ll K$. Hence the integral over $q$ is dominated by values of $q \simeq K\gg {\rm Max}\left(\,|a|^{-1} \, , \, \lambda_T^{-1}\,\right)$.
On the other hand, due to the factor $e^{-\beta\,E}$\,, the integral over $E$ is dominated by $E\lesssim 1/\beta$\,.
Hence $E-\frac{3q^2}{4m} \simeq -\frac{3q^2}{4m}$\,
{lies on the real negative axis, away from the branch cut of $t_2$\,,
so that
the only discontinuity of the integrand in Eq.~(\ref{Gamma3}) when $E$ crosses the positive real axis comes from $t_{3,l}$\,.}
Furthermore,    
$t_2\!\left(E-\frac{3q^2}{4m} \right) \simeq - \frac{8\pi}{\sqrt{3}\, m K}$\,.
Also using Eq.~(\ref{eq:t3l_im}), we obtain
\be
\Gamma^{(3)}(K,\beta^{-})
\ \simeq \ \,\frac{48\,z^3}{m \beta\,K^2}
\ \sum_{l=0}^\infty
\ \, \int_0^{\infty}
dq
\ e^{-\frac{3\,\beta(K-q)^2}{8 m}} 
\ \int_0^\infty\ \frac{dE}{\pi}\ e^{-\beta\,E}
\
\Im \, t_{3,l}(q,q \, ;E + i\,0^+)
\nn
\ee
From Eq.~(\ref{eqimt3l}), we have
$\Im\,t_{3,l}(q,q \, ;E+i0^{+})
\simeq \, m\ \mathcal{B}_l(m\,E\, ;\, a^{-1})/K^{\,2s_l+2}$\,.
       This gives a contribution $\propto \! 1/K^{\,2s_l+4}$ to $\Gamma^{(3)}(K,\beta^{-})$\,, so that the leading order comes from $l=1$.
       Moreover, to leading order, we can replace the lower bound of the integral over $q$ by $-\infty$.
       Finally,  we again use Eq.~(\ref{eq:boltz_eos}) to eliminate the fugacity $z$ in favor of $n$.
This yields
\be
\Gamma^{(3)}(K,\beta^{-})
\ \simeq \ 128\,\sqrt{3}\ \pi^5\ n^3\ \frac{\beta^3}{m^5}\ 
\dfrac{D_1}{K^{\,2s+4}}
\label{eq:gamm3n3}
\ee
   {Hence $\Gamma^{(3)}(K,\beta^{-})$ has a power-law decay at large $K$, which dominates over the gaussian decay of $\Gamma^{(2)}(K,\beta^{-})$ if we take the $K\to\infty$ limit before the non-degenerate limit.
     Furthermore, we expect that for $N\geq 4$, the order $z^N$ term $\Gamma^{(N)}(K,\beta^{-})$
     is negligible compared to $\Gamma^{(3)}(K,\beta^{-})$
     in the non-degenerate ($z\to0$) and large-$K$ limit,
     irrespectively of the order of the two limits.\footnote{Regarding $\Gamma^{(4)}(K,\beta^{-})$\,,
     which contains diagrams with a four-body T-matrix $T_4$ closed by two slashed
     single-particle lines propagating backwards in time, we expect that its large-$K$ asymptotic behavior contains
     a leading tail $ \propto \! z^4 / K^{\,2s+4}$,
     and a four-body tail  $ \propto \! z^4 / K^{\,2s^{(4)}+4}$
     (in agreement with footnote~10 of Article~I\,) which is subleading
     because, as already mentioned, $s^{(4)}>s$\,.}
     Using Eq.~(\ref{eqnpGamma}), we conclude that in the non-degenerate limit
(taken after the large $K$\, limit),
      \be
      \lim_{K\to\infty}\ K^{\,2s+4}\ n_P(K\,)  \
      \simeq \ -128\,\sqrt{3}\ \pi^5\ n^3 \ \left(\frac{\beta}{m}\right)^3\
      D_1
      \nn
      \ee
      This agrees with the general relation given by Eqs.~(\ref{eq:NP_tail}) and (\ref{eq:Mr}), with $\Cr_3$ given by the result from the previous section Eq.~(\ref{eqC3}).}

\section{Conclusion and outlook}\label{secconcl}
We have computed the three-body contact $\Cr_3$\,, a quantity describing three-body bunching in the BEC-BCS crossover, to leading order in the virial expansion. Using a wavefunction approach, we performed the calculation analytically at the unitary limit. We extended the computation to negative scattering length using the Feynman diagram technique. To do so, we clarified two important points: a cancellation between leading order two-body and three-body contributions, and a high-wavevector power-law behavior of the three-body T-matrix.
It seems likely that these two points will be key ingredients for any calculation of \,$\Cr_3$\, within the Feynman diagram technique.

The present work should be generalizable to the mass-imbalanced case below the threshold where the Efimov effect appears, which is relevant to experiments on mixtures of fermionic atoms.
Another natural extension would be to
treat positive scattering lengths. 
Furthermore, in the degenerate regime,
asymptotically exact computations should be possible in the weak-coupling BCS \cite{KaganGamma3Fermions} and strong-coupling BEC limits.

Computing $\Cr_3$ in the degenerate strongly correlated regime seems particularly challenging.
  Indeed,  a prerequisite for computing $\Cr_3$ is to obtain with good accuracy the exponent for the power-law scaling $g_3 \propto R^{2s-4}=R^{-0.455}$, since $\Cr_3$ is the prefactor of this power-law.
  BCS mean-field theory is far from meeting this requirement,
  since it yields
    \,$g_{3, {\rm BCS}} %(\rr_1,\rr_2,\rr_3)
  \propto
  R^{-2}$
in the superfluid phase\footnote{Within BCS mean-field theory, Wick's theorem is applicable, and yields
    \be
    g_{3, {\rm BCS}}(\rr_1,\rr_2,\rr_3)
 \ = \ n^3/8 \, + \, 
  \big[ \, |\Fr(r_{13})|^2 \, + \,  |\Fr(r_{23})|^2 \big] \, n/2
  \, - \,  g_1(r_{12})^2 \, n/2
  \, - \, 2 \, g_1(r_{12}) \ {\rm Re}\big[ \Fr(r_{13}) \Fr^*(r_{23}) \big]
  \nn
  \ee
  where $r_{ij} = \|\rr_i-\rr_j\|\,$, 
  $\ \ds \Fr(r)= \left\langle \hat{\psi}_\down(\rr) \, \hat{\psi}_\UP(\vn) \right\rangle\,$
  and \,$g_1(r) = \bigg\langle \hat{\psi}^\dagger_\UP(\rr) \, \hat{\psi}_\UP(\vn) \bigg\rangle$.
  For $r\to0$, we have
$\Fr(r)\sim m\,\Delta/(4\pi\hbar^2r)$~\cite{YvanBCS},
  and $g_1(r)\to n/2$\,.
Thus for $R\to0$,  
  $\,g_{3, {\rm BCS}}(\rr_1,\rr_2,\rr_3) \sim
  \big(1/r_{13}-1/r_{23}\big)^2\,m^2|\Delta|^2 n \, / \, (32\,\pi^2\hbar^4)
  = R^{-2} g(\Oo)$, with $\int d^5\Omega\ g(\Oo)>0\,$
  in the superfluid phase where $\Delta\neq0$.},
the same incorrect scaling than the one obtained in the non-degenerate limit by only keeping the two-body diagrams, see Eq.~(\ref{eq:g32_R-2}).
  In the fixed node quantum Monte Carlo approach~\cite{Giorgini,Carlson_C_relations,ForbesRange},
  in order to obtain the correct scaling of \,$g_3$\,,
it may be necessary to design new trial wavefunctions whose nodal surface is the one of the exact three-body wavefunction at small $R$.\,
A diagrammatic Monte Carlo computation~\cite{VanHouckeEOS,RossiEOS,RossiContact} might be feasible,
building on the insights from the present work on the emergence of the three-body correlations within the diagrammatic formalism.
  Another possibility would be to perform lattice quantum Monte Carlo computations
  and extrapolate to the continuum and thermodynamic limits~\cite{zhenyaPRL,Goulko_UFG_2016,AlhassidC,AlhassidPseudogapContinuum}.

The exact determination of \,$\Cr_3$\, in the non-degenerate limit presented here provides a unique possibility for benchmarking such challenging computations, as well as experimental determinations of \,$\Cr_3$\,
such as the one performed
at Heidelberg University
{\it via}\, the three-body loss rate~\cite{HeintzePrep}.
  Beyond cold atoms,
  our work may 
help computing
  the contribution of three-body forces to the equation of state of neutron matter in the dilute non-degenerate regime,
 with potential applications in astrophysics~\cite{HorowitzSchwenkVirial2006,CarboneSchwenkFiniteT_PRC2019,CarbonePRR2020,SchwenkHebelerFiniteT_PRC_2021,RivieccioNeutronVirial2025}.

\longthanks
We thank
F.~Chevy,
 S.~Endo,
  M.~Ga\l{}ka,
  C. Heintze,
S.~Jochim,
  P.~Lunt, C.~Salomon, and T.~Yefsah
for stimulating discussions.
We acknowledge support from
ANR through projects QSOFT (X.L., ANR-24-CE97-0007-01) and LODIS (F.W., ANR-21-CE30-0033).

\subsubsection*{Data availability}

A public repository~\cite{ZenodoC3II} contains
data files for the functions $f_3\left(\frac{\lambda_T}{a}\right)$ and $\,F\left(\frac{1}{k_0|a|}\right)\,$,
including their numerical errors and how they were obtained,
as well as the function
$q\mapsto t_{3,1}(q,q\,;E+i0^+)$ for relevant values of $\,\frac{1}{k_0|a|}\,$.

  \appendix

\section{Three-particle momentum distribution at large relative momenta}
\label{app:N3K}

In this appendix, we express the tail of the three-particle momentum distribution function in terms of \,$\Cr_3$\,.
The three-particle momentum distribution
is
  \be
  n_3(\,\kk_1,\kk_2 , \kk_3) \ = \ \big\langle \,
  \hat{c}^{\,\dagger}_{\kk_1 \UP} \,
  \hat{c}^{\,\dagger}_{\kk_2 \UP} \,
  \hat{c}^{\,\dagger}_{\kk_3 \down} \,
  \hat{c}_{\kk_3 \down} \,  
  \hat{c}_{\kk_2 \UP} \,
  \hat{c}_{\kk_1 \UP} 
  \, \big\rangle
\label{eq:def_n3}
\ee
or equivalently   \,$\big\langle \,
  \hat{c}_{\kk_1 \down}^{\,\dagger} \,
  \hat{c}^{\,\dagger}_{\kk_2 \down} \,
  \hat{c}^{\,\dagger}_{\kk_3 \UP} \,
  \hat{c}_{\kk_3 \UP} \,  
  \hat{c}_{\kk_2 \down} \,
  \hat{c}_{\kk_1 \down} 
\, \big\rangle$\,.
Here, $\hat{c}_{\kk \,\sigma}$ is the operator annihilating a particle
\vskip .1cm
\noindent of spin $\sigma$ and momentum $\kk$, {\it i.e.},
\be
\hat{c}_{\kk \,\sigma}  \ = \  \int d^3\!r \ \,  \frac{e^{-i\,\kk\cdot\rr}}{\sqrt{\Vr}} \ \, \hat{\psi}_\sigma(\rr)\,.
\label{eq:ck_vs_psir}
\ee
Accordingly, the normalization is
$\int n_3(\,\kk_1,\kk_2 , \kk_3) \,
d^3\!k_1 \, d^3\!k_2 \, d^3\!k_3
\,/\, (2\pi)^{9}
= n^3/8$\,.
The function $n_3$ may become experimentally accessible through high-resolution imaging in momentum space (for progress in this direction, see {\it e.g.} Refs.~\cite{Jochim_nk_noise,Clement_Mott_g2_g3_2021,Clement_g6_Mott-SF_2023,JochimImaging_PRL2025,WuQuench2025}).
If one measures the momenta of all the particles,
  then the number of $\UP\UP\down$ triplets such that
  the momentum of one of the $\UP$ particles is within a small volume $\delta_1$ containing $\kk_1$\,,
  the momentum of the other $\UP$ particle is within a small volume $\delta_2$ containing $\kk_2$\,,
and the momentum of the $\down$ particle is within a small volume $\delta_3$ containing $\kk_3$\,,
is
\,$n_3(\,\kk_1,\kk_2,\kk_3)\,\delta_1\delta_2\delta_3 / (2\pi)^9 \times \Vr^3/2$\,
(where the factor $1/2$ comes from the indistinguishability of the $\UP$ particles).

Given a triplet $(\,\kk_1,\kk_2,\kk_3)$ we define the total momentum
\be
\KK \ = \ \kk_1 + \kk_2 + \kk_3
\label{eq:def_KK}
\ee
and we collect the relative momenta into the 6-dimensional vector
\be
\kk \ = \ \Bigg( \ \tfrac{\kk_3-\kk_1}{2} \ , \
\tfrac{1}{\sqrt{3}}\,\bigg( \kk_2 - \tfrac{\kk_1+\kk_3}{2} \bigg) \  \Bigg)\,.
\label{eq:def_kk}
\ee
Let $\hat{k}=\kk/k$ be the unit vector corresponding to the direction of $\kk$.
We consider the three-particle momentum distribution function integrated over $\KK$ and $\hat{k}$\,,
\be
n_3(k) \ = \ \int \frac{d^3 K}{(2\pi)^3} \ \int \dhk \ \, n_3(\,\kk_1,\kk_2,\kk_3)\,.
\label{eq:n3k_def}
\ee
Here, $\dhk$\, is the differential solid angle on the unit-sphere in 6-dimensional space,
such that
\be
d^6\!k  \ = \  \dhk \ \, k^5 \ \dk\,.
\label{eq:def_dhk}
\ee
We find
\be
n_3(k)\  \Vr^2 \ \
  \underset{k\to\infty}{\sim} \ \
\Mr_3 \ \ \frac{\Cr_3}{k^{2s+8}}
\label{eq:N3K}
\ee
with the prefactor\footnote{The large exponent $2s+8 = 11.545449\ldots$ is unfavorable for the observability of the tail, but the large value of the prefactor $\Mr_3$ is favorable. These two effects essentially compensate each other, since
\,$\Mr_3/k^{2s+8} \,=\, (\Mr_3'\,/\,k\,)^{2s+8}$ where $\Mr_3' =  3.26\ldots$ is of order unity.}
\be
\Mr_3 \ = \ 48\sqrt{3}\ \pi^5 \, (s+1)\, \big[1+\cos(s\,\pi)\big]
\ \Gamma\!\left(s+\frac{3}{2}\right)^2
\ = \ 843140.7\ldots
\label{eq:Mr3}
\ee
 The factor \,$\Vr^2$\, in Eq.~(\ref{eq:N3K}) comes from the fact that the {\it total}\, number of triplets is \,$\propto \Vr^3$\,, whereas the number of triplets with large relative momenta is \,$\propto \Vr$\,.

To derive Eqs.~(\ref{eq:N3K}) and (\ref{eq:Mr3}), we start by substituting Eq.~(\ref{eq:ck_vs_psir}) into Eq.~(\ref{eq:def_n3}), which yields in first quantization
\begin{multline}
  n_3(\,\kk_1,\kk_2 , \kk_3) \  \Vr^3  \ = \ 
N_\UP\,(N_\UP-1)\,N_\down\,
\int d^3\!r_1 \, d^3\!r_2 \, d^3\!r_3 \,
d^3\!r'_1 \, d^3\!r'_2 \, d^3\!r'_3 \ 
\ \ \exp\!\Big[i\,\sum_{j=1}^3\kk_j \cdot (\rr_j-\rr'_j\,)\Big]
\\
\times \ \int d^3\!r_4 \ldots d^3\!r_N \ \, \psi^*(\rr_1, \rr_2, \rr_3, \rr_4, \ldots , \rr_N) \ 
\psi(\rr'_1, \rr'_2, \rr'_3, \rr_4, \ldots , \rr_N)\,.
\label{eq:n3_psi}
\end{multline}
We then change the integration variables to the center-of-mass and Jacobi coordinates,
$(\rr_1 , \rr_2 , \rr_3) \, \longrightarrow \, (\CC,\rr,\rrho)$
according to Eqs.~(\ref{eq:def_jaco}) and (\ref{eq:def_C}),
and similarly
$(\rr'_1 , \rr'_2 , \rr'_3) \, \longrightarrow \, (\CC',\rr',\rrho')$.
The relative positions of particles 1,2,3 are determined by the six-dimensional vectors \,$\RR=(\rr,\rrho)$\, and \,$\RR'=(\rr',\rrho')$\,.
Using
\,$\exp\!\big[i\sum_{j=1}^3\kk_j \cdot (\rr_j-\rr'_j)\big]
=
e^{i \, \KK \cdot (\CC-\CC')}
\, e^{i\,\kk\cdot(\RR-\RR')}
$\,,
we see that in the limit \,$k\to\infty$\, for fixed \,$\hat{k}$\, and \,$\KK$\,,
the integral Eq.~(\ref{eq:n3_psi}) is dominated by the small-hyperradius power-law singularity of the wavefunction, \,$\psi(\rr_1, \ldots , \rr_N) \propto R^{s-2}$\, and \,$\psi(\rr'_1, \rr'_2, \rr'_3, \rr_4, \ldots , \rr_N)\propto (R')^{s-2}$\,, see Eq.~(15) of Article~I. This yields
\begin{multline}
n_3(\,\kk_1,\kk_2 , \kk_3) \  \Vr^3  \ \underset{k\to\infty}{\sim} \ \ 
N_\UP\,(N_\UP-1)\,N_\down\ \left(\frac{\sqrt{3}}{2}\right)^6 \
\int d^3\!C \,
d^3\!C' \,
d^3\!r_4 \ldots d^3\!r_N \ \sum_{\md=-1}^1 \  \sum_{\md'=-1}^1
e^{i\,\KK\cdot(\CC-\CC')}
\\ \times B_{\md}(\CC;\rr_4 , \ldots , \rr_N)^*
\ B_{\md'}(\CC';\rr_4 , \ldots , \rr_N)
\ \int_0^\infty d\!R \ R^{s+3} \ \int d^5\Omega \ \, e^{i\,\kk\cdot\RR} \ \phi_{\md}(\Oo)^*
\\ \times \int_0^\infty d\!R' \ R'^{s+3} \ \int d^5\Omega' \ \, e^{-i\,\kk\cdot\RR'} \ \phi_{\md'}(\Oo')\,.
\end{multline}
By the stationary phase method,
the integrals over hyperangles are given by
\be
\int d^5\Omega \ \, e^{i\,\kk\cdot\RR} \ \phi_{\md}(\Oo)^* \ \ \underset{k R\to\infty}{\sim} \ \
\left(\frac{2\pi}{kR}\right)^{\frac{5}{2}} \ \left[ \,
  e^{i\,5\pi/4} \ \phi_{\md}\big(-\hat{k}\,\big)^* \ e^{-i k R}
  \ + \
  e^{-i\,5\pi/4} \ \phi_{\md}\big(\hat{k}\,\big)^* \ e^{i k R}
\,  \right]\,.
\ee
The integrals over $R$ and $R'$ are then given by the identity\footnote{This follows from Eqs.~4 and~9 in \S~3.761 of Ref.~\cite{Gradstein_7}.}
\,$\int_0^\infty du \, u^{s+1/2}\,e^{-i u} = \Gamma(s+3/2) \, e^{-i(s\pi/2 + 3\pi/4)}$\,.
After substitution into Eq.~(\ref{eq:n3k_def}),
the integration over \,$\KK$\, simply yields \,$\int d^3\!K\,e^{i\,\KK\cdot(\CC-\CC')}=(2\pi)^3\,\delta^3(\CC-\CC')$\,,
while the integration over \,$\hat{k}$\, is evaluated thanks to
the orthogonality relation
\,$\int d\hat{k} \  \phi_{\md}\big(\hat{k}\,\big)^*\,\phi_{\md'}\big(\hat{k}\,\big) = \delta_{\md,\md'}$\,
combined with \,$\phi_{\md}\big(-\hat{k}\,\big) = -\phi_{\md}\big(\hat{k}\,\big)$\,.
The result Eqs.~(\ref{eq:N3K}) and (\ref{eq:Mr3}) then follows from the expression of \,$C_{2,1}$\, in terms of the norm of \,$B_{\md}$\, given in Eq.~(17) of Article~I (given that \,$\Cr_3 = 2\,C_{2,1}/\Vr$\, for the homogeneous unpolarized gas).

\section{Three-body contact in the $l=0$ channel}
\label{app:C3_l0}

In addition to the leading-order expression of \,$N_3(\epsilon)$\, for $\epsilon\to0$ given by Eq.~(\ref{eq:N3}),
which comes from the $l=1$ sector of the three-body problem,
there is a subleading correction, coming from the
$l=0$ sector,
given by
\be
\frac{N'_3(\epsilon)}{\Vr'}
\ \underset{\epsilon\to0}{\sim} \ \ 
           {\Cr}'_3\ \, \epsilon^{\,2s'+2}
\label{eq:N'3}
\ee
where
\be
s'= s_{l=0} = 2.166221977\ldots
\label{eq:s'=}
\ee
and \,$\Cr'_3$\, is by definition the $l=0$ three-body contact density.

For a finite-range interaction supporting deeply bound dimers,
  the three-body loss rate, given to leading order by Eq.~(\ref{eq:Gamma3}),
includes a subleading term given by\footnote{This is derived in Section~3.1.1 of Article~I.
Note that Eq.~(\ref{eq:Gamma'_3}) corrects an error in the expression given in Article~I.}
  \be
  \frac{\Gamma'_3}{\Vr} \ \simeq \ -\,8 \, s' (s'+1) \ \frac{\hbar}{m}\ \Cr'_3\  \, {\rm Im} \, \bar{a}'_3\,.
  \label{eq:Gamma'_3}
  \ee
In the limit where the interaction range $b$ is small,
the $l=0$ contribution Eq.~(\ref{eq:Gamma'_3}) is negligible
\vskip .03cm
\noindent
compared to the $l=1$ contribution Eq.~(\ref{eq:Gamma3}),
because $\Gamma'_3/\Gamma_3 \propto {\rm Im} \, \bar{a}'_3 \, / \, {\rm Im} \, \bar{a}_3$\, scales like $b^{2(s'-s)}$.

For the unitary gas in the non-degenerate limit,
\,$\Cr'_3$\, can be computed analytically, by following the same steps than in Sec.~\ref{waveful}.
We obtain
\be
\Cr'_3(\mu,T) \ \simeq \ e^{3\beta\mu} \ \frac{(m \, k_B T)^{s'+5/2}}{\hbar^{2s'+5}} \times \frac{3^{3/2}}{2^{2 s'+5/2}\pi^{3/2} \, \Gamma(s'+2)}\
\ \ \ \ \ {\rm for} \ \ \ 
e^{\beta\mu}\to0
\label{eq:Cr'3}
\ee
and thus
\be
\Cr'_3(n,T) \ \simeq \ n^3 \ \left(\frac{m \, k_B T}{\hbar^2}\right)^{s'-2} \ \frac{3^{3/2}\pi^3}{2^{2s'+1}\, \Gamma(s'+2)}
\ \ \ \ \ {\rm for} \ \ \
n\,\lambda_T^3\to0.
\label{eq:Cr'3_can}
\ee
The $l=0$ result Eq.~(\ref{eq:Cr'3_can})
can be obtained
from the $l=1$ result Eqs.~(\ref{eq:C3_f}) and (\ref{eq:f(0)})
by replacing $s$ with $s'$,
and dividing by three,
due to the three values $\md\in\{-1,0,1\}$ for $l=1$.

While
\,$\Cr_3(n,T)\propto T^{s-2}$\,
    is a slowly {\it decreasing}\, function of $T$,
\,$\Cr'_3(n,T)\propto T^{s'-2}$\,
    is a slowly {\it increasing}\, function of~$T$.
This can be intuitively explained as follows.
    In the $l=1$ channel (resp.~\,$l=0$ channel), we have $s<2$ (resp.~\,$s'>2$). This implies that we have three-body bunching (resp.~anti-bunching):
    The wavefunction at small hyperradius diverges like $R^{s-2}$
    (resp.~vanishes like $R^{s'-2}$\,).
    When $T$ increases, both the bunching and anti-bunching effects become weaker: The prefactor of $R^{s-2}$ becomes {\it smaller} (weaker bunching, the wavefunction is {\it large} in a smaller range of $R$),
    whereas the prefactor of $R^{s'-2}$ becomes {\it larger} (weaker anti-bunching, the wavefunction is {\it small} in a smaller range of $R$).

\section{STM equation}
\label{app:STM}

In this appendix, we provide a derivation of the STM equation for the 3-body T-matrix,
together with some basic properties
(see also, {\it e.g.}, Refs.~\cite{STM,BedaqueVanKolckPLB1998,LeyronasBrodskyJETP2005,PetrovLesHouches2010}).
We use the convention 
    \be
    \hat{f}(E\,)  \ =  \ \int_0^\infty d\tau \ e^{E\tau}\ f(\tau)
    \nn
    \ee
    for the Laplace transform of  a function $f(\tau)$.
The Laplace transform of \,${T}_2({\bf p},\tau)$ is
 \be
 \hat{T}_2({\bf p},E\,) \ = \ t_2\left(E- \tfrac{p^2}{4m} \right)
\label{eq:T2_t2}
 \ee
with the expression
\be
t_2(E\,)\ = \ \frac{4\pi}{m\big(a^{-1}-\sqrt{-m E}\ \big)}
\label{eq:t2}
\ee
in terms of the principal square root.

\begin{figure}[h]
\begin{center}
\includegraphics[width=0.8\linewidth]{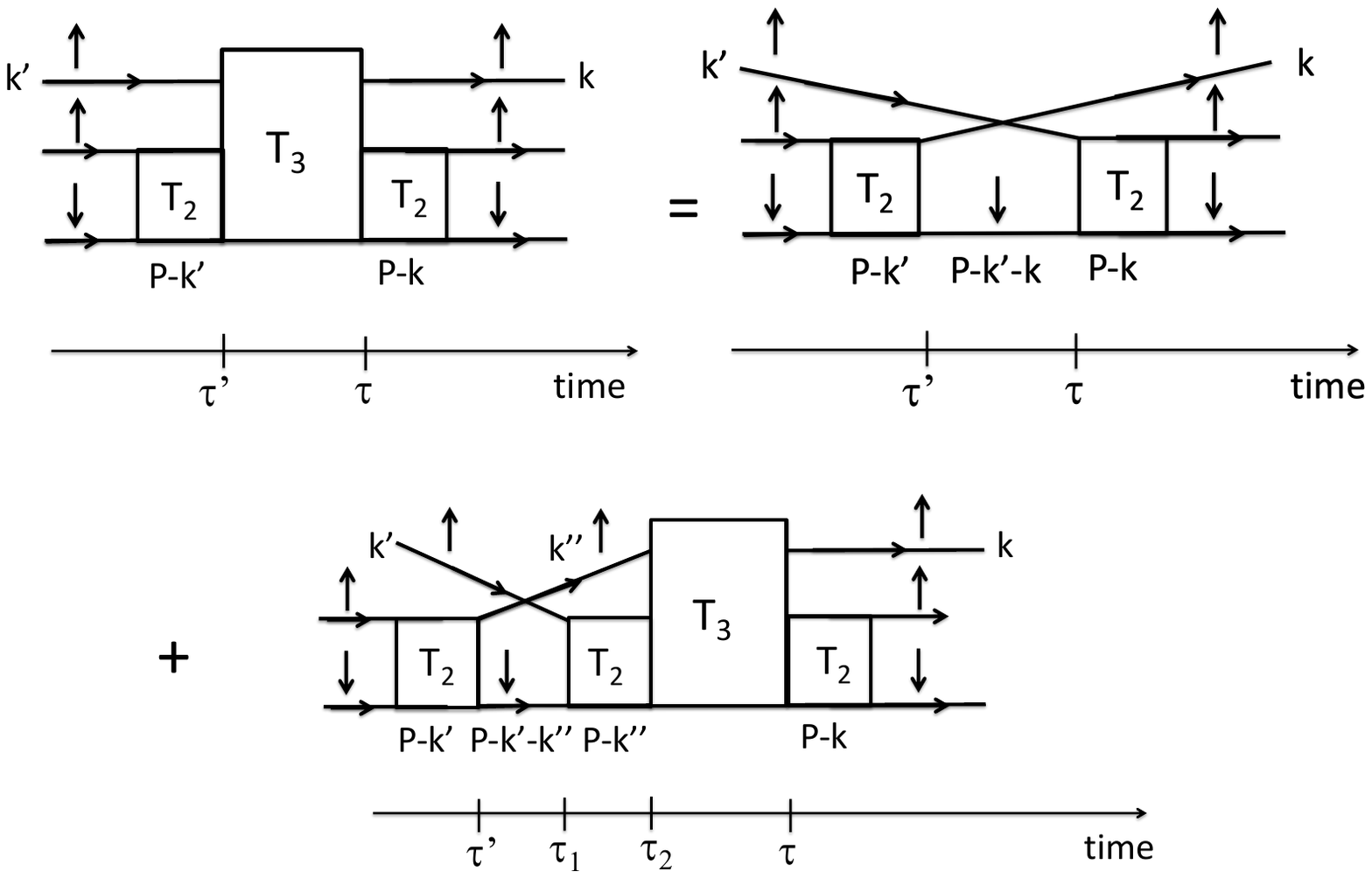}
\caption{
  Diagrammatic representation
  of the STM equation, which relates the two-body and three-body T-matrices $T_2$ and $T_3$\,.}
\label{figSTMequation}
\end{center}
\end{figure}

$T_3$ is given in terms of $T_2$ by a diagrammatic geometric series, see Fig.~\ref{fig:T3_serie}, which can be summed up as shown in Fig.~\ref{figSTMequation}.
The corresponding expression
in imaginary-time representation
reads
  \be
  T_3(\,{\bf k},{\bf k}' ; {\bf P} \, ; \, \tau-\tau')
  \ = \
  T_3^{B}(\,{\bf k},{\bf k}';{\bf P} \, ; \, \tau-\tau')
  \ + \
T_3^{\,{\rm int}}(\,{\bf k},{\bf k}';{\bf P}\, ; \, \tau-\tau')  
  \nn
  \ee
where the source term
(also known as the Born approximation) is
\be
T_3^{B}(\,{\bf k},{\bf k}';{\bf P}\,;\,\tau-\tau')
\ = \ 
-\,G^{(0,0)}(\,{\bf k'},\tau-\tau') \ G^{(0,0)}(\,{\bf k},\tau-\tau') \ G^{(0,0)}(\,{\bf P}-{\bf k'}-{\bf k} \, , \, \tau-\tau')
\label{eq:TB_G0}
\ee
and the integral term is
 \bea
 T_3^{\,{\rm int}}(\,{\bf k},{\bf k}';{\bf P}\,;\,\tau-\tau')
\ &=& \ 
-\int_{\mathcal{D}}
d\tau_1\,d\tau_2\,
  \int\frac{d^3k''}{(2\pi)^3}
  \ G^{(0,0)}(\,{\bf k'},\tau_1-\tau') \ G^{(0,0)}(\,{\bf k''},\tau_2-\tau')
 \nn
 \\
&& \times  G^{(0,0)}(\,{\bf P}-{\bf k'}-{\bf k''},\tau_1-\tau')\ 
 T_2(\,{\bf P}-{\bf k''},\tau_2-\tau_1)\ T_3(\,{\bf k},{\bf k''};{\bf P}\,;\,\tau-\tau_2)
 \label{eq:Tint_G0}
 \eea
 with the time integration domain $\mathcal{D}=\big\{\,
 (\tau_1,\tau_2)
 \ \big| \ \tau'<\tau_1<\tau_2<\tau \, \big\}$.

Using the expression
 $G^{(0,0)}({\bf p},\tau)=-e^{-\varepsilon_{{\bf p}}\tau}\ \theta(\tau)$ of the propagator of a particle of momentum~${\bf p}$ in vacuum during imaginary time $\tau$, with $\varepsilon_{{\bf p}}= \frac{p^2}{2m}$\, and $\theta$ the Heaviside function, Eqs.~(\ref{eq:TB_G0},\ref{eq:Tint_G0}) become
\bea
 T_3^{B}(\,{\bf k},{\bf k}';{\bf P}\,;\,\tau-\tau')
 \ &=& \
 e^{-(\varepsilon_{{\bf k'}}+\varepsilon_{{\bf k}}+\varepsilon_{{\bf P}-{\bf k'}-{\bf k}})(\tau-\tau')}
 \ \theta(\tau-\tau')
 \\
 T_3^{\,{\rm int}}(\,{\bf k},{\bf k}';{\bf P}\,;\,\tau-\tau')
\ &=& \ 
\int_{\mathcal{D}}
d\tau_1\,d\tau_2
\ \int\frac{d^3k''}{(2\pi)^3} \ 
e^{-(\varepsilon_{{\bf k'}}+\varepsilon_{{\bf k''}}+\varepsilon_{{\bf P}-{\bf k'}-{\bf k''}})(\tau_1-\tau')}
 \nonumber\\
 &&\times \ 
 \big(e^{-\varepsilon_{{\bf k''}}(\tau_2-\tau_1)}T_2(\,{\bf P}-{\bf k''},\tau_2-\tau_1)\big)
 \ \times \ T_3(\,{\bf k},{\bf k''};{\bf P}\,;\,\tau-\tau_2)
 \label{eq:T3int_tau}
 \end{eqnarray}
  As a function of the time difference $\tau-\tau'$,
  both $T_3^{B}$ and $T_3^{\,{\rm int}}$ are retarded,
  and we take their Laplace transforms
  with respect to $\tau-\tau'$.
  The Laplace transform of $T_3^{B}$ is
 \begin{eqnarray}
 \hat{T}_3^{B}(\,{\bf k},{\bf k}';{\bf P}\,;E\,)&=&\frac{1}{\varepsilon_{{\bf k'}}+\varepsilon_{{\bf k}}+\varepsilon_{{\bf P}-{\bf k'}-{\bf k}} -E}
 \end{eqnarray}
 In Eq.~(\ref{eq:T3int_tau}), the integration over times is a convolution product, so that its Laplace transform is the product of the Laplace transforms of the different terms,
 by the convolution theorem for Laplace transforms: 
\be
  \hat{T}_3^{\,{\rm int}}(\,{\bf k},{\bf k}';{\bf P}\,;E\,)
  \ = \ \int\frac{d^3k''}{(2\pi)^3}
\ \frac{1}{\varepsilon_{{\bf k'}}+\varepsilon_{{\bf k''}}+\varepsilon_{{\bf P}-{\bf k'}-{\bf k''}}-E} \ 
t_2\left(E-\tfrac{\|{\bf P}-{\bf k''}\|^2}{4m}-\varepsilon_{{\bf k''}} \right)
\, \hat{T}_3(\,{\bf k},{\bf k''};{\bf P}\,;E\,)
\nn
 \ee
 where we
used Eq.~(\ref{eq:T2_t2}).
This yields the integral equation for $\hat{T}_3$:
 \begin{equation}
   \hat{T}_3(\,{\bf k},{\bf k'};{\bf P}\,;E\,)
   \ = \ \frac{1}{\varepsilon_{{\bf k'}}+\varepsilon_{{\bf k}}+\varepsilon_{{\bf P}-{\bf k'}-{\bf k}}-E}
   \ + \ \int\frac{d^3k''}{(2\pi)^3} \ \,
\frac{t_2\left(E-\frac{\|{\bf P}-{\bf k''}\|^2}{4m}-\varepsilon_{{\bf k''}}\right)}{\varepsilon_{{\bf k'}}+\varepsilon_{{\bf k''}}+\varepsilon_{{\bf P}-{\bf k'}-{\bf k''}}-E}
\  \, \hat{T}_3(\,{\bf k},{\bf k''};{\bf P}\,;E\,)
 \end{equation}
 Finally, we {go to the center-of-mass} reference frame thanks to the change of variables
 ${\bf q}={\bf k}-{\bf P}/3$, ${\bf q'}={\bf k'}-{\bf P}/3$,
{${\bf q''}={\bf k''}-{\bf P}/3$,}
$E'=E-P^2/(6m)$ and the change of function\footnote{$T_3$ can be viewed
as a T-matrix for the collision between a single atom and a
pair of nearby atoms.
The variables in \,$t_3({\bf q},{\bf q'};E')$\, then have the following interpretation:
$E'$ is the energy ({\it i.e.} the frequency) in the center-of-mass frame,
${\bf q}$ is the outgoing momentum of the single atom
in the center-of-mass frame
({\it i.e.} half of the relative momentum between the single atom and the
pair),
and ${\bf q'}$ is the analogous incoming momentum.}
\be\hat{T}_3(\,{\bf k},{\bf k'};{\bf P}\,;E\,) \ = \  t_3({\bf q},{\bf q'};E')
\nn\ee
Renaming $E'$ as $E$, we arrive at the STM equation for $t_3$\,:
 \be
   t_3({\bf q} \, ,{\bf q'};E\,) \  =  \
   \frac{1}{\frac{q^2+q'^{\phantom{.}2}+ \, {\bf q}\cdot{\bf q'}}{m}-E}
   \ \, + \ \, \int\frac{d^3q''}{(2\pi)^3}
   \ \ \frac{t_2\left(E-\frac{3q''^{\phantom{.}2}}{4m} \right)}{\frac{q'^{\phantom{.}2}+q''^{\phantom{.}2}+ \, {\bf q'}\cdot\,{\bf q''}}{m}-E} \ \ 
   t_3({\bf q} \, ,{\bf q''};E\,)
   \label{eq:STMt3}
   \ee
      We assume that $t_3({\bf q} \, ,{\bf q''};E\,)$ is an analytic function of $E$ outside the real positive axis.
This is a plausible assumption because {\it (i)} we consider $1/a\leq0$, so that the three-body scattering continuum starts at $E=0$, and {\it (ii)} there are no three-body bound states, as shown in~\cite{kartavtsev_malyk_JPhysB}.

By rotational invariance, $t_3({\bf q}\,,{\bf q}';E\,)$ only depends on the angle between ${\bf q}\,$ and ${\bf q}'$ (besides $q$, $q'$ and $E$), so that
        one can expand it onto the basis of Legendre polynomials $P_l$\,,
\be
t_3({\bf q}\,,{\bf q}';E\,) \ = \ \sum_{l=0}^\infty \ P_l(\,\hat{q}\cdot\hat{q}\,'\,)\ t_{3,l}(q,q';E\,)
\label{eqt3Pl}
\ee
where $\hat{q}={\bf q}/q$ and $\hat{q}\,'={\bf q'}/q'$.
Injecting the decomposition Eq.~(\ref{eqt3Pl}) into Eq.~(\ref{eq:STMt3}), one arrives at decoupled STM equations for $t_{3,l}$\, which read~\cite{Leyronas_b3,TheseAlzetto,Petrov3fermions,PSS_PRA,EndoNaidonUeada_atom_dimer,castin_tignone,AlzettoAtomDimer}
  \begin{multline}
  t_{3,l}(q,q';E\,) \ = \ - \, \frac{m \, (2l+1)}{q\,q'}\
   \ Q_l\left(\frac{m E-q^2-q'^{\phantom{.}2}}{q q'}   \right)\\
\\ -
\,
 \frac{m}{q'}\ 
\int_0^{\infty} \,
\dfrac{d q''}{2\pi^2} \ \,q'' \
t_{3,l}(q,q'';E\,)
\ t_2\left(E- \frac{3q''^{\phantom{.}2}}{4m}\right)
\ Q_l\left(\frac{m E -q''^{\phantom{.}2} -q'^{\phantom{.}2}}{q' q''}\right) 
 \label{eq:STMt3l_gauche}
  \end{multline}
where $Q_l$ is the Legendre function of the second kind \cite{Abramowitz},
defined by
\be
Q_l(z) \ = \ \frac{1}{2} \ \int_{-1}^{1}\,dv\ \ \frac{P_l(v)}{z-v}
\label{eq:defQl}
\ee
for $z$ outside of the branch cut $[-1 ; 1]$.
Note that for $x$ real,
$Q_l(x+i\,0^+)$ is real for $|x| \geq 1$.
By~changing the directions of the propagators in Fig.~\ref{figSTMequation} and following the same derivation, one arrives at another form of the STM
equation,
 \be
   t_3({\bf q} \, ,{\bf q'};E\,) \   =  \
   \frac{1}{\frac{q^2+q'^{\phantom{.}2}+ \, {\bf q}\cdot{\bf q'}}{m}-E}
   \ \, + \ \, \int\frac{d^3q''}{(2\pi)^3}
   \ \ \frac{t_2\left(E-\frac{3q''^{\phantom{.}2}}{4m} \right)}{\frac{q^{\phantom{.}2}+q''^{\phantom{.}2}+ \, {\bf q}\cdot\,{\bf q''}}{m}-E} \ \ 
   t_3({\bf q''}, {\bf q'} ;E\,)
   \label{eq:STMt3_swap}
   \ee
leading to
\begin{multline}
  t_{3,l}(q,q';E\,) \ = \ - \, \ \frac{m\,(2l+1)}{q\,q'}\
   \ Q_l\left(\frac{m E-q^2-q'^{\phantom{.}2}}{q\,q'}   \right)\\
\\ -
\,
 \frac{m}{q}\ 
\int_0^{\infty} \,
\dfrac{d q''}{2\pi^2} \ \,q'' \
 Q_l\left(\frac{m E-q^2-q''^{\phantom{.}2}}{q\,q''}\right)
 \ t_2\left(E-
 \frac{3q''^{\phantom{.}2}}{4m}\right)
 \ t_{3,l}(q'',q';E\,)
 \label{eq:STMt3l}
\end{multline}

Taking the complex conjugate of Eq.~(\ref{eq:STMt3l}), and using the uniqueness of the solution of that equation, we obtain the identity $t_{3,l}(q,q' ; E^*) \ = \ t_{3,l}(q,q' ; E\,)^*$\,.
This implies that for real $E$,
        \be
        t_{3,l}(q,q' ; E+ i \, 0^+)
        \ - \ t_{3,l}(q,q' ; E+ i \, 0^-)
        \ = \ 2\,i\ \Im t_{3,l}(q,q' ; E+ i \, 0^+)
        \label{eq:t3l_im}
        \ee
Another useful identity is that $t_{3,l}$ is symmetric,
  \be
  t_{3,l}(q,q' ; E\,) \ = \ t_{3,l}(q',q ; E\,)
  \label{eq:t3_sym}
  \ee
       To justify this, let us start from Eq.~(\ref{eq:STMt3l_gauche}), and set
       \,$\tilde{t}_{3,l}(q',q ; E) \, := \, t_{3,l}(q , q' ; E)$.
       By exchanging the dummy variables $q$ and $q'$\, we find that $\tilde{t}_{3,l}(q,q' ; E)$\, solves Eq.~(\ref{eq:STMt3l}). Hence by uniqueness of the solution of Eq.~(\ref{eq:STMt3l}),
       $\tilde{t}_{3,l}(q,q' ; E) \,=\, t_{3,l}(q,q' ; E)$.

 \section{Derivation of the expressions of $\,\mathcal{G}_3^{(2a)}$ and $\,\mathcal{G}_3^{(3a)}$ in terms of $t_2$ and $t_3$}

 \label{app:G3}

 In this appendix we provide the derivations of the expressions Eqs.~(\ref{eqG32a}) and (\ref{eqmG33a}) for the contributions $\,\mathcal{G}_3^{(2a)}$ and $\,\mathcal{G}_3^{(3a)}$ to the three-body propagator.

  \begin{figure}[h]
\begin{center}
\includegraphics[width=0.7\linewidth, trim={0 5cm 0 4cm}, clip]{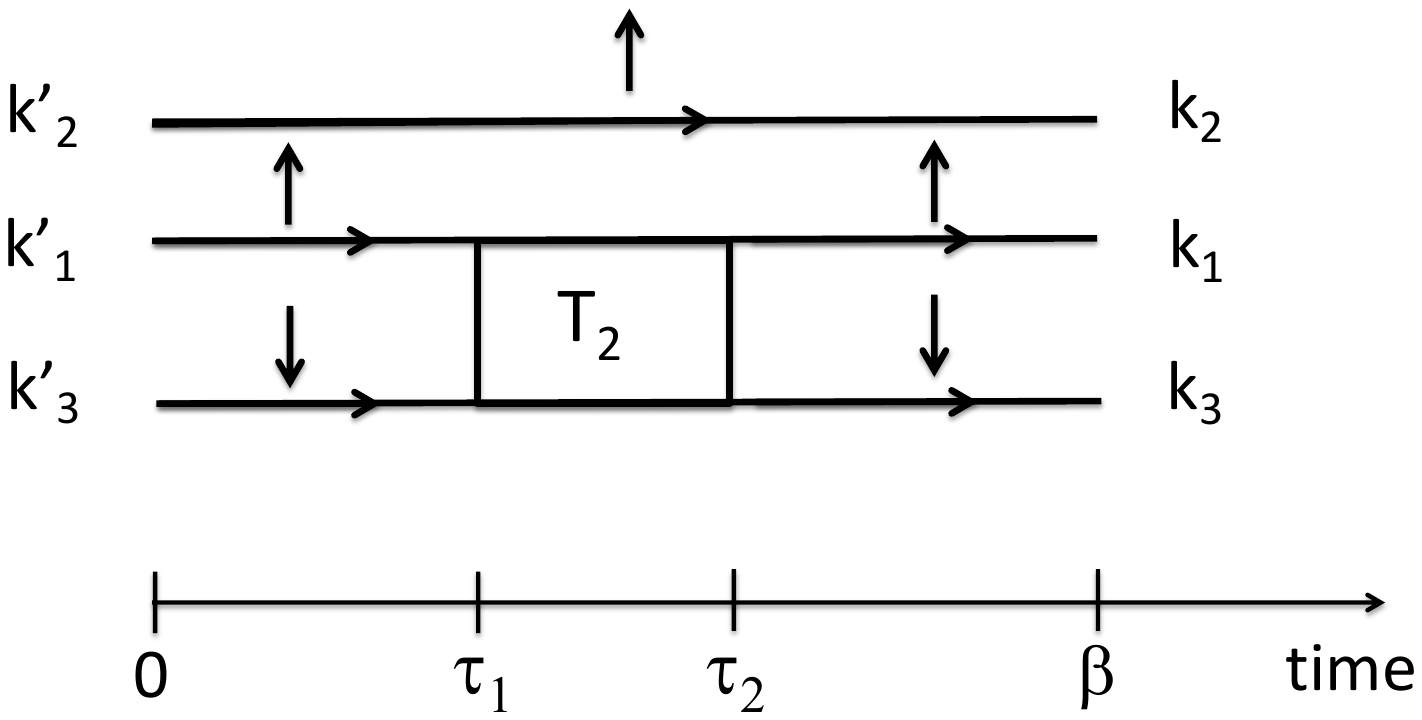}
\caption{Diagrammatic representation 
  in imaginary time
  of the contribution $\mathcal{G}_3^{(2a)}$\, to the three-body propagator.
}
\label{figG32atime}
\end{center}
\end{figure}

  From Figure~\ref{figG32atime},
the expression for $\mathcal{G}_3^{(2a)}$ in the imaginary time representation is\footnote{The factor $\Vr^{-1}$ comes from the interaction hamiltonian which is of the form
$\frac{g_0}{\Vr}\sum' c^{\dagger}_{\cdot\uparrow}c^{\dagger}_{\cdot\downarrow}c_{\cdot\downarrow}c_{\cdot\uparrow}$ where $g_0$ is the bare coupling constant and the primed sum involves a UV cutoff.
Note that, contrary to self-energy diagrams, there is no summation over an internal momentum which cancels the $1/\Vr$ factor.}
\begin{eqnarray}
\mathcal{G}_3^{(2a)}(\,{\bf K},{\bf K'};\beta) \ &=& \ z^3 \ \delta_{{\bf k'}_2,{\bf k}_2}\ \Vr^{-1} \ 
\int_{\mathcal{D}} d\tau_1\, d\tau_2\ 
G^{(0,0)}(\,{\bf k}'_2\,,\beta)\,G^{(0,0)}(\,{\bf k}'_1\,,\tau_1)\,G^{(0,0)}(\,{\bf k}'_3\,,\tau_1)\nonumber\\
& & \times \ T_2(\,{\bf k}'_1+{\bf k}'_3 \, , \, \tau_2-\tau_1)
\ G^{(0,0)}(\,{\bf k}_1,\beta-\tau_2)\,G^{(0,0)}(\,{\bf k}_3\,,\beta-\tau_2)\nonumber\\
&=& \ z^3 \ \delta_{{\bf k}'_2,{\bf k}_2}\ \Vr^{-1} \ 
\int_{\mathcal{D}} d\tau_1\, d\tau_2\ 
e^{
-(\varepsilon_{{\bf k}'_1}+\varepsilon_{{\bf k}'_2}+\varepsilon_{{\bf k}'_3})\tau_1
}
 \ \big(
e^{
\varepsilon_{{\bf k}'_2}(\tau_2-\tau_1)
}
\,T_2(\,{\bf k}'_1+{\bf k}'_3\,,\tau_2-\tau_1)
\big)\nonumber\\
&&\times  \ e^{
-(\varepsilon_{{\bf k}_1}+\varepsilon_{{\bf k}_2}+\varepsilon_{{\bf k}_3})(\beta-\tau_2)
}
\end{eqnarray}
where the time integration domain is $\mathcal{D}= \big\{ \, (\tau_1,\tau_2) \ \big| \ 0<\tau_1<\tau_2<\beta \, \big\}$.
The last expression is a convolution product of three functions, evaluated at imaginary time $\beta$. Therefore it is the inverse Laplace transform, evaluated at time $\beta$, of the product of the Laplace transform of the three functions,
\be
\mathcal{G}_3^{(2a)}(\,{\bf K},{\bf K}';\beta)
\ = \ z^3\ \Vr^{-1} \ \delta_{{\bf k}_2,{\bf k}_2'}
\ \int_{\mathcal{C}_{\gamma}}\, \frac{dE}{2\pi i}
\,\,
e^{-\beta\,E}
\ \frac{ \hat{T}_2\left(\,{\bf P}-{\bf k}_2 \, , \, E-\frac{{k}_2^2}{2m}\right)}
        {\left(E-\tfrac{k_1'^{\phantom{.}2}+k_2'^{\phantom{.}2}+
k_3'^{\phantom{.}2}}{2m}\right)
          \left(E-\tfrac{k_1^2+k_2^2+k_3^2}{2m}\right)}\label{eqG32a1}
\ee
where
$\mathcal{C}_{\gamma}$ is the Bromwich contour in the complex $E$ plane defined by $\Re(E\,)=\gamma$ with increasing imaginary part,
such that the integrand is an analytic function of $E\,$ for $\Re(E\,)<\gamma$.
We then use the expression~(\ref{eq:T2_t2}) of $\hat{T}_2$ in terms of $t_2$.
We also change from the momentum variables
$\,\kk_1\,, \kk_2\,, \kk_3\,,
\kk'_1\,, \kk'_2\,, \kk'_3\,$
to the total momentum $\PP$ and the
relative momenta
$\pp,\qq\,,\pp',\qq'$
defined in Eqs. (\ref{eq:defP}), (\ref{eqdefpq}) and (\ref{eqdefpq'}), and
we introduce the energy of the three particles in their center-of-mass frame
\be
E' \ = \ E-\dfrac{{P}^2}{6m}
\label{eqdefEprime}
\ee
Equation~(\ref{eqG32a1}) then reduces to Eq.~(\ref{eqG32a}),
after renaming the integration variable from $E'$ to $E$.
We have deformed the integration contour from a vertical Bromwich contour to the contour $\Cr_+$ surrounding the real positive axis (see Fig.~\ref{fig:C+}),
which is justified because the integrand is analytic outside of the real positive axis, and the contribution from the large quarter-circles vanishes since the modulus of the integrand is  $O\big(|E|^{-3/2}\big)$.

\begin{figure}[h]
\begin{center}
  \includegraphics[width=0.7\linewidth, trim={0 5cm 0 4cm}, clip]{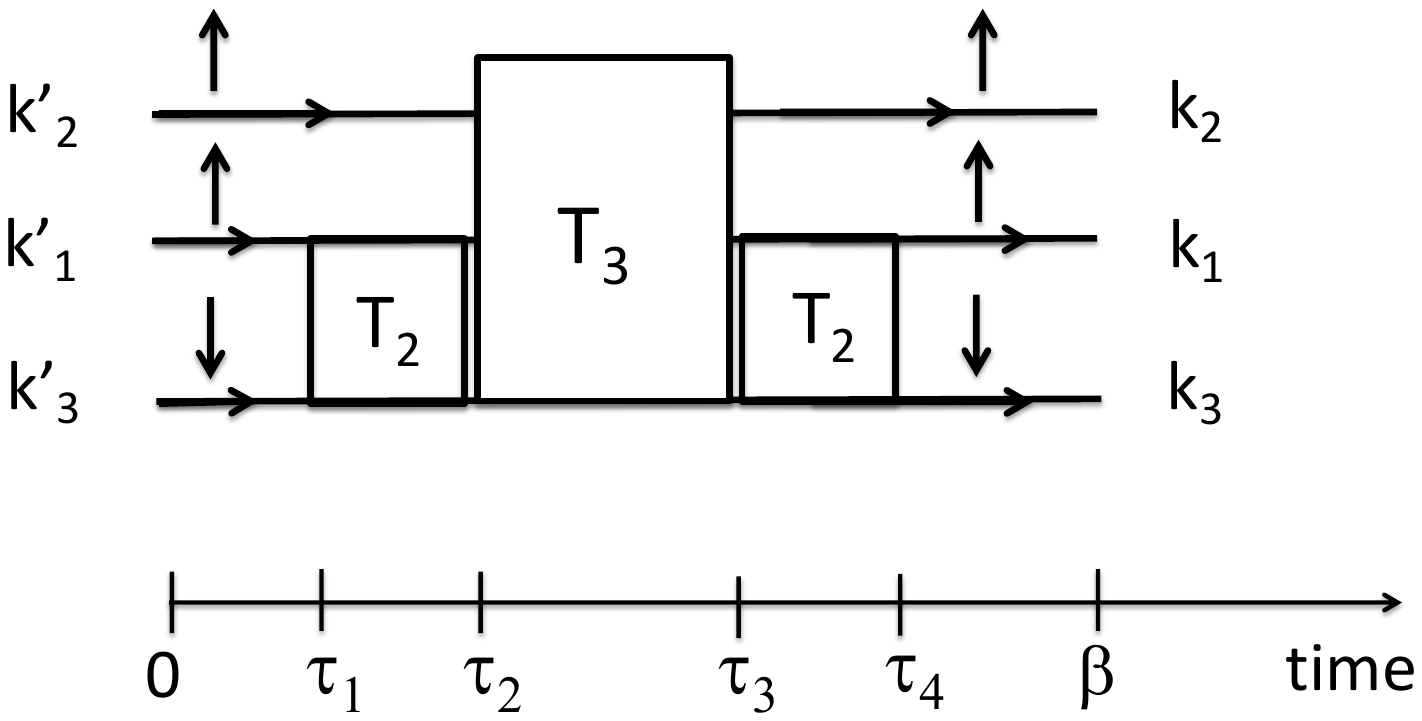}
       % trim={<left> <lower> <right> <upper>}
\caption{Diagrammatic representation 
  in imaginary time
  of the contribution $\mathcal{G}_3^{(3a)}$\, to the three-body propagator.}
\label{figG33atime}
\end{center}
\end{figure}
The expression of $\mathcal{G}_3^{(3a)}$ is (see Fig.~\ref{figG33atime})
\begin{eqnarray}
  \mathcal{G}_3^{(3a)}(\,{\bf K},{\bf K}';\beta)&=&z^3\,\Vr^{-2}\int_{\mathcal{D}}
  d\tau_1\,
  d\tau_2\,
  d\tau_3\,
  d\tau_4\ 
  G^{(0,0)}(\,{\bf k}'_2\,,\tau_2)
G^{(0,0)}(\,{\bf k}'_1\,,\tau_1)G^{(0,0)}(\,{\bf k}'_3\,,\tau_1)\nonumber\\
&&\times\ T_2(\,{\bf k}'_1+{\bf k}'_3\,,\tau_2-\tau_1)T_3(\,{\bf k}_2\,,{\bf k}'_2;{\bf P}\,;\tau_3-\tau_2)
T_2(\,{\bf k}_1+{\bf k}_3\,,\tau_4-\tau_3)\nonumber\\
&&\times\ G^{(0,0)}(\,{\bf k}_2\,,\beta-\tau_3)G^{(0,0)}(\,{\bf k}_1\,,\beta-\tau_4)G^{(0,0)}(\,{\bf k}_3\,,\beta-\tau_4)\nonumber\\
&=&z^3\,\Vr^{-2}\int_{\mathcal{D}}
  d\tau_1\,
  d\tau_2\,
  d\tau_3\,
  d\tau_4\ 
e^{
-(
\varepsilon_{{\bf k}'_1}+\varepsilon_{{\bf k}'_2}+\varepsilon_{{\bf k}'_3}
)
(\tau_1)
}
\times \big(
e^{-\varepsilon_{{\bf k}'_2}(\tau_2-\tau_1)}
T_2(\,{\bf k}'_1+{\bf k}'_3\,,\tau_2-\tau_1)
\big)\nonumber\\
&&\times \ T_3(\,{\bf k}_2,{\bf k}'_2\,;{\bf P}\,;\tau_3-\tau_2)
\times\big(e^{-\varepsilon_{{\bf k}_2}(\tau_4-\tau_3)}T_2(\,{\bf k}_1+{\bf k}_3\,,\tau_4-\tau_3)\big)
\times e^{
-(\varepsilon_{{\bf k}_1}+\varepsilon_{{\bf k}_2}+\varepsilon_{{\bf k}_3})
(\beta-\tau_4)
}\nonumber
\end{eqnarray}
where the time integration domain is $\mathcal{D}= \big\{ \, (\tau_1,\tau_2,\tau_3,\tau_4) \ \big| \ 0<\tau_1<\tau_2<\tau_3<\tau_4<\beta\, \big\}$,
and the factor $\Vr^{-2}$ comes from the two external $T_2$'s.
The last expression is a convolution product of five functions, evaluated at imaginary time $\beta$. Therefore it is the inverse Laplace transform, evaluated at time $\beta$, of the product of the Laplace transform of the five functions,
 \be
   \mathcal{G}_3^{(3a)}(\,{\bf K},{\bf K}';\beta) \ = \
   \frac{z^3}{\Vr^{2}}
\  \int_{\mathcal{C}_{\gamma}}\ \frac{dE}{2\pi i}\ e^{-\beta E} \
\dfrac{\hat{T}_2\Big({\bf P}-{\bf k}_2 \, , \, E-\frac{k_2^2}{2m}\Big)\, \hat{T}_3(\,{\bf k}_2 \, , \, {\bf k}'_2 \, ; \, {\bf P} \, ; \, E)\,
 \hat{T}_2\Big({\bf P}-{\bf k}'_2 \, , \, E-\frac{k_2'^{\phantom{.}2}}{2m}\Big)
}{\Big(E-\frac{k_1^2+k_2^2+k_3^2}{2m}\Big)
\  \Big(E-\frac{k_1'^{\phantom{.}2}+k_2'^{\phantom{.}2}+k_3'^{\phantom{.}2}}{2m}\Big)}
\label{eq:G3_3a}
 \ee
Here, $\hat{T}_3(\,{\bf k}_2 \, , \, {\bf k}'_2 \, ; \, {\bf P} \, ; \, E  )$ is the Laplace transform of
   $T_3(\,{\bf k}_2 \, , \, {\bf k}'_2 \, ; \, {\bf P} \, ; \, \tau )$\,, with ${\bf k}_2$ (resp.~${\bf k}'_2$\,) the momentum of the outgoing (resp.~ingoing) single-particle propagator,
   and ${\bf P}$ the total momentum of the three particles.
   Expressing once again the result in terms of the variables ${\bf P}$, ${\bf q}$, ${\bf q}'$, ${\bf p}$, ${\bf p}'$ and $E'$ defined in Eqs.~(\ref{eq:defP}), (\ref{eqdefpq}), (\ref{eqdefpq'}) and (\ref{eqdefEprime}),
and using
Eq.~(\ref{eq:T2_t2}) as well as $\hat{T}_3(\,{\bf k}_{2}\,, {\bf k}'_{2}\,; {\bf P}\,; E)=t_3({\bf q},{\bf q}';E')$\,,
we obtain Eq.~(\ref{eqmG33a}),
after renaming the integration variable $E'$ as $E$.
We have again deformed the integration contour from a vertical Bromwich contour to the contour $\Cr_+$ surrounding the real positive axis, 
which is justified because the integrand is analytic outside of the real positive axis, and assuming that $t_3$ does not grow too quickly as a function of $E$ so that there is no contribution from infinity when deforming the contour.

\section{ Low-wavevector and mixed contributions to \,$\gt_3^{(3)}$}\label{app:g3tilde_low+mixed}

In this appendix,
we show that the low-wavevector contribution
       \,$\gt_3^{(3)}\big|_{\rm low}$\, and the mixed contribution \,$\gt_3^{(3)}\big|_{\rm mixed}$\,
are negligible compared to the high-wavevector contribution \,$\gt_3^{(3)}\big|_{\rm high}\propto R^{2s-4}$.
   
   \subsection{Mixed contribution} \label{sec:g33mixed}
   
   In this subsection we consider
   $\gt_3^{(3)}\big|_{\rm mixed}$\,,
   given by Eq.~(\ref{eq:gt3mixed}),
    which we decompose over angular momentum,
    \be
    \gt_3^{(3)}\big|_{\rm mixed}
     \ = \  \sum_{l=1}^\infty \ \gt_3^{(3,l)}\big|_{{\rm mixed}}
     \nn
     \ee
 Let us first consider
 \{$q<\Lambda$\,, $q'>\Lambda$\}.
Since $q<\Lambda$,
we can consider
that $q\rho , \, |\tilde{q}\,|\,r \ll 1$,
so that in Eq.~(\ref{eq:def_Phi_q}), we can approximate the exponential by 1,
 and the spherical Bessel function by its asymptotic behavior
 \be
 j_l(t) \ \underset{t\to0}{\sim} \ \, c_l\ t^l
 \label{eq:jl_small}
\ee
with $c_l$ a constant.
On the other hand, since $q'>\Lambda$,
as in Section~\ref{subsec:gtilde33},
we have $t_2\big(E-\tfrac{3q'^{\phantom{.}2}}{4m}\big) \simeq -\,8\pi/(\sqrt{3}m\,q')$,
and we can set $\tilde{q}\,'\simeq q'$ in the expression~(\ref{eq:def_Phi'_q'}) of \,$\Phi'_{q'}$\,.
Furthermore, we replace $t_{3,l}$\, by its asymptotic behavior\footnote{
We have numerically checked Eq.~(\ref{eq:t3_q'_large}) for $a=\infty$ and $l=1$.
We also provide the following partial justification of Eq.~(\ref{eq:t3_q'_large}).
Recall that $t_{3,l}$\, is a symmetric function of $(q,q')$,
see Eq.~(\ref{eq:t3_sym}).
Thus Eq.~(\ref{eq:t3_q'_large}) is equivalent to
\,$t_{3,l}(q,q';x+i\,0^+) \, \sim \, A_l(q';x)/q^{s_l+1}$\, for \,$q\to\infty$\,.
To justify the latter asymptotic behavior, let us check that the ansatz
\,$t_{3,l}(q,q';x+i\,0^+) \, = \, A_l(q';x)/q^{s_l+1}$\,
solves the STM equation Eq.~(\ref{eq:STMt3l})
to leading order in the limit \,$q\to\infty$\,.
  We restrict to the case $s_l<l+1$ for simplicity.
 Then the first term
  on the RHS of Eq.~(\ref{eq:STMt3l}), which
  by~(\ref{eq:Ql_large_z})
  behaves as $1/q^{l+2}$,
  is negligible compared to the LHS of Eq.~(\ref{eq:STMt3l}).
 In the integral term, we consider the 
 domain $q''>\Lambda$,
 where $\Lambda$ is such that
 $q\gg \Lambda \gg k_0, \,|a|^{-1}$.
   To leading order,
 we can replace $k_0$ and $a^{-1}$ by zero,
 and we find for the integral term
$A_l(q';x)$ multiplied by
\be
     \frac{4}{\sqrt{3}\,\pi \,  q} \, \int_{\Lambda}^{\infty}
\, \frac{Q_l\big(-q/q''-q''/q \, \big)}{(q'')^{s_l+1}}
     \ dq''
     \ = \ 
     \frac{4}{\sqrt{3}\,\pi\,q^{s_{ l} +1}} \  \int_{\Lambda/q}^{\infty}
     \ \frac{Q_l(-t-1/t)}{t^{s_l+1}}\ dt
     \nn
     \ee
where we can replace the lower bound by 0 since  $q\gg \Lambda$.
The LHS of
(\ref{eq:STMt3l})
  is $A_l(q\,;x)/q^{s_l+1}$.
  Thus (\ref{eq:STMt3l}) reduces to
  the equation
\begin{equation}
  1 \ = \ \frac{4}{\sqrt{3}\,\pi} \, \int_0^{\infty}
 \ \frac{Q_l(-t-1/t)}{t^{s_l+1}}
  \ dt
\end{equation}
Using the change of variable $t=e^u$,
and writing $e^{-s_l u}={\rm cosh}(s_l u)-{\rm sinh}(s_l u)$, the contribution of the $\sinh$ vanishes by parity,
and we are left with
\begin{equation}
  1\ = \ \frac{4}{\sqrt{3}\,\pi} \, \int_{-\infty}^{\infty} \, Q_l(-2 \,\cosh u ) \ {\rm cosh}(s_l u) \, du
  \label{eq:eqs}
\end{equation}
This equation indeed holds: Comparing to Eq.~(\ref{eq:lambdal}), we see that the RHS of (\ref{eq:eqs}) is $\lambda(i s_l)$\,,
which equals 1, see the discussion around Eqs.~(\ref{eq:stilde_l0})
and~(\ref{eq:stilde_l>0}).
}
\be
t_{3,l}(q,q';x+i\,0^+) \ \underset{q'\to\infty}{\sim} \ \ \frac{A_l(q;x)}{(q')^{s_l+1}}
\label{eq:t3_q'_large}
\ee
with  $A_l(q;x)$ which behaves as $q^{-(s_l+1)}$ for large $q$.
This yields
   \begin{equation}
 \gt_3^{(3,l)}\big|_{{\rm mixed}} \ \propto \     R^{l-2}\int_0^{\Lambda}d q \, q^{2+l}\ t_2\left(E-
\frac{3q^2}{4m}
     \right)\ A_l(q;E\,) \ 
  \int_{\Lambda}^{\infty}d q'
\   \exp\left(-\,\frac{\sqrt{3}}{2}q'\,r\right)\ j_l\left(\frac{\sqrt{3}}{2}q'\,\rho\right)\ (q')^{-s_l}
   \end{equation}
where we omitted a prefactor which depends on $\alpha$ and $\hat{\rho}$\,. The change of variable $t=\frac{\sqrt{3}}{2}q' R$\, yields
  \begin{multline} 
 \gt_3^{(3,l)}\big|_{{\rm mixed}} \ \propto \     R^{s_l-3+l}
    \,\left[ \, \int_0^{\Lambda}d q \ q^{2+l}\
      t_2\left(E-\frac{3q^2}{4m}
      \right)
      \ A_l(q;E\,) \right]
    \left[\int_{\frac{\sqrt{3}}{2} \, \Lambda\,R}^{\infty}
      \, d t\,
    t^{-s_l}\ e^{-t\,\sin \alpha}\ j_l\big(t\,\cos\alpha\big)
    \right]
    \\
    \label{eqprod2int}
  \end{multline}
  which we
have to estimate for $\Lambda\to\infty$ and $R\,\Lambda\to 0$.
  We distinguish two cases.
  \begin{itemize}
  \item For $1+l-s_l>0$,
    the integral over $q$
      diverges like $\Lambda^{1+l-s_l}$ for $\Lambda\to\infty$,
    whereas the integral over $t$ converges in the neighborhood of $t=0$ so that we can replace it by a constant. Hence
    $ \gt_3^{(3,l)}\big|_{{\rm mixed}} \propto  R^{s_l-3+l}\,\Lambda^{1+l-s_l}$. Compared to the high-wavevector contribution $\gt_3^{(3)}\big|_{\rm high}\propto \! R^{\,2s_l-4}$, we find a ratio $\gt_3^{(3,l)}\big|_{{\rm mixed}}
\, / \, \gt_3^{(3)}\big|_{\rm high} \propto
    (\Lambda\,R)^{1+l-s_l} \ll 1$.

  \item For $1+l-s_l<0$,
     the integral over $q$ converges for $\Lambda\to\infty$, whereas
    the integral over $t$ diverges as $(\Lambda\,R)^{1+l-s_l}$ for $\Lambda R\to0$. Hence
   $ \gt_3^{(3,l)}\big|_{{\rm mixed}} \propto  R^{\,2l-2}\,\Lambda^{1+l-s_l}$.
 We have $l\geq2$ (indeed,
 $l\neq0$ by definition of $\gt_3$,
     and  $l\neq1$ because the condition
$1+l-s_l<0$ is violated for $l=1$).
   Hence
   $\gt_3^{(3,l)}\big|_{{\rm mixed}}
     \, / \, \gt_3^{(3)}\big|_{\rm high}
   \propto  R^{\,2+2l-2s_1}\,\Lambda^{1+l-s_l}$, which tends to zero, since $R^{\,2+2l-2s_1}$ and $\Lambda^{1+l-s_l}$ both tend to zero.
 \end{itemize}
Hence the contribution of the $\{q<\Lambda$\,, $q'>\Lambda\}$ domain is negligible compared to the high-wavevector contribution.
A similar analysis leads to the same conclusion for the $\{q'<\Lambda$\,, $q>\Lambda\}$ domain.

 \subsection{Low-wavevector contribution} 

We turn to $\gt_3^{(3)}\big|_{\rm low}$\,, 
given by Eq.~(\ref{eq:gt3low}).
Let $\gt_3^{(3,l)}\big|_{\rm low}$
\big(resp.~$\gt_3^{(3,l)}\big|_{\rm high}$\,\big)
denote the angular-momentum $l$ contribution to 
$\gt_3^{(3)}\big|_{\rm low}$
\big(resp.~$\gt_3^{(3)}\big|_{\rm high}$\,\big).
In Eq.~(\ref{eq:def_Phi_q}),
we can replace
(as in Sec.~\ref{sec:g33mixed})
the exponentials by $1$ and the spherical Bessel function by its asymptotic behavior Eq.~(\ref{eq:jl_small}).
We can do the same in Eq.~(\ref{eq:def_Phi'_q'}).
Thus
$\gt_3^{(3,l)}\big|_{\rm low}$\,
  equals a function of \,$\Oo$ times
 \begin{multline}
 R^{\,2\,l-2}\    \int_0^\infty\, \frac{d E}{\pi}
 \
e^{-\beta E}
 \ \int_0^{\Lambda}d q\ q^{l+2}\
 \int_0^{\Lambda}d q'\,(q')^{l+2}
 \\ \times \  \Im\left[t_2\left(E+i\,0^{+}-
\,      \tfrac{3q^2}{4m}\right)
\,t_2\left(E+i\,0^+-
 \,   \tfrac{3q'^{\phantom{.}2}}{4m}\right)\,
 t_{3,l}\left(q,q';E+i\,0^+\right) \right]
  \nn
 \end{multline}
 We then need to study the behavior of this integral for $\Lambda\to\infty$.
 At large enough $q$ and $q'$, the $t_2$'s are real, and  scale like $1/q$ and $1/q'$ respectively, while the imaginary part of $t_{3,l}$ scales like $(q\,q')^{-(s_l+1)}$ by Eq.~(\ref{eqimt3l}).
 The integrand thus behaves like $(q\,q')^{l-s_l}$,
 which is integrable at infinity for $l-s_l<-1$.
We thus distinguish between two cases:
  \bi
\item If \,$s_l > l+1$, 
  we can replace $\Lambda$ by $\infty$,
  so that $g_3^{(3, l)} {\big|_{{\rm low}}} \propto R^{\,2\,l-2}$.
  This is negligible compared to
  {$g_3^{(3, l=1)} {\big|_{{\rm high}}} \propto \! R^{\,2s-4}$} (because $l+1-s \, \geq \, 2-s \, > \,  0$).
\item If \,$s_l < l+1$, the
integral over $q$ and $q'$ is dominated by large $q$ and $q'$,
and scales like $\Lambda^{2(l+1-s_l)}$,
hence $g_3^{(3, l)} {\big|_{{\rm low}}} \propto (\Lambda\,R)^{2\,l-2}\,\Lambda^{4-2\,s_l}$.
On the other hand, we have in this case $g_3^{(3, l)} {\big|_{{\rm high}}} \propto R^{\,2\,s_l-4}$ (see Section~\ref{subsec:gtilde33}),
which yields a ratio
$g_3^{(3, l)} {\big|_{{\rm low}}} \ / \
g_3^{(3, l)} {\big|_{{\rm high}}}
\propto (\Lambda\,R)^{2\,(l+1-\,s_l)}\ll 1$. 
    \ei

\section{Sum rule for $t_{3,0}$}\label{appsumrule}
In this appendix, we derive the sum rule Eq.~(\ref{sumrulel=0}).
We start from the STM equation for $t_{3,l}$\,, Eq.~(\ref{eq:STMt3l}), in the case $l=0$.
We consider the limit $q\to \infty$\,, for fixed $q'$ {and $E$}.
From Eq.~(\ref{eq:defQl}),
we get
\be
Q_0(z) \ \underset{|z|\to\infty}{\sim} \ \ \frac{1}{z}
\label{eq:Q0_z_inf}
\ee
Therefore, the source term on the RHS of Eq.~(\ref{eq:STMt3l}) is $\simeq m/q^2$.
The integral term, for the same reason, is $\simeq m\, I_0(q',E\,)/q^2$ with
$I_0(q',E\,)  =  \int_0^{\infty} \,
t_2\!\left(E-\frac{3q''^{\phantom{.}2}}{4m}\right)\ t_{3,0}(q'',q';E\,)
\,q''^{\phantom{.}2}\,d q''/(2\pi^2)$\,.
Let us show that $1+I_0(q',E\,)=0$ by contradiction.
      If we had $1+I_0(q',E\,)\neq 0$, then by the above analysis,
 $t_{3,0}(q,q';E\,)\sim m\,[1+I_0(q',E\,)]\,/q^2$ for $q\to\infty$\,.
  This would imply that in the STM equation Eq.~(\ref{eq:STMt3l}),
  in the integral over $q''$,
  the integrand
  is $\propto 1/q''$ for $q'' \to \infty$,
  so that the integral would diverge, and Eq.~(\ref{eq:STMt3l}) would not hold,
  in contradiction with the definition of $t_{3,0}$\,.
  Hence we must have $1+I_0(q',E\,)=0$.
 Using the symmetry property Eq.~(\ref{eq:t3_sym}) of $t_3$\,, we obtain Eq.~(\ref{sumrulel=0}).

  \section{Small-$R$\, behavior of \,$g_3^{(2)} \, + \, g_3^{(3,l=0)}$} \label{app:g32+g330}

  In this appendix, we analyze the $R\to0$ behavior of \,$g_3^{(2)} \, + \, g_3^{(3,l=0)}$, and we show that it is  bounded.
We start from Eq.~(\ref{eqg32-g33l=0})
[recall that $\Phi_q$ and $\chi_q$ are given by Eqs.~(\ref{eq:Phi_q_l0}) and~(\ref{eq:chi_q_l0})].
We will need to go beyond the leading-order behavior Eq.~(\ref{eq:chiq_phiq_sim}), because the leading-order result vanishes
(as we have seen in the main text, see also below).
  We thus introduce
$\Delta \chi_q$ and \,$\Delta \Phi_q$\, such that
\be
\chi_q \ = \ \left(\frac{1}{r} \ - \ \frac{1}{r'} \, \right)^2 \ + \ \Delta \chi_q \ \ \ {\rm and}
\ \ \ \ 
\Phi_q \ = \ \frac{1}{\sqrt{4\,\pi}} \ \left(\frac{1}{r} \ - \ \frac{1}{r'} \,\right) \ + \ \Delta \Phi_q 
\nn
\ee
Inserting this into Eq.~(\ref{eqg32-g33l=0})
and expanding yields four contributions
\be
g_3^{(2)} \ + \ g_3^{(3,l=0)} \ = \ g_3^{\rm(I)} \ + \
\ g_3^{\rm(II)} \ + \
\ g_3^{({\rm III})} \ + \
\ g_3^{\rm(IV)}
\nn
\ee
which we consider separately in what follows.
\bi 
\item The first contribution \,$g_3^{\rm(I)}$\, is given by Eq.~(\ref{eqg32-g33l=0bis}),
  which equals zero due to the sum rule Eq.~(\ref{sumrulel=0}).
\item 
The second contribution is
\be
g_3^{\rm(II)}
 \ = \ -c \    \int_{\mathcal{C}_+}\,\frac{d E}{2\pi i}
  \ e^{-\beta\,E}
  \ \int_0^{\infty}\frac{dq\,q^2}{2\,\pi^2} \ \ t_2\left(E-\frac{3q^2}{4m}\right) \ \Delta\chi_q
  \nn
  \ee
  Exchanging the order of integration, 
  deforming the integration contour from $\Cr_+$ to the contour surrounding the line $E>\frac{3q^2}{4 m}$, and performing the change of variable $E=\frac{3q^2}{4 m}+x$, 
  we get
\be
g_3^{\rm(II)}
\ = \ -c \  \int_0^{\infty}\frac{dq\,q^2}{2\,\pi^2} \ e^{-\beta\,\frac{3q^2}{4m}} \ \int_0^{\infty}\frac{dx}{\pi}\ e^{-\beta\,x}\ \Im\left[\,t_2\left(x+i\,0^+\right)\,\Delta\chi_q \,\right]
\nn
 \ee
 Due to the exponential factors, the integral is dominated by the region
 \,$q\lesssim\sqrt{m/\beta}, \ x\lesssim 1/\beta$\,,
 so that we can safely take the limit $R\to0$  under the integral.
 $\Delta\chi_q$\,, and hence also \,$g_3^{\rm(II)}$\,, are of order $R^0$, {\it i.e.}, they have a finite limit when $R\to0$.\footnote{More explicitly, $\Delta\chi_q=
3\,\tilde{q}^2-\big(
3\,\tilde{q}^2(\sin\alpha+\sin\alpha'\,)^2-\,q^2\,\|\cos\alpha\,\hat{\rho}-\cos\alpha'\,\hat{\rho}' \|^2
\big)/(4\,\sin\alpha \, \sin\alpha'\,)$ where $\tilde{q}=-i\,2\sqrt{m\,x/3}$.
Also using \,$t_2(x+i\,0^+)=(4\,\pi/m)\big(a^{-1}+i\sqrt{m\,x}\,\big)^{-1}$\,, we get the product of two converging integrals over $q$ and $x$.}
\item The third contribution 
   \,$g_3^{({\rm III})}$\, is the sum of two cross terms,
    obtained by replacing
      $\Phi_q\,\Phi_{q'}$
        by
    $(\Delta\Phi_q\,+\,\Delta\Phi_{q'})\,(1/r-1/r')\,/\sqrt{4\pi}$ in Eq.~(\ref{eqg32-g33l=0}). Due to the symmetry property Eq.~(\ref{eq:t3_sym}) of $t_3$, the two terms are equal, hence
\begin{multline}
 g_3^{({\rm III})} \ = \ -\,\sqrt{4\pi}\,c  
   \,\left(
\frac{1}{r}-\frac{1}{r'}
\right)\ 
\int_{\mathcal{C}_+}\,\frac{d E}{2\pi i}
  \ e^{-\beta\,E}
  \ \int_0^{\infty}\frac{dq\,q^2}{2\,\pi^2} \ \ t_2\left(E-\frac{3q^2}{4m}\right)
  \ \Delta\Phi_q
  \\ \times\  \int_0^{\infty}\frac{dq'\,q'^{\phantom{.}2}}{2\,\pi^2}
\ t_{3,0}( q, q';E)
\ t_2\left(E-\frac{3q'^{\phantom{.}2}}{4m}
\right)
\nn
\end{multline}
Due to the sum rule Eq.~(\ref{sumrulel=0}), the integral over $q'$ equals $-1$,
so that
\be
g_3^{({\rm III})} \ = \ \sqrt{4\pi}\,c
\ \left(
\frac{1}{r}-\frac{1}{r'}
\right) \ 
\int_{\mathcal{C}_+}\,\frac{d E}{2\pi i}
  \ e^{-\beta\,E}
  \ \int_0^{\infty}\frac{dq\,q^2}{2\,\pi^2} \ \ t_2\left(E-\frac{3q^2}{4m}\right) \ \Delta\Phi_q
  \nn
  \ee
Performing the same manipulations as for the second contribution, we get
\be g_3^{({\rm III})} \ = \ \sqrt{4\pi} \,c
\ \left(
\frac{1}{r}-\frac{1}{r'}
\right) \ 
\int_0^{\infty}\frac{dq\,q^2}{2\,\pi^2} \ e^{-\beta\,\frac{3}{4}\,\frac{q^2}{m}} \int_0^{\infty}\frac{dx}{\pi}\,e^{-\beta\,x}\,\Im\left[\,t_2\left(x+i\,0^+\right)\,\Delta\Phi_q \, \right]
\nn
\ee
   Using 
     Eq.~(\ref{eq:Phi_q_l0}), we get $\Delta\Phi_q=O(R)$ for $R\to0$.
   Hence \,$g_3^{({\rm III})}=O(R^0)$\,.
 \item
   The fourth contribution is 
  \begin{multline}
 g_3^{({\rm IV})} \ = \ -\, 4\pi c
  \int_{\mathcal{C}_+}\,\frac{d E}{2\pi i}
  \ e^{-\beta\,E}
  \ \int_0^{\infty}\frac{dq\,q^2}{2\,\pi^2} \ \ t_2\left(E-\frac{3q^2}{4m}\right)\Delta\Phi_{q}
  \\ \times \  \int_0^{\infty}\frac{dq'\,q'^{\phantom{.}2}}{2\,\pi^2}
\ t_{3,0}( q, q';E)
\ t_2\left(E-\frac{3q'^{\phantom{.}2}}{4m}
\right)
\ \ \Delta\Phi_{q'} \nn
  \end{multline}
We introduce a cutoff $\Lambda$ as in Eqs.~(\ref{eq:<<Lambda<<}) and (\ref{eq:Lambda_double_lim}),
we split the integral over $q$ and $q'$ into four contributions,
and we show that each contribution tends to zero.
 \bi
 \item[(a)] $\{q<\Lambda \, , \, q'<\Lambda\}$.
   In the integrals over $q$ and $q'$, we use the $R\to0$ asymptotic behavior $\Delta\Phi_{q} \simeq \big\{-m\,E(r-r')/2+q^2 \big[3(r-r')-(\rho^2/r-\rho'\,^2/r') \big] /8 \,\big\} /\sqrt{4\pi}$
     and similarly for $\Delta\Phi_{q'}$.
   This scales like the hyperradius $R$. The integrals 
are dominated by large values of $q$ and $q'$, because they diverge for $\Lambda\to\infty$. Indeed, for $q,\,q'\,\gg \max(|a|^{-1},\lambda_T)$, all quantities except $t_{3,0}$ are independent of $E$, and only the imaginary part of $t_{3,0}$ contributes. At large wave vectors, $\Im t_{3,0}$ scales like $\mathcal{B}_0(m\,E,a^{-1})/(q\,q')^{s_0+1}$.
  Therefore the integrand
  is $\propto q^{2-s_0}$,
    whose integral over $q$ is $\propto \Lambda^{3-s_0}$,
    which diverges for $\Lambda\to\infty$ since $s_0<3$.
    This gives
(after multiplying by $R$)
    a result
    $\propto R\,\Lambda^{3-s_0}$
    for the integral over $q$. Thus the double integral over $q$ and $q'$ is
  $ \propto R^2\Lambda^{6-2s_0}$.
  Since $3-s_0\leq 1$,
  we have $R^2\Lambda^{6-2s_0}=O(R^2\Lambda^2)$,
  which tends to $0$ since $\Lambda R \to 0$.
  In conclusion, contribution~(a) tends to $0$.
\item[(b)]
  $\{q>\Lambda \,,\, q'<\Lambda\}$.
We replace $\Delta \Phi_{q'}$ by its $R\to 0$ asymptotic behavior as in case (a). 
In the integral over $q'$,
for $q'\gg \max(|a|^{-1},\lambda_T)$, 
the integrand
scales like
    $(q')^{2-1-(s_0+1)+2} = (q')^{2-s_0}$. This gives a diverging integral of order $\Lambda^{3-s_0}$. After multiplying by $R$, this gives a contribution of order
$R\,\Lambda^{3-s_0}$, as in case (a).
  In the integral over $q$, using Eq.~(\ref{eq:qtilde->q}),
  we get $\Delta\Phi_q
 \simeq
  \,\tilde{\Phi}(q\,R,\Oo)/R$\, with
\be
\frac{\tilde{\Phi}(q\,R,\Oo)}{R} \ = \ 
\frac{1}{\sqrt{4\pi}} \, \left[
\frac{e^{-\frac{\sqrt{3}}{2}q\,r}\,j_0(\frac{\sqrt{3}}{2}q\,\rho)-1}{r} \ - \ \big(r\to r',\rho\to \rho'\big)\right]\label{eq:DeltaPhiqlim}
\ee
The change of variable $q=t/R$\, yields a contribution 
\be
R^{s_0-2}\int_{R\,\Lambda}^{\infty}dt\ \, \tilde{\Phi}(t,\Oo) \ t^{-s_0}
\nn\ee
Due to the $t^{-s_0}$ term, this integral converges at infinity. In the $t\to0$ limit, $\tilde{\Phi}(t,\Oo)$ is $O(t^2)$, and the integrand scale like $t^{2-s_0}$, therefore the integral converges. The double integral over $q$ and $q'$ scales like
$R\,\Lambda^{3-s_0}\times R^{s_0-2}=
  (\Lambda R)^{3-s_0} R^{2s_0-4}$,
which tends to 0.
\item[(c)] 
  $\{q'>\Lambda \, , \, q<\Lambda\}$.
      By symmetry, this contribution is equal to contribution (b).

\item[(d)] $\{ q> \Lambda \, , \, q' >\Lambda \,\}$.
 The reasoning is similar to the one above Eq.~(\ref{eq:g3high_Il}).
We can replace $t_2\big(E-3q^2/(4m)\big)$ by a term of order $1/q$, 
replace $\Delta\Phi_q$ by Eq.~(\ref{eq:DeltaPhiqlim}), and perform similar replacements for the functions of $q'$.
These do not depend on the energy $E$, so we are left with the imaginary part of $t_{3,0}$ which scales like $\mathcal{B}_0(m\,E,a^{-1})/(q\,q')^{s_0+1}$. Therefore the integrals over $q$ and $q'$ factorize. The integral over $E$ gives 
$\int_0^{\infty}dE/\pi e^{-\beta\,E}\mathcal{B}_0(m\,E,a^{-1})$, which is
finite.
The integral over $q$ gives 
\be
\int_{\Lambda}^{\infty}dq\,
q^{-s_0} \ 
\left[
  \frac{e^{-\frac{\sqrt{3}}{2}q\,r} \, j_0\left(\tfrac{\sqrt{3}}{2}q\,\rho \right)-1}{r}
  \ - \ \big(r\to r',\rho\to \rho' \big)
\right]
\nn\ee
After the change of variable $q=t/R$, the lower bound is $\Lambda\,R\to0$.
For $t\to0$, the integrand scales like $t^{2-s_0}$ so that the integral converges. At high $t$, the integrand decays like $t^{-s_0}$ and the integral converges as well.  
We thus find that this integral scales like $R^{s_0-2}$.
In conclusion, the contribution (d) is of order $R^{2s_0-4}$, which
tends to 0 for $R\to0$.
\ei
\ei   
 \section{STM equations on the {positive} real energy axis}\label{appSTMreal}

In this appendix, we consider the STM equation Eq.~(\ref{eq:STMt3l}) for $E$ on the positive real axis. We set $E=k_0^2/m + i\,0^+$ with $k_0$ real and positive.
Let
\be
f(q,q') \ = \ -\, {\frac{q q'}{m(2l+1)}} \ t_{3,l}(q,q'; k_0^2/m + i\,0^+)
\label{eq:defft3l}
\ee
where we suppress the dependence of $f$ on $k_0$ and $l$\,.
 The STM equation in terms of $f$ is
 \begin{equation}
   f(q,q') \ = \ Q(q/k_0,q'/k_0)
  \ - \ \frac{2}{\pi} \ \int_0^{\infty}\,d q''
 \ \, \frac{Q(q/k_0 \, , \, q''/k_0)}{a^{-1}-\sqrt{-k_0^2+\frac{3}{4}q''^{\phantom{.}2}-i\,0^+}\ }
 \ \, f(q'',q'\,)
 \label{eq:STM_f}
 \end{equation}
 with $Q(x,x')=Q_l\left(\dfrac{1+\,i\,0^{+}-x^2-(x')^2}{x\,x'}\right)$\,.
We decompose $f$ and $Q$ into real and imaginary parts,
\be
f \ = \ X\,+\,i\,Y \ \ \ , \ \ \  \ \ \  Q \ = \ R\,+\,i\,I_{\mathcal{D}}
\nn
\ee
We have the identity
 \begin{equation}
   \frac{1}{a^{-1}-\sqrt{-k_0^2+\frac{3}{4}q''^{\phantom{.}2} {- i\,0^+}}}
   \ = \
 \frac{\Theta_+(q'')}{a^{-1}-\sqrt{-k_0^2+\frac{3}{4}q''^{\phantom{.}2}}}
 \ + \
 \frac{
 \Theta_{-}(q'')
 \left(a^{-1}-i\,\sqrt{k_0^2-\frac{3}{4}q''^{\phantom{.}2}}\ \right)
 }
 {
 a^{-2}+k_0^2-\frac{3}{4}q''^{\phantom{.}2}
 }
 \end{equation}
 where $\Theta_+(q'')=\theta\!\left(q''-2 k_0 / \sqrt{3}\,\right)$ and
 $\Theta_-(q'')=\theta\!\left(-q''+ 2 k_0 / \sqrt{3} \, \right)$\,, with $\theta$ the Heaviside step function.
 The function
$I_{\mathcal{D}}(q/k_0,q'/k_0)$
is non-zero
for \,$| (k_0^2-q^2-q'^{\phantom{.}2})(q q') | < 1$,
{\it i.e.}
in the domain 
 \begin{equation}
 \mathcal{D}=\left\{\,(q,q')\ \big| \ \,q^2+q'^{\phantom{.}2}+q\,q'-k_0^2 \ > \ 0\ \, \textrm{and}\ \,
 q^2+q'^{\phantom{.}2}-q\,q'-k_0^2 \ < \ 0\ \right\}
 \label{eq:def_Dr}
 \end{equation}
We have
\be
 \Dr \ \subset \ \left[ \, 0 \ ; \ \frac{2}{\sqrt{3}}\,k_0 \, \right]^2
 \label{eq:D_subset}
 \ee
Hence $I_{\mathcal{D}}(q/k_0\,,q''/k_0)\,\Theta_+(q'')=0$.
As a result, the real and imaginary parts of Eq.~(\ref{eq:STM_f}) read
 \begin{multline}
   X(q,q') \ =
   \  R\left(\frac{q}{k_0},\frac{q'}{k_0}\right)
 \ - \
 \frac{2}{\pi}\ \int_0^{\frac{2}{\sqrt{3}}k_0}
 d q''
 \
 \left[
 \frac{
   R\left(\frac{q}{k_0},\frac{q''}{k_0}\right)a^{-1}
   \ + \
 I_{\mathcal{D}}\left(\frac{q}{k_0},\frac{q''}{k_0}\right)\sqrt{k_0^2-\frac{3}{4}q''^{\phantom{.}2}}
 }
 {
 a^{-2}+k_0^2-\frac{3}{4}q''^{\phantom{.}2}
 }
 \ X(q'',q')
 \right.
 \\
  +
\left.  \frac{-
 I_{\mathcal{D}}\left(\frac{q}{k_0},\frac{q''}{k_0}\right)a^{-1}+
R\left(\frac{q}{k_0},\frac{q''}{k_0}\right)\sqrt{k_0^2-\frac{3}{4}q''^{\phantom{.}2}}
 }
 {
 a^{-2}+k_0^2-\frac{3}{4}q''^{\phantom{.}2}
 } \ Y(q'',q')
\right]
\ - \ \frac{2}{\pi} \ \int_{\frac{2}{\sqrt{3}}k_0}^{\infty}\,d q''\
\frac{R\left(\frac{q}{k_0},\frac{q''}{k_0}\right)\ X(q'',q')}{a^{-1}-\sqrt{-k_0^2+\frac{3}{4}q''^{\phantom{.}2}}}
\\
\phantom{\Big(}\label{eqXBCS}
 \end{multline}
 and
 \begin{multline}
   Y(q,q')  \ =
   \ I_{\mathcal{D}}\left(\frac{q}{k_0},\frac{q'}{k_0}\right)
\ - \ \frac{2}{\pi} \ \int_0^{\frac{2}{\sqrt{3}}k_0}d q'' \
\left[
 \frac{
 R\left(\frac{q}{k_0},\frac{q''}{k_0}\right)a^{-1}+
 I_{\mathcal{D}}\left(\frac{q}{k_0},\frac{q''}{k_0}\right)\sqrt{k_0^2-\frac{3}{4}q''^{\phantom{.}2}}
 }
 {
 a^{-2}+k_0^2-\frac{3}{4}q''^{\phantom{.}2}
 }
 \ Y(q'',q')
 \right.
 \\
 +
\left.  \frac{
 I_{\mathcal{D}}\left(\frac{q}{k_0},\frac{q''}{k_0}\right)a^{-1}-
R\left(\frac{q}{k_0},\frac{q''}{k_0}\right)\sqrt{k_0^2-\frac{3}{4}q''^{\phantom{.}2}}
 }
 {
 a^{-2}+k_0^2-\frac{3}{4}q''^{\phantom{.}2}
 }\ X(q'',q')
\right]
 \ - \ \frac{2}{\pi} \ \int_{\frac{2}{\sqrt{3}}k_0}^{\infty}d q''\
\frac{R\left(\frac{q}{k_0},\frac{q''}{k_0}\right)\ Y(q'',q')}{a^{-1}-\sqrt{-k_0^2+\frac{3}{4}q''^{\phantom{.}2}}}
\\
\phantom{\Big(}\label{eqYBCS}
 \end{multline}

 \section{Behavior of the imaginary part of \,$t_{3,l}$\, at large wavevectors}\label{AppImt3largek}

 In this appendix, we derive the power-law large-momentum asymptotic behavior Eq.~(\ref{eqimt3l}) 
 of the three-body T-matrix at fixed real energy. Furthermore, we obtain an expression for $\Br_1$.
 For $a=\infty$, we will check numerically that this expression of $\Br_1$ agrees with a fit
 by Eq.~(\ref{eqimt31})
 of the numerically computed $t_{3,1}$.
 
   Our general strategy will be to decompose the STM equation into two sectors corresponding to low and high momenta, and to diagonalize the restriction to the large-momentum sector of the kernel of the STM equation (this diagonalization is the subject of Appendices~\ref{app:eigUL} and~\ref{Appeigenvec}).
   We treat in detail the case $a=\infty$ and $l=1\,$
   in Subsec.~\ref{AppImt3UL}.
 We generalize to arbitrary negative~$a$ and arbitrary~$l$ in Subsec.~\ref{AppImt3BCS}.

  \subsection{Unitary limit}\label{AppImt3UL}

\subsubsection{Decomposition into low and high wavevector sectors} \label{sssec:decomp_sect}
  
  We define the domains
  \be
  \Dr_1 \, = \, \left[0 \ ; \  \tfrac{2}{\sqrt{3}} \,  k_0 \right[
   \ \ \ \ \ \  {\rm and} \ \ \ \ \ \ 
    \Dr_2 \, = \, \left[\tfrac{2}{\sqrt{3}} \, k_0 \ ; \  +\infty \right[
      \nn
  \ee
In the STM equations~(\ref{eqXBCS}) and~(\ref{eqYBCS}), we change from the variable $q''$ to a new variable, $\theta$ or $u$ depending on the domain:
\bea
\bullet \, \textrm{For}\ q'' \in \Dr_1\,, \quad
q''&=& \, \tfrac{2}{\sqrt{3}}\ k_0 \ \sin\, \theta
\ \ \ \ {\rm with} \ \ \theta\in\left[0 \ ; \ \tfrac{\pi}{2}\right[
\nn\\
\bullet \, \textrm{For}\ q'' \in \Dr_2\,, \quad
q''&=& \,
\tfrac{2}{\sqrt{3}}\ k_0 \ \cosh\, u
\ \ \ \ {\rm with} \  u \in[0 \ ; \ +\infty[
    \nn
\eea
    Similarly, we change from the variable $q$ to $\,\theta_1$ or $u_1$ according to
      \be
      q = \,
\tfrac{2}{\sqrt{3}}\ k_0
\, \sin\,\theta_1 \ \ {\rm for}\ \ q\in\Dr_1\,, \ \ \ \ {\rm and}\ \ \ \ q = \,
\tfrac{2}{\sqrt{3}}\ k_0
\, \cosh\,u_1\ \ {\rm for} \ \ q\in\Dr_2
      \nn\ee
      and we change from $q'$ to $\,\theta'_1$ or $u'_1$ according to
      \be
      q' = \,
\tfrac{2}{\sqrt{3}} \ k_0 \, 
\sin\,\theta'_1 \ \ {\rm for}\ \ q'\in\Dr_1\,, \ \ \ \ {\rm and}\ \ \ \ q' = \, \tfrac{2}{\sqrt{3}}\ k_0 \, \cosh\,u'_1\ \ {\rm for} \ \ q'\in\Dr_2
      \nn\ee

    The domain $\Dr$ defined in Eq.~(\ref{eq:def_Dr}) is included in $\Dr_1 \times \Dr_1$\,,
    as already stated in Eq.~(\ref{eq:D_subset}).
In terms of the new variables
$\theta_1$ and $\theta'_1$, the domain \,$\Dr$ corresponds to the rectangle defined by the conditions
\be
|\theta_1-\theta'_1| < \,\pi/3
\ \ \ \ {\rm and} \ \ \ \ \pi/3 < |\theta_1+\theta'_1| < \,2\pi/3
\label{eq:D_angles}
\ee
To derive Eq.~(\ref{eq:D_angles}), we proceed as follows.
Using the identity\footnote{To derive Eq.~(\ref{eq:identity_angles}),
we start from
$\sin^2(\theta_1) = \frac{1 - \cos(2\,\theta_1)}{2}$\, and
$\sin^2(\theta'_1) = \frac{1 - \cos(2\,\theta'_1)}{2}$\,,
we use
\be\cos(2\,\theta_1) + \cos(2\,\theta'_1) \ = \
\cos\left[ \big(\theta_1 + \theta'_1\big) + \big(\theta_1 - \theta'_1\big) \right]
\ + \
\cos\left[ \big(\theta_1 + \theta'_1\big) - \big(\theta_1 - \theta'_1\big) \right]
\nn\ee
and we expand the cosines using the angle addition and subtraction formulae.
}
\be
\sin^2(\theta_1) + \sin^2(\theta'_1) \ = \ 1 - \cos(\theta_1-\theta'_1) \, \cos(\theta_1+\theta'_1)
\label{eq:identity_angles}
\ee
we rewrite
the conditions in Eq.~(\ref{eq:def_Dr}) as
\bea
\left[ \, \tfrac{1}{2} \, - \, \cos(\theta_1 + \theta'_1) \right]
\ \left[ \, \tfrac{1}{2} \, + \, \cos(\theta_1 - \theta'_1) \right] \ &>& \ 0
\label{eq:cond_E}
\\
\left[ \, \tfrac{1}{2} \, + \, \cos(\theta_1 + \theta'_1) \right]
\ \left[ \, \tfrac{1}{2} \, - \, \cos(\theta_1 - \theta'_1) \right] \ &<& \ 0
\label{eq:cond_E'}
\eea
Since $\theta_1 \, , \, \theta'_1 \, \in \left[ 0 \, ; \, \frac{\pi}{2} \, \right]$\, we have
\,$\theta_1 + \theta'_1 \, \in \, [ 0 \, ; \, \pi \, ]$
and
\,$\theta_1 - \theta'_1 \, \in \, \left[ - \,  \frac{\pi}{2} \, ; \, \frac{\pi}{2} \, \right]$\,.
Hence $\frac{1}{2} \, + \,\cos(\theta_1 - \theta'_1) > 0$
and Eq.~(\ref{eq:cond_E}) simplifies to
$\frac{1}{2} \, - \, \cos(\theta_1 + \theta'_1) > 0$\,,
{\it i.e.}
\be
\theta_1 + \theta'_1 \ > \ \pi/3
\label{eq:cond_1}
\ee
On the other hand, Eq.~(\ref{eq:cond_E'}) is equivalent to:
\bea
| \theta_1 - \theta'_1|  \ > \  \pi/3
\ \ &{\rm and}& \ \
\theta_1 + \theta'_1 \ > \ 2\pi/3
\label{eq:E'1}
\\
 & {\rm or} & \nn
\\
| \theta_1 - \theta'_1|  \ < \  \pi/3
\ \ &{\rm and}& \ \
\theta_1 + \theta'_1 \ < \ 2\pi/3
\label{eq:E'2}
\eea
In the allowed range, Eq.~(\ref{eq:E'1}) never holds.
We conclude that \,$\Dr$ is defined by Eqs.~(\ref{eq:cond_1}) and (\ref{eq:E'2}),
{\it i.e.} by Eq.~(\ref{eq:D_angles}).

For $m,n=1$ or \,2,
let $x_{m\,n}$ denote the restriction of $X$ to \,$\Dr_m \times \Dr_n$\,
expressed in terms of the new variables,
{\it i.e.}\,, 
\bea
x_{1 1}(\theta_1\,,\theta'_1) \, = \, X\big(
\, \tfrac{2}{\sqrt{3}}\ k_0 \ \sin \theta_1\,,
\ \tfrac{2}{\sqrt{3}}\ k_0 \ \sin \theta'_1 \big)\ &;&
\ \ \ 
x_{1 2}(\theta_1\,, u'_1) \, = \, X\big( 
\ \tfrac{2}{\sqrt{3}} \ k_0 \ \sin \theta_1\,,
\, \tfrac{2}{\sqrt{3}} \ k_0 \ \cosh u'_1 \big)
\nn \\
x_{2 1}(u_1\,,\theta'_1) \, = \,  X\big( 
\ \tfrac{2}{\sqrt{3}}\  k_0 \ \cosh u_1\,,
\,  \tfrac{2}{\sqrt{3}} \ k_0 \ \sin \theta'_1 \big)\ &;&
\ \ \ 
x_{2 2}(u_1\,, u'_1) \, = \, X\big( 
\,  \tfrac{2}{\sqrt{3}} \ k_0 \ \cosh u_1\,,
\,  \tfrac{2}{\sqrt{3}}\ k_0 \ \cosh u'_1 \big)
\nn
\eea
Similarly, let $y_{m n}$ (resp.~$r_{m n}$)
denote the restriction
of $Y$ (resp.~$R/A$) to \,$\Dr_m \times \,\Dr_n$\,
expressed in terms of the new variables,
where
\be
A \ = \ \frac{\sqrt{3}\ \pi}{4}
\label{eq:def_A}
\ee
In particular,
     \be
        y_{22}(u,u') \ = \ {\rm Im} \,f(q,q') \,, \ \ \ \ \ \ {\rm with} \ \ q \,=\, \frac{2}{\sqrt{3}} \ k_0 \,\cosh u \ \ {\rm and} \ \
        q' \,=\, \frac{2}{\sqrt{3}}\ k_0 \,\cosh u' \,.
        \label{eq:rel_y22_Imf}
   \ee
Finally, let \,$i_\Dr(\theta_1 \, , \theta'_1) = I_\Dr\big(\, \tfrac{2}{\sqrt{3}}\ k_0 \, \sin \theta_1\,,
\, \tfrac{2}{\sqrt{3}} \ k_0 \, \sin \theta'_1 \big) \, / A$\,.
    Recall that $I_\Dr(q,q')$ vanishes for $(q,q')\notin \Dr_1\times\Dr_1$. This will allow us to simplify the expression of $r_{21}$\,,
        \be
        r_{21}(u,\theta) \ =\  A^{-1} \ Q_l\left( \frac{3/4 - \cosh^2 u - \sin^2 \theta}{\cosh u \ \sin\theta} \right)\,.
        \label{eq:r21}
        \ee 
We have
\bea
dq'' \ &=& \  \tfrac{2}{\sqrt{3}}\ k_0\,
 \cos\theta
\  d\theta\,
\ \ \ \ \ \ \ \ \ {\rm for}\ \  q''\in\Dr_1\,,
\nn\\
dq'' \ &=& \  \tfrac{2}{\sqrt{3}}\ k_0\,
 \sinh u
\  du\,
\ \ \ \ \ \ \ {\rm for}\ \  q''\in\Dr_2\,.
\nn
\eea
The STM equations for $X$ and $Y$, Eqs.~(\ref{eqXBCS}) and (\ref{eqYBCS}), hence take the form
\bea
x_{m n} \ &=&
\ A\ r_{m n}-r_{m\,1}y_{1\,n}-\delta_{m\,1} \, {i_\Dr}x_{1\,n}
+r_{m\,2}x_{2\,n}
\label{eq:STM_x}
\\
y_{m n} \ &=&
\ A\ \delta_{m\,1}\delta_{n\,1} \, {i_\Dr}+r_{m\,1}x_{1\,n}-\delta_{m\,1} \, {i_\Dr}y_{1\,n}
+r_{m\,2}y_{2\,n}
\label{eq:STM_y}
\eea
Here, operator products are implied:
  $r_{11} x_{11}$ stands for
  $\int_0^{\pi/2} d\theta\ r_{11}(\theta_1, \theta)\  x_{11}(\theta,\theta'_1)$\,,
while  $r_{12} x_{22}$ stands for
$\int_0^{\infty} du\ r_{12}(\theta_1, u)\  x_{22}(u,u'_1)$\,,
and similarly for the other combinations.

 In the remainder of Subsection~\ref{AppImt3UL}, we set the real part of the energy and the mass to unity,
 \be
 E=1+i\,0^+ \ \ \ \ {\rm and}\ \ \  m=1
 \label{eq:E=1}
 \ee
which does not imply any restriction due to the scale invariance at the unitary limit.
In the remainder of Subsection~\ref{sssec:decomp_sect}, we will express
  $x_{2\,2}$ and $y_{2\,2}$
  in terms of
  $x_{11}$ and $y_{11}$  
  ({\it i.e.}, we will express the real and imaginary parts of $t_{3,l}(q,q'; 1+0^+)$ when both $q$ and $q'$ are in the large-momentum sector,
in terms of the same quantities when both $q$ and $q'$ are in the low-momentum sector). We start from
Eqs.~(\ref{eq:STM_x}) and~(\ref{eq:STM_y}) for $m=n=2$, which yield
\bea
x_{2\,2} \ &=& \ A\ r_{2\,2}-r_{21}y_{12}+r_{2\,2}x_{2\,2}\nn\\
y_{2\,2} \ &=& \ r_{21}x_{12}+r_{2\,2}y_{2\,2}\nn
\eea
We then assume  that the operator $\mathds{1}_2-r_{2\,2}$ is invertible.
This assumption can be checked
for $l=0$ and $l=1$:
As shown in Appendix~\ref{app:eigUL},
the eigenvalues of \,$\mathds{1}_2-r_{2\,2}$\,
  are given by $1-\lambda$, with $\lambda$ given by
  Eq.~(\ref{eq:lambda_l0}) for $l=0$ and Eqs.~(\ref{eq:lambda_l1}) and~(\ref{eq:lambda_ds}) for $l=1$,
and one can check that $1-\lambda>0$.
  This allows us to express $x_{2\,2}$ and $y_{2\,2}$  in terms of $x_{1 2}$ and $y_{1 2}$\,,
\bea
x_{2\,2}&=&\left(\mathds{1}_2-r_{2\,2}\right)^{-1}\left(A\,r_{2\,2}-r_{2 1}y_{1 2}\right)\label{eqx221}\\
y_{2\,2}&=&\left(\mathds{1}_2-r_{2\,2}\right)^{-1}\left(r_{2 1}x_{1 2}\right)\label{eqy221}
\eea
  Next, we express $x_{2 1}$ and $y_{2 1}$
    in terms of $x_{1 1}$ and $y_{1 1}$\,:
Eqs.~(\ref{eq:STM_x}) and~(\ref{eq:STM_y}) for $m=2$ and $n=1$ yield
\bea
x_{2 1}&=&A\,r_{2 1}-r_{2 1}y_{1 1}+r_{2\,2}x_{2 1}\nn\\
y_{2 1}&=&r_{2 1}x_{1 1}+r_{2\,2}y_{2 1}\nn
\eea
and thus
\bea
x_{2 1} \ &=& \ \left(\mathds{1}_2-r_{2\,2}\right)^{-1}\left(A\,r_{2 1}-r_{2 1}y_{1 1}\right)\label{eq:x21}\\
y_{2 1} \ &=& \ \left(\mathds{1}_2-r_{2\,2}\right)^{-1} r_{2 1}x_{1 1}
\label{eq:y21}
\eea
      Let $a^t$ be the transpose of an operator $a$,
for example $x_{2 1}^t(\theta,u) = x_{2 1}(u,\theta)$.
      Since $t_{3,l}(q,q' ; 1+i\,0^+)$ and $Q(q,q')$ are
 symmetric functions of $(q,q')$,
we have
\bea
& x_{2 1}^{t}=x_{1 2}\ \, ,
\ \, y_{2 1}^{t}=y_{1 2}\ \, ,
\ \, r_{2 1}^{t}=r_{1 2}
\label{eq:prop_transp_xyr}
\\
& x_{nn}^t = x_{nn}\ \,, \ \, y_{nn}^t = y_{nn}\ \, ,\ \, r_{nn}^t=r_{nn}
\label{eq:prop_sym_xyr}
\eea
Taking the transpose of Eqs.~(\ref{eq:x21}) and (\ref{eq:y21}) thus yields
\bea
x_{1 2} \ &=& \ \left(A\,r_{1 2}-y_{1 1}r_{1 2}\right)\left(\mathds{1}_2-r_{2\,2}\right)^{-1}\nn\\
y_{1 2} \ &=& \  x_{1 1}r_{1 2} \left(\mathds{1}_2-r_{2\,2}\right)^{-1}\nn
\eea
Substituting these expressions in Eqs.~(\ref{eqx221}) and (\ref{eqy221}) finally yields  the desired expression of 
$y_{2\,2}$ in terms of
$x_{1 1}$ and $y_{1 1}$,
\bea
y_{2\,2} \ &=& \ \left(\mathds{1}_2-r_{2\,2}\right)^{-1}
\,r_{2 1}\left(A \ \mathds{1}_1-\,y_{1 1} \right)\,r_{1 2}
\left(\mathds{1}_2-r_{2\,2}\right)^{-1}
\label{eqy22vec}
\eea
\big[although we will not use it in this article, we note that there is a similar expression for $x_{22}$\,,
\,$x_{2\,2} \ = \ \left(\mathds{1}_2-r_{2\,2}\right)^{-1}A\,r_{2\,2}
-\left(\mathds{1}_2-r_{2\,2}\right)^{-1}
r_{2 1}x_{1 1}r_{1 2}
\left(\mathds{1}_2-r_{2\,2}\right)^{-1}$\,\big].

\subsubsection{Large momentum limit}\label{AppasympUL}

We restrict to $l=1$ in this subsection (the general case of arbitrary $l$ and negative $a$ will be treated in Subsec.~\ref{AppImt3BCS}).
Our goal is to determine the asymptotic behavior of
${\rm Im}\,t_{3,1}$\, at large momenta, {\it i.e.},
  \,$y_{2\,2}(u_1,u'_1)$\, for \,$u_1,u'_1\to\infty$\,.
  To this end, we will combine Eq.~(\ref{eqy22vec}) with the
  diagonalization of the operator $r_{22}$ carried out in Appendix~\ref{app:eigUL}.

We first rewrite Eq.~(\ref{eqy22vec}) as
\be
y_{22}=\left(\mathds{1}_2-r_{2\,2}\right)^{-1}O\left(\mathds{1}_2-r_{2\,2}\right)^{-1}
\label{eq:y22_temp}
\ee
with $O=r_{2 1}\left(A \ \mathds{1}_1-\,y_{1 1} \right)\,r_{1 2}$\,.
 In App.~\ref{app:eigUL}, we show that the operator $r_{22}$ has
a continuous spectrum of eigenvalues $\lambda(\omega)$ parameterized by a real positive wavevector $\omega$, as well as a discrete eigenvalue $\lambda_{\rm ds}$,
and that the corresponding eigenvectors $g_\omega$ and $g_{\rm ds}$ satisfy
the closure relation Eq.~(\ref{eq:closure}).
Inserting this closure relation into Eq.~(\ref{eq:y22_temp}) yields
\begin{equation}
y_{22}(u_1,u'_1) \ = \ \int_0^\infty dw \ \int_0^\infty dw'\ B(u_1,w) \ O(w,w') \ B(u'_1,w')\label{eq:y22BOB}
\end{equation}
with
\be
B(u,w) \ = \ \int_0^\infty \, \frac{g_{\omega}(u)\,g_{\omega}(w)}{1-\lambda(\omega)}\ d\omega
 \ \, + \ \, \frac{g_{\rm ds}(u)\,g_{\rm ds}(w)}{1-\lambda_{\rm ds}}\label{eq:Buw}
\ee
and 
\be
O(w,w') \ = \ \int_0^{\frac{\pi}{2}}d\theta\int_0^{\frac{\pi}{2}}d\theta'\ r_{21}(w,\theta) \ \bigg(A\ \delta(\theta-\theta')-y_{11}(\theta,\theta') \bigg) \ r_{21}(w',\theta')\label{eq:Owwp}
\ee
Substituting Eq.~(\ref{eq:Owwp}) into Eq.~(\ref{eq:y22BOB}) yields
\be
y_{22}(u_1,u'_1) \ = \ \int_{0}^{\frac{\pi}{2}}d\theta\int_{0}^{\frac{\pi}{2}}d\theta'\ C_{21}(u_1,\theta) \ \bigg(
A\ \delta(\theta-\theta')-y_{11}(\theta,\theta')
\bigg)
\ C_{21}(u'_1,\theta')\label{eq:y22C21}
\ee
where
\be
C_{21}(u,\theta) \ = \ \int_0^\infty dw\ B(u,w)\,r_{21}(w,\theta)
\label{eq:C21}
\ee
Inserting Eq.~(\ref{eq:g_om_gbar}) into Eq.~(\ref{eq:Buw})
yields
\be
B(u,w) \ = \ \frac{1}{2\pi} \ \int_{-\infty}^{\infty} \ \frac{\bar{g}_{\omega}(u)\,\bar{g}_{\omega}(w)-\bar{g}_{\omega}(u)\bar{g}_{-\omega}(w)}{(\omega^2+1) \, [1-\lambda(\omega)]}\ d\omega \ + \ \frac{g_{\rm ds}(u)\,g_{\rm ds}(w)}{1-\lambda_{\rm ds}}
\label{eq:Buw2}
\ee
where we have used the fact
that $\lambda(\omega)$ is an even function of $\omega$
by Eq.~(\ref{eq:lambda_l1}).
The same Eq.~(\ref{eq:lambda_l1})
implies that
as a function of the complex variable $\omega$, the integrand in Eq.~(\ref{eq:Buw2}) has simple poles for \,$\omega=\pm i$\, and  
\,$\omega=\pm\,i\,s_{1,n}\,$ with $n\in\mathbb{N}$\,, and no other singularity.\footnote{ Indeed, when $\omega\to\pm i$,
$1-\lambda(\omega)\to
1-\lambda_{\rm ds} \neq 0$
[see Eqs.~(\ref{eq:lambda_ds}) and (\ref{eq:lambda_i})],
  and when $\omega\to0$,
$1-\lambda(\omega)\to 7/3 - 8/(\sqrt{3}\pi)\neq0$.}
The integral over $\omega$
contains two terms, which we evaluate
by closing the integration contour and applying the residue theorem:
\bi
\item
For the first term
  [proportional to
  \,$\bar{g}_{\omega}(u)\,\bar{g}_{\omega}(w)$\,],
we can close the integration contour in 
the upper half plane; this yields the contributions from the poles
$\omega = +i$\, and $\omega = +i\,s_{1,n}$\, given by
\be
-i \, \sum_{n=0}^\infty \, \frac{\bar{g}_{i\,s_{1,n}}(u) \ \bar{g}_{i\,s_{1,n}}(w)}{\big(1-s_{1,n}^2\big)\,\lambda'(i\,s_{1,n})} 
 \ - \ \frac{g_{\rm ds}(u)\,g_{\rm ds}(w)}{2 \, (1-\lambda_{\rm ds})}
\nn\ee
 where we have used
Eqs.~(\ref{eq:gbar_i}) and (\ref{eq:lambda_i}).
\item
  For the second term  [proportional to
  \,$\bar{g}_{\omega}(u)\,\bar{g}_{-\omega}(w)$\,], we have to distinguish the cases $u>w$ or $u<w$,
where we can close the contour in the upper or lower half plane respectively.
For $u>w$, this yields
\be
i\, \sum_{n=0}^\infty\frac{\bar{g}_{i\,s_{1,n}}(u) \ \bar{g}_{-i\,s_{1,n}}(w)}{\big(1-s_{1,n}^2\big) \, \lambda'(i\,s_{1,n})} 
 \ - \ \frac{g_{\rm ds}(u)\,g_{\rm ds}(w)}{2 \, (1-\lambda_{\rm ds})}
\nn\ee
whereas for $u<w$, we get
\be
i\,\sum_{n=0}^\infty\frac{\bar{g}_{-i\,s_{1,n}}(u) \ \bar{g}_{i\,s_{1,n}}(w)}{\big(1-s_{1,n}^2\big)\,\lambda'(i\,s_{1,n})} 
 \ - \ \frac{1}{2} \ \frac{g_{\rm ds}(u)\,g_{\rm ds}(w)}{1-\lambda_{\rm ds}}
\nn\ee
 where we have used $\lambda'(-\omega)=-\lambda'(\omega)$.
 \ei
 Summing all contributions to $B(u,w)$,
 the three terms containing \,$g_{\rm ds}$ cancel out, and we are left with
\begin{multline}
 B(u,w) \ = \ 
 -i\, \sum_{n=0}^\infty \ \frac{
 \ \bar{g}_{i\,s_{1,n}}(u) \ \big[\bar{g}_{i\,s_{1,n}}(w) \, - \, \theta(u-w) \, \bar{g}_{-i\,s_{1,n}}(w)\big] 
 \, - \, \theta(w-u) \, \bar{g}_{-i\,s_{1,n}}(u)\,\bar{g}_{i\,s_{1,n}}(w)}{\big(1-s_{1,n}^2\big) \  \lambda'(i\,s_{1,n})} 
 \label{eq:Buw3}
\end{multline}

  As a next step, we insert Eq.~(\ref{eq:Buw3}) into Eq.~(\ref{eq:C21})
  and take the limit $u\to \infty$.
  Exchanging the sum and the integral, we get
\be
  C_{21}(u, \theta) \ = \ -i \ \sum_{n=0}^\infty \ \frac{I_1^{(n)}
    \ + \ I_2^{(n)} \ + \ I_3^{(n)} }
    {
    \big( 1 - s_{1,n}^2 \big)
     \ \lambda'(i \, s_{1,n})
     }
\nn\ee
where
\begin{align*}
  I_1^{(n)} \ &= \ \bar{g}_{i\,s_{1,n}}(u) \ \int_0^\infty dw \ \, \bar{g}_{i\,s_{1,n}}(w) \ r_{21}(w,\theta)
  \\
  I_2^{(n)} \ &= \  -\,\bar{g}_{i\,s_{1,n}}(u) \ \int_0^u dw \ \, \bar{g}_{-i\,s_{1,n}}(w)\ r_{21}(w,\theta)
    \\
    I_3^{(n)} \ &= \ -\,\bar{g}_{-i\,s_{1,n}}(u) \ \int_u^\infty dw \ \, \bar{g}_{i\,s_{1,n}}(u) \ r_{21}(w,\theta)
  \end{align*}
From the expression~(\ref{eq:def_gbar_omega}) of $\bar{g}_\omega$ it follows that, for $u\to\infty$,
$\bar{g}_{i\,s_{1,n}}(u) \, \sim \, {\rm Constant} \times e^{-s_{1,n} u}$\, 
and thus also $I_1^{(n)} \, \sim \, {\rm Constant} \times e^{-s_{1,n} u}$\,;
hence the leading contribution to $C_{21}(u, \theta)$ from the $I_1^{(n)}$'s
is the $n=0$ term, $\propto e^{-s u}$.
The contribution of $I_3^{(n)}$ is negligible:
Using Eq.~(\ref{eq:r21}),
as well as the asymptotic behavior
\be
Q_1(z) \ \underset{|z|\to\infty}{\sim} \ \ \frac{1}{3 z^2}
\label{eq:Q1_z_inf}
\ee
[which follows from Eq.~(\ref{eq:defQl})],
one finds
$r_{21}(w,\theta) \,\sim\, {\rm Constant} \times \sin^2\theta \ e^{-2w}$\,
 for $w\to\infty$; thus for $u\to\infty$,
$I_3^{(n)} \,\sim\, {\rm Constant}\times \sin^2\theta \ e^{s_{1,n}u}\ \int_u^\infty dw \ e^{-(s_{1,n}+2)\,w}
\,=\, O(e^{-2u})$\,,
which is negligible compared to $e^{-su}$ since $s<2$.
Finally, for $I_2^{(n)}$\,, the integrand is $\propto e^{(s_{1,n}-2)w}$ for $w\to\infty$\,, and thus there are two cases:
For $n=0$, we have $s_{1,n} = s <2$\,, so that one can replace the upper bound $u$ of the integral by $\infty$, which yields
$I_2^{(0)} \sim \ -\,e^{su}\,i\,(s+1)\ \int_0^\infty dw \ \bar{g}_{-i\,s_{1,n}}(w)\ r_{21}(w,\theta)$\,;
for $n\geq1$, we have $s_{1,n} > 2$\,, so that the integral is dominated by large values of $w$, and we get
$I_2^{(n)} \,=\, O(e^{-2u})$\,, which is again negligible compared to $e^{-su}$.
We conclude that
\be
C_{21}(u,\theta)
 \ \underset{ u \to \infty}{\sim} \ \frac{A_{21}(s,\theta)}{i\,(1-s)\,\lambda'(i\,s)}\,e^{-s\,u}\label{eq:C21asympt}
\ee
where 
\be
A_{21}(s,\theta) \ = \ i\int_0^{\infty}r_{21}(u,\theta) \ \big[\bar{g}_{i\,s}(u)-\bar{g}_{-i\,s}(u)\big]\ du\label{eq:A21}
\ee
We obtain the analytical expression
     \be
     A_{21}(s,\theta) \ = \      \frac{8}{\sqrt{3}}  \ \frac{\sqrt{3}\ \sin(s\,\tfrac{\pi}{6}) - s \ \cos(s \,\tfrac{\pi}{6}) }{(1-s^2)  \  \sin\left(s \,\tfrac{\pi}{2}\right)} \  \bigg\{ s \, \cos\left[s \left(\tfrac{\pi}{2}-\theta\right)\right] \, + \, {\rm cotan}\, \theta \ \sin\left[s \left(\tfrac{\pi}{2}-\theta\right)\right] \, \bigg\}
     \label{eq:A21=}
\ee
based on the results of App.~\ref{app:eigUL}:
Using Eqs.~(\ref{eq:r21}), (\ref{eq:def_gbar_omega}), (\ref{eq:I1=F1}) and (\ref{eq:def_F1}), we express $A_{21}$ in terms of~$\Ir_1$\,,
\be
A_{21}(s,\theta) \ = \ i \
\big[ \, \Ir_1(\sin\theta, i s)
  \ - \
  \Ir_1(\sin\theta, -i s) \, \big]\,,
\nn\ee
and then we use the expression of $\Ir_1$ given by Eqs.~(\ref{eq:I1=}) and (\ref{eq:lambda_l1}), analytically continued to \,$u_1 = i \,(\pi/2 - \theta)$.

Inserting Eq.~(\ref{eq:C21asympt}) into Eq.~(\ref{eq:y22C21}) yields,
for $u_1\to\infty$ and $u'_1\to\infty$,
\be
y_{22}(u_1,u'_1) \ \sim \
\frac{
A \, \int_0^{\pi/2}d\theta\, A_{21}(s,\theta)^2-\int_0^{\pi/2}d\theta\int_0^{\pi/2}d\theta' A_{21}(s,\theta)
y_{11}(\theta,\theta')A_{21}(s,\theta')
}{|\lambda'(is)|^2\ (1-s)^2}\ e^{-s(u_1+u'_1)}\label{eq:y22asympUL}
\ee
 where we used that fact that $\lambda'(is)$ is imaginary.
 Finally,
by Eqs.~(\ref{eq:defft3l}) and (\ref{eq:rel_y22_Imf}), and
 given that $e^{u_1}\sim \sqrt{3}\, q_1$ and $e^{u'_1}\sim \sqrt{3}\, q'_1$ for $q_1\,,\,q'_1\to\infty$, we obtain
   \be
\Im \ t_{3,1}\left(q_1,q'_1\,;E=1+i\,0^+ \right) \ \sim \ -\,\frac{C_1}{\left(q_1\,q'_1\right)^{s+1}} \ \ \ {\rm for} \ \, q_1 \,, \ q'_1 \ \to \ \infty
\nn\ee
where the constant $C_1$ is given by 
\be
C_1 \ = \frac{3^{1-s}}{|\lambda'(is)|^2\ (1-s)^2}
\left[
A\int_0^{\frac{\pi}{2}}d\theta \ A_{21}(s,\theta)^2 \ - \ \int_0^{\frac{\pi}{2}}d\theta\int_0^{\frac{\pi}{2}}d\theta' \, A_{21}(s,\theta) \, 
y_{11}(\theta,\theta') \, A_{21}(s,\theta')
\right]
\nn\ee
which we evaluate numerically:
Using Eq.~(\ref{eq:A21}),
we get \,$A\,\int_0^{\pi/2}d\theta\,A_{21}(s,\theta)^2= 13.86$;
for the second integral, using the numerical solution $y_{11}$ of the STM equations, we obtain $4.94$;
also using $|\lambda'(i\,s)|=4.676377731$,
we finally get $C_1= 0.292(1)$.
Noticing that $C_1=F(0)$ where $F$ is defined in Eq.~(\ref{eq:def_F}),
we can use Eq.~(\ref{eqC3nUL}) to get $f_3(0)=4.55(2)$,
in agreement with the analytical result Eq.~(\ref{eq:f(0)}).
This is a non trivial consistency check.

%This results agrees with the one we obtained from a direct fit of the numerical solution of the STM equation $y_{22}(u,u')$ at large $u$ and $u'$ (following the procedure described in Section~\ref{subsubsec:C3_a<0}).
%This is a non trivial cross-check of the two methods. 

\subsection{Negative scattering length and arbitrary $l$}\label{AppImt3BCS}

We now extend our analysis to a finite negative scattering length,
  $a^{-1}\leq\,0$\,
and arbitrary $l$.
In the STM equations~(\ref{eqXBCS}) and~(\ref{eqYBCS}), 
we perform the
 same changes of variables and of functions than in Subsection~\ref{sssec:decomp_sect}.
The resulting equation for $y_{22}$ is
\be
y_{22}(u_1, u'_1) \ = \ \int_0^\infty du \ \bar{r}_{22}(u_1,u)\,y_{22}(u,u'_1) \ + \ \int_0^{\pi/2} d\theta \, \big[ r'_{21}(u_1,\theta)\,y_{12}(\theta, u'_1) \ + \ \bar{r}_{21}(u_1,\theta)\,x_{12}(\theta,u'_1)\,\big]
\nn\ee
{\it i.e.}
\be
y_{22} \  = \  \bar{r}_{22}\,y_{22} \ + \ r'_{21}\,y_{12} \ + \ \bar{r}_{21}\,x_{12}
\label{eq:STM_y22}
\ee
where
\be
r'_{21}(u_1,\theta) \ = \  \frac{
(k_0 |a|)^{-1}
  \ \cos\theta}{
 (k_0 a)^{-2}
  +\cos^2 \theta}
\ r_{21}(u_1,\theta)\  \ \ ,
\ \ \ \ \ \bar{r}_{21}(u_1,\theta) \ = \ \frac{\cos^2 \theta}{
  (k_0 a)^{-2}
  +\cos^2\theta}
\ r_{21}(u_1,\theta)\ \ \ ,
\nn\ee
and
\be
\bar{r}_{22}(u_1,u) \ = \
W(u)
\,r_{22}(u_1,u)
\nn\ee
with
\be
W(u)
\ = \ \frac{\sinh(u)}{\sinh(u)+(k_0 |a|)^{-1}} \,.
\label{eq:def_w_alpha}
\ee
$\bar{r}_{22}$ is not a symmetric kernel. We prefer to work with the symmetric kernel
\be
\tilde{r}_{22}(u_1,u'_1) \ = \ \sqrt{W(u_1)} \ r_{22}(u_1,u'_1) \ \sqrt{W(u'_1)}
\nn
\ee
We also define the new kernels
\begin{align*}
\tilde{r}_{21}(u_1,\theta'_1)& \ = \ \sqrt{W(u_1)} \ r_{21}(u_1,\theta'_1)
\\ \tilde{r'}_{21}(u_1,\theta'_1)& \ = \ \sqrt{W(u_1)} \ r'_{21}(u_1,\theta'_1)
\\ \tilde{\bar{r}}_{21}(u_1,\theta'_1)& \ = \ \sqrt{W(u_1)} \ \bar{r}_{21}(u_1,\theta'_1)
\end{align*}
and accordingly, the new functions
\begin{align*}
  \tilde{y}_{22}(u_1,u'_1) \ & = \ \sqrt{W(u_1)} \, y_{22}(u_1,u'_1) \, \sqrt{W(u'_1)}
  \\
\tilde{y}_{12}(\theta_1,u'_1) \ & = \ y_{12}(\theta_1,u'_1) \, \sqrt{W(u'_1)}
\\
\tilde{y}_{21}(u_1,\theta'_1) \ & = \ \sqrt{W(u_1)}\,y_{21}(u_1,\theta'_1)
\end{align*}
and similarly for $x_{12}$
 and $x_{21}$.
 In term of these new quantities, Eq.~(\ref{eq:STM_y22}) becomes
 \be
\tilde{y}_{22} \ =\  \tilde{r}_{22}\,\tilde{y}_{22}\, + \, \tilde{r'}_{21}\,\tilde{y}_{12}\, + \, \tilde{\bar{r}}_{21}\,\tilde{x}_{12}
\label{eq:ytilde_22=}
 \ee
and projecting the STM equations (\ref{eqXBCS}) and~(\ref{eqYBCS}) onto the 21 sector yields
 \bea
\tilde{y}_{21} \ &=& \ \tilde{r}_{22}\,\tilde{y}_{21}\, + \, \tilde{r'}_{21}\,y_{11}\, + \, \tilde{\bar{r}}_{21}\,x_{11}
\label{eq:ytilde_21=}
\\
\tilde{x}_{21} \ &=& \  A\ \tilde{r}_{21}\, + \, \tilde{r}_{22}\,\tilde{x}_{21}\, + \, \tilde{r'}_{21}\,x_{11}
-
\tilde{\bar{r}}_{21}\,y_{11}
\label{eq:xtilde_21=}
\eea

Similarly to the $a=\infty$ case, we then express $\tilde{y}_{22}$ (which corresponds to large momenta) in terms of $x_{11}$ and $y_{11}$ (which correspond to low momenta).
We define $\tilde{r}_{12}$\,, $\tilde{r'}_{12}$ and $\tilde{\bar{r}}_{12}$ analogously to \,$\tilde{r}_{21}$\,, $\tilde{r'}_{21}$ and $\tilde{\bar{r}}_{21}$\,, that is,
$\tilde{r}_{12}(\theta,u) = r_{12}(\theta,u)\,\sqrt{W(u)}$\,,
      $\tilde{r'}_{12}(\theta,u) = \tfrac{(k_0 |a|)^{-1} \,\cos\theta}{(k_0 a)^{-2} + \cos^2 \theta} \, r_{12}(\theta,u)\,\sqrt{W(u)}$ and
      $\tilde{\bar{r}}_{12}(\theta,u) = \sqrt{W(u)} \, r_{12}(\theta,u)\,
      \tfrac{\cos^2\theta}{(k_0 a)^{-2} + \cos^2 \theta}$\,.
We assume that $\tilde{r}_{22}$ is invertible.
From Eq.~(\ref{eq:prop_transp_xyr}), it follows that
$\tilde{x}_{2 1}^{t}=\tilde{x}_{1 2}\ \, ,
\ \, \tilde{y}_{2 1}^{t}=\tilde{y}_{1 2}\ \, ,
\ \, \tilde{r}_{2 1}^{t}=\tilde{r}_{1 2}\ \, ,
\ \, \tilde{r'}_{2 1}^{t}=\tilde{r'}_{1 2}\ \, ,
\ {\rm and} \ \tilde{\bar{r}}_{2 1}^{t}=\tilde{\bar{r}}_{1 2}$\ .
Injecting the transpose of Eqs.~(\ref{eq:ytilde_21=}) and~(\ref{eq:xtilde_21=}) into Eq.~(\ref{eq:ytilde_22=}), we obtain
\bea
\tilde{y}_{22}&=&\left(\mathds{1}_2-\tilde{r}_{2\,2}\right)^{-1}S_{22}\left(\mathds{1}_2-\tilde{r}_{2\,2}\right)^{-1}
\label{eqytilde22}
\eea
with
\begin{equation}
S_{22}=\tilde{r'}_{21}\,\left(y_{11}\,\tilde{r'}_{12} +x_{11}\,\tilde{\bar{r}}_{12}\right)
+\tilde{\bar{r}}_{21}\,\left(A \, \tilde{r}_{12}+x_{11}\,\tilde{r'}_{12} -y_{11}\,\tilde{\bar{r}}_{12}\right)
\label{eq:S22=}
\end{equation}
This expression is the sum of five terms, which we write in the form
\be
S_{22} \ = \ \sum_{j=1}^5\,a^{\j}_{21}\,b^{\j}_{11}\,c^{\j}_{12}
\label{eq:def_abc}
\ee
where $(a^{\j}_{21}\,b^{\j}_{11}\,c^{\j}_{12})(u,u')\, := \, \int_0^{\pi/2}d\theta\int_0^{\pi/2}d\theta'\,a^{\j}_{21}(u,\theta)\,b^{\j}_{11}(\theta,\theta')\,c^{\j}_{12}(\theta',u')$ for $u$ and $u'$ positive. 
Injecting Eq.~(\ref{eq:def_abc}) into (\ref{eqytilde22}) yields
\be
\tilde{y}_{22}(u,u') \ = \ \sum_{\j=1}^5 \ \int_0^{\frac{\pi}{2}}d\theta\int_0^{\frac{\pi}{2}}d\theta' \ 
\mathcal{A}^{\j}_{21}(u,\theta)\,b^{\j}_{11}(\theta,\theta')\,\mathcal{C}^{\j}_{12}(\theta,u')
\label{eq:y22A21C21}
\ee
where 
\be
\mathcal{A}^{\j}_{21}(u,\theta) \ = \ \int_0^{\infty}dw\ \left(\mathds{1}_2-r_{2\,2}\right)^{-1}\!(u,w)\ a^{\j}_{21}(w,\theta)
\nn\ee
and
\be
\mathcal{C}^{\j}_{12}(\theta,u) \ = \ \int_0^{\infty}dw\ c^{\j}_{12}(\theta,w)\ \left(\mathds{1}_2-r_{2\,2}\right)^{-1}\!(w,u)
\nn\ee

 Using the diagonalization of $\tilde{r}_{22}$\,, discussed in Appendix~\ref{Appeigenvec}, we get
 \be
 \left(\mathds{1}_2-\tilde{r}_{2\,2}\right)^{-1}(u,w) \ = \ \int_0^{\infty}d\omega\ \frac{g_{\omega}(u)\,g_{\omega}(w)}{1-\lambda(\omega)}
\ +\ 
\sum_{i=1}^{N_{\rm ds}}\  \frac{g_{{\rm ds}, i}(u)\,g_{{\rm ds},i}(w)}{1-\lambda_{{\rm ds},i}}
\nn\ee
    where $N_{\rm ds}$ is the number of discrete states ({\it e.g.} $N_{\rm ds}=0$ for $l=0$
    and $N_{\rm ds}=1$ for $l=1$).
    Hence
    \be
    \mathcal{A}^{\j}_{21}(u,\theta) \ =\
\mathcal{A}^{\j}_{21, {\rm cont}}(u,\theta) \ +\
\mathcal{A}^{\j}_{21, {\rm ds}}(u,\theta)
\label{eq:A=cont+ds}
\ee
where the contribution of the continuous spectrum is
\be
\mathcal{A}^{\j}_{21, {\rm cont}}(u,\theta) \ = \ \int_0^\infty dw \ \int_0^\infty d\omega \ \frac{g_{\omega}(u)\,g_{\omega}(w)}{1-\lambda(\omega)} \ a_{21}^{\j}(w,\theta)
\nn\ee
while the contribution of the discrete states is
\be
\mathcal{A}^{\j}_{21, {\rm ds}}(u,\theta) \ = \ \sum_{i=1}^{N_{\rm ds}} \ \int_0^\infty dw \ \frac{g_{{\rm ds}, i}(u)\,g_{{\rm ds}, i}(w)}{1-\lambda_{{\rm ds}, i}} \ a_{21}^{\j}(w,\theta)
\nn\ee
Let us simplify $\mathcal{A}^{\j}_{21, {\rm cont}}$ in the large-$u$ limit.
  $g_\omega$ and $\Delta(\omega)$ are {\it a priori} defined for $\omega\geq0$, and we extend them to $\omega<0$ by setting $g_{-\omega}(u)=g_{\omega}(u)$ and accordingly $\Delta(-\omega)=-\Delta(\omega)$.
The integrand is then an even function of $\omega$, so that
we can extend the integral over $\omega$ to $]-\infty,+\infty[$\, and divide by~2.
     In the limit $u\to\infty$, we use Eq.~(\ref{eq:gomegaasympt});
      we then write the cosine as sum of two exponentials,
      and make the change of variable $\omega \to -\omega$ in the contribution of the second exponential, which yields
\be
\mathcal{A}^{\j}_{21,  {\rm cont}}(u,\theta) \ \underset{ u\to\infty}{\sim} \  \frac{1}{\sqrt{2\,\pi}} \ \int_0^{\infty}dw\,\int_{-\infty}^{\infty}d\omega\ \frac{e^{i\omega u}e^{i\Delta(\omega)}g_{\omega}(w)}{1-\lambda(\omega)}\ a_{21}^{\j}(w,\theta)
     \label{eq:mathcalA21uinf}
\ee

In what follows we will use the property
\be
a_{21}^{\j}(w,\theta) \  \underset{w\to\infty}{\sim} \ \ f^j(\theta) \ e^{-(l+1)\,w}
\label{eq:a21_asym}
\ee
To derive this, we start from the definition of $a_{21}^{\j}$\,, given by Eqs.~(\ref{eq:S22=}) and (\ref{eq:def_abc}), which yields
  \,$a_{21}^{\j} = \tilde{r'}_{21}$\, for $\j\in\{1,2\}$
  and
  \,$a_{21}^{\j} = \tilde{\bar{r}}_{21}$\, for $\j\in\{3,4,5\}$\,.
  By expanding the integrand in Eq.~(\ref{eq:defQl}) in powers of \,$1/z$\, and using the fact that $\int_{-1}^1 dv \ P_l(v) \, v^k$ is zero for $k\in[|0,l-1|]$ and non-zero for $k=l$, we get
  \be
  Q_l(z) \  \underset{|z|\to\infty}{\sim} \  \ \frac{{\rm Constant}}{z^{l+1}}
  \label{eq:Ql_large_z}
  \ee
Thus by Eq.~(\ref{eq:r21}),
$r_{21}(w,\theta) \sim {\rm function}(\theta) \, e^{-(l+1)\,w}$ for $w\to\infty$\,.
Hence \,$\tilde{r'}_{21}(w,\theta)$ and \,$\tilde{\bar{r}}_{21}(w,\theta)$\, also \,$\sim {\rm function}(\theta) \, e^{-(l+1)\,w}$\, for $w\to\infty$.
This concludes the derivation of (\ref{eq:a21_asym}).

Since $g_{\omega}(w)$ has the large-$w$ asymptotic behavior Eq.~(\ref{eq:gomegaasympt}), we can
 define $\tilde{g}_{\omega}$ such that \be
  g_{\omega}(w) \ = \ \tilde{g}_{\omega}(w) \ + \ \tilde{g}_{-\omega}(w)
  \label{eq:g_split}
\ee
 and \,$\tilde{g}_{-\omega}(w) = \tilde{g}_{\omega}(w)^*$\, for real $\omega$,
with the asymptotic behavior
\be
\tilde{g}_{\omega}(w) \ \underset{w\to\infty}{\sim} \ \
  \frac{e^{i\,[\omega\,w \,+\, \Delta(\omega)]}}{\sqrt{2\pi}}
 \label{eq:gtilde_asym}
\ee
   Inserting (\ref{eq:g_split}) into~(\ref{eq:mathcalA21uinf}) yields,
    after exchanging the order of integration,
     \be
     \mathcal{A}^{\j}_{21,  {\rm cont}}(u,\theta) \ \underset{ u\to\infty}{\sim} \  \, I^j_+(u,\theta)   \ + \ I^j_-(u,\theta)
     \label{eq:A21=I+I}
     \ee
     where
    \be
    I^j_\pm(u,\theta)    \ = \ \frac{1}{\sqrt{2\,\pi}} \
   \int_{-\infty}^{\infty}d\omega\ 
    \frac{e^{i\omega u}e^{i\Delta(\omega)}}{1-\lambda(\omega)}\
     \int_0^{\infty}dw\  \tilde{g}_{\pm\omega}(w) \ a_{21}^{\j}(w,\theta)
\label{eq:Ipm}
\ee
Let us first consider $I^j_+(u,\theta)$. In the integral over $\omega$, we close the contour in the upper half plane,
after analytically continuing the integrand to complex $\omega$.
  Applying the residue theorem yields a sum over the singularities of the integrand in the upper half plane.
   
   The function \,$1/[1-\lambda(\omega)]$, and thus the integrand in Eq.~(\ref{eq:Ipm}), has poles at $\omega=i\,\tilde{s}_{l,n}$\, for $n\in\mathbb{N}$, 
where $\tilde{s}_{l,0} < \tilde{s}_{l,1} < \ldots $\, are the real positive solutions of the equation $\lambda(i\,s)=1$.
One can show\footnote{For $l=0$ and $l=1$, this follows from the analytical expressions of $\lambda(\omega)$, Eqs.~(\ref{eq:lambda_l0}) and~(\ref{eq:lambda_l1}).
For arbitrary $l$, this can be shown by comparing the expression~(\ref{eq:lambdal_int_inf}) of $\lambda(\omega)$  with an expression in Ref.~\cite{castin_tignone}.
Specifically, the scaling exponents $s_{l,n}$ are the solutions of
$\Lambda_l(s)=0$ where
the function $\Lambda_l$ is given by Eq.~(35) of~\cite{castin_tignone}, and by changing the order of integration and comparing with our Eq.~(\ref{eq:lambdal_int_inf}), one finds $\Lambda_l(s) = \frac{\sqrt{3}}{2} \, [1-\lambda(is)]$ for imaginary $s$, and thus for any $s$ by analytic continuation.}
that
\bea
&& \tilde{s}_{0,0}
\ = \ 2  \ , 
\ \ \ \ \ \ \ \ \ \ \ 
\tilde{s}_{0,n} \ = \ s_{0, \, n-1}  \ \ \ {\rm for} \ \ \ n>0\,, \ \ \ \ \ \ \ \ \ \ \ \ 
\label{eq:stilde_l0}
\\
&& \tilde{s}_{l,n}
\ = \
s_{l, n}  \ \ \ {\rm for} \ \ \ l > 0
\label{eq:stilde_l>0}
\eea
where the $s_{l,n}$ are the scaling exponents of the unitary three-body problem given in Refs.~\cite{Werner3corpsPRL,WernerThese,C3_I}.
   
We assume that 
the contribution of other singularities
to
\,$I^j_+(u,\theta)$\,
(and hence to $\mathcal{A}^{\j}_{21,{\rm cont}}$\,)
   cancels out
   with the contribution \,$\mathcal{A}^{\j}_{21, {\rm ds}}$ coming from the discrete states,
as we have explicitly shown at the unitary limit for $l=1$
[see the derivation of Eq.~(\ref{eq:Buw3})].
This yields
\be
I^j_+(u,\theta)
\  + \ \mathcal{A}^{\j}_{21, {\rm ds}}(u,\theta) \ = \ \sqrt{2\pi}\ \sum_{n= 0}^\infty \ e^{- \,\tilde{s}_{l,n} \, u \ }\frac{e^{i\,\Delta(i\,\tilde{s}_{l,n})}}{i\,\lambda'(i\,\tilde{s}_{l,n})}
\ \int_0^{\infty}dw\ \tilde{g}_{i\,\tilde{s}_{l,n}}(w)\ a_{21}^j(w,\theta)\label{eq:Ij1}
\ee

We turn to $I^j_-(u,\theta)$.
    From Eqs.~(\ref{eq:a21_asym}) and~(\ref{eq:gtilde_asym}), we have
    \be
    \tilde{g}_{-\omega}(w)\,a_{21}^j(w,\theta)
    \ \underset{w\to\infty}{\sim} \ 
\frac{f^j(\theta)}{\sqrt{2\pi}} \ 
e^{i[-w\,\omega+\Delta(-\omega)]}\ e^{-(l+1)\,w}
\label{eq:ga_asym}
\ee
In contrast to the case of $I_+$\,, we cannot directly close the $\omega$-integration contour, neither in the upper nor on the lower half plane,
because substituting (\ref{eq:ga_asym}) into~(\ref{eq:Ipm}) yields a factor $e^{i\omega(u-w)}$ and the sign of $u-w$ is not definite.
To proceed,
we define the difference \,$\delta_\omega^j(w,\theta)$\, between $\tilde{g}_{-\omega}(w)\,a_{21}^j(w,\theta)$ and its asymptotic behavior, {\it i.e.}
\be
\tilde{g}_{-\omega}(w)\,a_{21}^j(w,\theta)
    \ = \ 
\frac{f^j(\theta)}{\sqrt{2\pi}} \ 
e^{i[-w\,\omega+\Delta(-\omega)]}\ e^{-(l+1)\,w}
\ + \
\delta_\omega^j(w,\theta)
\label{eq:ga_decomp}
\ee
Substituting (\ref{eq:ga_decomp}) into~(\ref{eq:Ipm}),
and using $\Delta(-\omega)=-\Delta(\omega)$,
we get
\be
I^j_-(u,\theta) \ = \ I^j_{-,{\rm asym}}(u,\theta) \ + \ \, \delta I^j_-(u,\theta)
\label{eq:dI-=sum}
\ee
where
\be
I^j_{-,{\rm asym}}(u,\theta) \ = \ \frac{f^j(\theta)}{2\pi}
\ \int_{-\infty}^{\infty}d\omega \ 
\frac{e^{i\omega \,u}}{1-\lambda(\omega)}
 \ \int_0^{\infty}dw \ 
e^{-i\omega\,w}e^{-(l+1)w}
\label{eq:I-asym}
\ee
and
\be
\delta I^j_{-}(u,\theta) \ = \
\frac{1}{\sqrt{2\,\pi}} \
    \int_{-\infty}^{\infty}d\omega\ 
    \frac{e^{i\omega u}e^{i\Delta(\omega)}}{1-\lambda(\omega)}\
    \int_0^{\infty}dw\ \, \delta^j_\omega(w,\theta)
    \label{eq:dI-}
    \ee
Let us first evaluate Eq.~(\ref{eq:I-asym}). The integral over $w$ equals $1/(i\omega+l+1)$.
Closing the $\omega$ contour in the upper half plane, the integrand has poles at \,$\omega=i\,\tilde{s}_{l,n}$ for $n\in\mathbb{N}$,
but not at $\omega=i\,(l+1)$ because the pole
$1/(i\,\omega+l+1)$ is compensated by the divergence of $\lambda(\omega)$ for $\omega\to i\,(l+1)$, as shown in Section~\ref{Applambdal}.
This yields
    \be
    I^j_{-,{\rm asym}}(u,\theta) \ = \ f^j(\theta) \ \sum_{n=0}^\infty \ \frac{e^{-\tilde{s}_{l,n}\,u}}{i\,\lambda'(i\,\tilde{s}_{l,n})(l+1-\tilde{s}_{l,n})} 
    \label{eq:Ij2asymp}
    \ee
    
    We also need to evaluate the contribution \,$\delta I^j_{-}(u,\theta)$\,.
We assume that the difference between $\tilde{g}_{-\omega}(w)$ and its asymptotic behavior [\,given by (\ref{eq:gtilde_asym})] has the form
\be
\tilde{g}_{-\omega}(w) \ - \  \frac{e^{i[-\omega\,w \, + \, \Delta(-\omega)]}}{\sqrt{2\pi}}
\ \underset{w\to\infty}{\sim} \ \,
           {\rm function}(\omega)\ \, e^{-i\,\omega\,w} \ e^{-2w}
           \label{eq:g-_asym}
           \ee
 where ${\rm function}(\omega)$ is a function of $\omega$ we do not need to specify. This is true for $l=0$ and $l=1$ at the unitary limit.\footnote{Indeed, for $l=0$,
we have $g_\omega(w)=\sqrt{2/\pi}\ \cos(\omega w)$ (see App.~\ref{eigenvr22l0})
and thus $\tilde{g}_\omega(w)=e^{i\omega w}/\sqrt{2\pi}$,
while for $l=1$,
Eq.~(\ref{eq:g_omega_l1_norm}) implies $\tilde{g}_\omega(w)=(\omega+i\,\tanh w)\,e^{i\omega w}
/\sqrt{2\pi(\omega^2+1)}$
and $e^{i\Delta(\omega)}=(\omega+i)/\sqrt{\omega^2+1}$.
 }
 Furthermore, the difference between $a_{21}^j$ and its asymptotic behavior [given by~(\ref{eq:a21_asym})] has the form
 \be
 a_{21}^j(w,\theta) \ - \ f^j(\theta)\ e^{-(l+1)\,w}
 \ \underset{w\to\infty}{\sim} \ {\rm function}(\theta) \ \, e^{-(l+3)\,w}
 \label{eq:a21-asym}
 \ee
 Combining Eqs.~(\ref{eq:g-_asym}) and~(\ref{eq:a21-asym}) yields
\be
 \delta_\omega^j(w,\theta)
 \ \underset{w\to\infty}{\sim} \  {\rm function}(\omega,\theta)\ e^{-i\,\omega\,w} \ e^{-(l+3)w}
 \label{eq:delta_om_asym}
 \ee
Using this, let us now estimate Eq.~(\ref{eq:dI-}), by splitting the integral over $w$ in two parts, $\int_0^{\infty}dw=\int_0^{u}dw+\int_u^{\infty}dw$.
\bi
\item
  For the first part
  \big(\,$\int_0^{u}dw$\big)\,, we have $u>w$,
  and we close the $\omega$ integration contour in the upper half plane, where $e^{i\omega(u-w)}\to0$. Keeping only the contributions coming from the poles at $\omega=i\,\tilde{s}_{l,n}$\,, we get
\be
\sqrt{2\pi} \ \sum_{n=0}^\infty \ \frac{e^{-\tilde{s}_{l,n}\,u}\ e^{i\,\Delta(i\,\tilde{s}_{l,n})}}{i\,\lambda'(i\,\tilde{s}_{l,n})}\ 
\int_0^{u}dw\  \, \delta_{i\,\tilde{s}_{l,n}}^j(w,\theta)
\nn\ee
By Eq.~(\ref{eq:delta_om_asym}), the integrand behaves as $e^{[\tilde{s}_{l,n}-(l+3)]w}$ for $w\to\infty$.

If \,$\tilde{s}_{l,n}<l+3$,
we can replace the upper integration limit $u$ by \,$\infty$
(at leading order in the $u\to\infty$ limit), which
yields
\be
\sqrt{2\pi} \ \sum'_{n} \ \frac{e^{-\tilde{s}_{l,n}\,u}\ e^{i\,\Delta(i\,\tilde{s}_{l,n})}}{i\,\lambda'(i\,\tilde{s}_{l,n})}\ 
\int_0^{\infty}dw\ \,
\delta_{i\,\tilde{s}_{l,n}}^j(w,\theta)
\nn\ee
where the sum $\sum'_{n}$ is restricted to the values of $n$ such that $\tilde{s}_{l,n}<l+3$.

If $\tilde{s}_{l,n}>l+3$
(which is always the case for large $n$), the integrand diverges for $w\to\infty$, and the final contribution, after multiplication by $e^{-\tilde{s}_{l,n}\,u}$, is  $O\big(e^{-(l+3)\,u}\big)$.
\item For the second part
  \big(\,$\int_u^{\infty}dw$\big)\,, we close the $\omega$ integration contour in the lower half plane, which yields
\be
 - \sqrt{2\pi} \ \sum_{n=0}^\infty \ \frac{e^{\tilde{s}_{l,n}\,u}\ e^{i\,\Delta(-i\,\tilde{s}_{l,n})}}{i\,\lambda'(-i\,\tilde{s}_{l,n})}\ 
\int_{u}^\infty  dw\  \, \delta_{-i\,\tilde{s}_{l,n}}^j(w,\theta)
\nn\ee
By Eq.~(\ref{eq:delta_om_asym}),
we have \,$\delta_{-i\,\tilde{s}_{l,n}}^j(w,\theta)\propto e^{-(\tilde{s}_{l,n}+l+3)w}$\, for $w\to\infty$,
and thus the integral over $w$ is of order $e^{-(\tilde{s}_{l,n}+l+3)u}$,
which yields a final contribution $O\big(e^{-(l+3)\,u}\big)$.
\ei
As a result,
\be
\delta I^j_{-}(u,\theta)
\ \underset{u\to\infty}{=} \
\sqrt{2\pi} \ \, \sum'_{n} \ \frac{e^{-\tilde{s}_{l,n}\,u}\ e^{i\,\Delta(i\,\tilde{s}_{l,n})}}{i\,\lambda'(i\,\tilde{s}_{l,n})}\ 
\int_0^{\infty}dw\ \,
\delta_{i\,\tilde{s}_{l,n}}^j(w,\theta)
\ + \ O\big(e^{-(l+3)u}\big)
\label{eq:dIji=}
\ee

By substituting Eqs.~(\ref{eq:Ij2asymp}) and (\ref{eq:dIji=}) into (\ref{eq:dI-=sum}), we get
\be
I^j_{-}(u,\theta) \ \underset{u\to\infty}{=} \ \,
\sum'_{n} \frac{e^{-\tilde{s}_{l,n}\,u}}{i\,\lambda'(i\,\tilde{s}_{l,n})}\left(
\frac{f^j(\theta)}{l+1-\tilde{s}_{l,n}}
+\sqrt{2\pi} \ e^{i\Delta(i\,\tilde{s}_{l,n})}\,
\int_0^{\infty}dw\
\delta_{i\,\tilde{s}_{l,n}}^j(w,\theta)
\right)
\ + \ O\big(e^{-(l+3)u}\big)
\label{eq:I-_asym}
\ee
By combining Eq.~(\ref{eq:I-_asym}) with Eqs.~(\ref{eq:A=cont+ds}), (\ref{eq:A21=I+I}) and (\ref{eq:Ij1}), we arrive at
\begin{multline}
  \mathcal{A}^j_{21}(u,\theta)
\ \underset{u\to\infty}{=} \ \
\sum'_{n} \frac{e^{-\tilde{s}_{l,n}\,u}}{i\,\lambda'(i\,\tilde{s}_{l,n})}\left[
\frac{f^j(\theta)}{l+1-\tilde{s}_{l,n}}
\right.
\\
\left.
+\sqrt{2\pi} \ e^{i\Delta(i\,\tilde{s}_{l,n})}\,
\int_0^{\infty}dw\
\left(
\tilde{g}_{i\,\tilde{s}_{l,n}}(w)\,a_{21}^j(w,\theta)
\, + \,
\delta_{i\,\tilde{s}_{l,n}}^j(w,\theta)
\, \right)
\, \right]
\ + \ O\big(e^{-(l+3)u}\big)
\label{eq:A21_asym_sum}
\end{multline}
For the final steps, we distinguish two cases.
\bi
\item
{\it Case} $l\geq1$.
      In the sum over $n$ in Eq.~(\ref{eq:A21_asym_sum}), we only need to keep
      the leading order behavior given by the $n=0$ term,
which is $\propto e^{-s_l \,u}$. The $O\big(e^{-(l+3)u}\big)$ contribution is negligible, because we have
      $s_l \, < \, l+3$~\cite{Werner3corpsPRL,WernerThese}.
Hence the asymptotic behavior of \,$\mathcal{A}^j_{21}$\, takes the form
\be
\mathcal{A}^j_{21}(u,\theta)
\ \underset{u\to\infty}{\sim} \ 
e^{-s_l\,u} \ h^j(\theta)
\label{eq:A21_asym}
\ee
Similarly, we find that the asymptotic behavior of \,$\mathcal{C}^j_{21}$\, takes the form
\be
\mathcal{C}^j_{21}(\theta',u')
\ \underset{u'\to\infty}{\sim} \ 
e^{-s_l\,u'}\,k^j(\theta')
\label{eq:C21_asym}
\ee
Substituting this into Eq.~(\ref{eq:y22A21C21}) gives the asymptotic behavior of $\tilde{y}_{22}$\,, which is equivalent to $y_{22}$ since $W(u)$ tends to 1
by Eq.~(\ref{eq:def_w_alpha}). This yields
\be
y_{22}(u,u')
\ \underset{u,u'\to\infty}{\sim} \ 
 e^{-s_l(u+u')}\,\sum_{j=1}^5 \ \int_0^{\frac{\pi}{2}}d\theta\int_0^{\frac{\pi}{2}}d\theta'
\,h^j(\theta)\ 
b^j_{11}(\theta,\theta')\ 
k^j(\theta')
\nn 
\ee
By Eqs.~(\ref{eq:defft3l}) and (\ref{eq:rel_y22_Imf}) we conclude that
\be
\Im t_{3,l} \big(q,q';E=\tfrac{k_0^2}{m}+i\,0^+\big)
\ \sim \ 
-\frac{m\,(2l+1)\,3^{-s_l}\,k_0^{2\,s_l}}{(q\,q')^{s_l+1}}\sum_{j=1}^5 \ \int_0^{\frac{\pi}{2}}d\theta\int_0^{\frac{\pi}{2}}d\theta'
\,h^j(\theta)
\ b^j_{11}(\theta,\theta')
\ k^j(\theta')
\label{eq:Imt3_asym_hk}
\ee
for \,$q,q'\to\infty$.
\item
  {\it Case} $l=0$.
  There is a special integer value of $\tilde{s}_{0,0}$\,, see Eq.~(\ref{eq:stilde_l0}), which however does not contribute to the final results, as we will see.
By following the same reasoning as for $l\geq1$, now keeping not only the leading $n=0$ term
\big($\propto e^{-\tilde{s}_{0,0}\,u} = e^{-2u}$\,\big)
but also the subleading $n=1$ term
\big($\propto e^{-\tilde{s}_{0,1}\,u} = e^{-s_0\,u}$\,\big),
we get
\begin{align}
y_{22}(u,u')
\ \underset{u,u'\to\infty}{\sim} \ \ \ 
\,\sum_{j=1}^5 \ \int_0^{\frac{\pi}{2}}d\theta \ \int_0^{\frac{\pi}{2}}d\theta' \ 
& \left[ \, e^{-2u} \, h^j_0(\theta)\ + \
  e^{-s_0\,u} \, h^j_1(\theta)\, \right] 
\nn\\
& \ \ \ 
b^j_{11}(\theta,\theta')\
\left[ \, e^{-2u'} \, k^j_0(\theta)\ + \
  e^{-s_0\,u'} \, k^j_1(\theta)\, \right]
\label{eq:y22_l0_gen}
\end{align}
We find that our numerical solution of the STM equation has the
asymptotic behavior \,$y_{22}(u,u) \sim {\rm Constant} \times e^{-2\,s_0 u}$\, for $u\to\infty$. This implies that in Eq.~(\ref{eq:y22_l0_gen}), the $e^{-2u-2u'}$, $e^{-2u-s_0 u'}$ and $e^{-s_0 u-2u'}$ terms have vanishing prefactors,
{\it i.e.},
$\sum_{j=1}^5 \, \int_0^{\pi/2} d\theta \ \int_0^{\pi/2}d\theta' \ 
h^j_n(\theta)\ 
b^j_{11}(\theta,\theta')\
k^j_{n'}(\theta) \,=\,0$\, for $(n,n')\in \big\{\, (0,0)\, , (0,1) \, , (1,0) \, \big\}$.
We have checked this numerically for $a=\infty$, and we assume that it also holds for finite~$a$.
Setting $h^j(\theta):= h^j_1(\theta)$ and \,$k^j(\theta):=k^j_1(\theta)$\,,
we conclude that Eq.~(\ref{eq:Imt3_asym_hk}) holds also for $l=0$.
\ei

\subsection{Pole of \,$\lambda(\omega)$ for arbitrary $l$} \label{Applambdal}

In this section, we derive the expression (\ref{eq:lambdaX}) of $\lambda(\omega)$, and we check that it has a pole at $\omega=i(l+1)$.
The expression for the eigenvalue of $r_{22}$ for any $l\geq 0$ is given by Eq.~(\ref{eq:lambdal}), {\it i.e.}
\begin{equation}
  \lambda(\omega) \ =
  A^{-1} \ \int_{-\infty}^{\infty}du\ \, Q_l(-2\cosh u)\ \exp(i\,\omega\,u)
  \label{eq:lambdal_int_inf}
\end{equation}
where
$Q_l$ and $A$ are defined in (\ref{eq:defQl})
and
(\ref{eq:def_A}).
In the complex $u$ plane, we consider the rectangular contour $\Gamma$\,
connecting the points $-R$, $R$, $R+i\,\pi$ and $-R+i\,\pi$, with $R\to\infty$.
From $\cosh(u+i\pi)=-\cosh(u)$ and $Q_{l}(-z)=(-1)^{l+1}\,Q_{l}(z)$, it follows that 
\begin{equation}
\lambda(\omega) \ = \ \frac{A^{-1}}{1+(-1)^{l}\exp(-\pi\,\omega)} \ \int_{\Gamma}du\ \, Q_l(-2\cosh u )\ \exp(i\,\omega\,u)
\end{equation}
The function $1/[1+(-1)^{l}\exp(-\pi\,\omega)]$ has a pole at  $\omega=i(l+1)$. 
In order to conclude that
$\lambda(\omega)$ has a pole at  $\omega=i(l+1)$, it remains to check that
the integral along the contour $\Gamma$ has a finite non-zero limit when $\omega\to i(l+1)$.

To do so,
we replace $Q_l$ by its expression (\ref{eq:defQl}),
and exchange the integrals over $u$ and $v$. For the integration over $u$,
 the function $1/(-2\,\cosh u-v)$ 
 has a pole
 for $u=u_0$\, such that $\cosh u_0=-v/2$, which is equivalent to $\cos(i\,u_0)=-v/2$. The solutions are
 $u_0=\pm i\,\arccos(-v/2)+i\,2\pi k$\,
 with $k\in\mathbb{Z}$.  Since $-1/2\leq v/2\leq 1/2$, we have $ \arccos(1/2)=2\pi/3\geq \arccos(-v/2)\geq \arccos(-1/2)=\pi/3$. Therefore the only solution inside $\Gamma$, {\rm i.e.} in the strip $\{ 0< {\rm Im}\,u < \pi \}$, is $u_0=i\,\arccos(-v/2)$. Thus by Cauchy's residue theorem
\be
\int_{\Gamma} \ \frac{\exp(i\,\omega u)}{(-2\,\cosh u-v)}\ du \ = \ 2\pi i\ \frac{\exp(i\,\omega u_0)}{(-2\,\sinh u_0)}
\nn\ee
We have $\sinh u_0 \,=\,i\,\sqrt{1-v^2/4}$\,, and thus
\be
\lambda(\omega) \ = \
-\,\frac{\pi}{2A}
 \ \left[1+(-1)^{l}\exp(-\pi\,\omega)\right]^{-1} \ \int_{-1}^{1} \, P_l(v)\ \frac{\exp[-\omega\arccos(-v/2)] }{\sqrt{1-v^2/4}} \ dv
\nn\ee
 The change of variables $X=\arccos(-v/2)$ yields
 \begin{equation}
  \lambda(\omega)\ = \ -\,\frac{4}{\sqrt{3}}
  \ \left[\,1+(-1)^{l}\,e^{-\pi\,\omega} \, \right]^{-1}\ \int_{\frac{\pi}{3}}^{\frac{2\,\pi}{3}}P_l(-2\cos X) \ e^{-\omega\,X}\ dX
  \label{eq:lambdaX}
\end{equation}
 [as a consistency check,
   performing the integral for $l=0$ and $l=1$, we recover Eqs.~(\ref{eq:lambda_l0}) and~(\ref{eq:lambda_l1})].
 We checked numerically for $0\leq l \leq 20$ that the integral in Eq.~(\ref{eq:lambdaX}) is non-zero for $\omega=i(l+1)$.

\section{Diagonalization of \,$r_{2\,2}$\, at the unitary limit}\label{app:eigUL}

We set ${\rm Re}\,E$ and the mass to unity, as in Eq.~(\ref{eq:E=1}).
The action of the operator $r_{2\,2}$ on any function $g$ is
  \be
  \left(r_{2\,2}\,g\right)(u_1) \ = \ \frac{4}{\pi\sqrt{3}}
  \ \int_0^{\infty}d u\ \
  Q_l\left(
  \frac{\frac{3}{4}-\cosh^2(u)-\cosh^2(u_1)}{\cosh(u)\cosh(u_1)}
  \right)\ g(u)\label{eqdefr22}
  \ee
Let us first explain how we have guessed the form of the eigenvectors of $r_{2\,2}$\,.
  We take inspiration from the analytical solution of the three-body problem at unitarity~\cite{Efimov}: 
  The exact solution in hyperspherical coordinates is $\Psi({\bf R})=\phi_0(\Oo)F(R)/R^2$, where $\phi_0$ is the unitary hyperspherical wavefunction, and $F(R)=J_s(R)$
  with $J_s$ the Bessel function of order $s$
(taking energy equal 1 and magnetic quantum number $\md=0$).
Let  \, $\tilde{f}(\rrho) \, := \, \underset{r\to 0}{\textrm{lim}}\ r\, \Psi({\bf R})\, \propto \, J_s(\rho) \, \rho^{-1} \, Y_l^0\!\left(\hat{\rho}\right)$\,,
    and $\tilde{f}(\qq) \, := \, \int d^3\!\rho \, e^{-i\pp\cdot\rrho}\,\tilde{f}(\rrho)$.
Using  Eq.~(\ref{eq:plane-wave_Ylm}), we get
        $\tilde{f}(\pp) 
\propto \int d\rho \, \rho \, J_s(\rho)\,j_l(p\rho)\,Y_l^0(\hat{\rho})$,
an integral that can be evaluated analytically.
\footnote{This procedure
  for solving the STM equation analytically was also followed
  in Ref.~\cite{Pricoupenko_3body_res_2025} in a different context
  (at negative energy and in presence of a three-body resonance).}
On the other hand, $\tilde{f}(\pp)$ solves an integral equation, of the form
\be
    \sqrt{-E+p^2}\ \ \tilde{f}(\pp) \ \, + \ \, \frac{1}{\sqrt{3}\,\pi^2} \, \int \, \frac{\tilde{f}({\bf p'})}{p^2 + p'^{\phantom{.}2} + \pp\cdot{\bf p'} - 3E/4}\  d^3\!p'
    \ \, = \ \, {\rm source\ term},
\nn\ee
hereafter called STM equation for $\tilde{f}$
    \big(see Eqs.~(3.14) and~(3.16) of Ref.~\cite{PetrovLesHouches2010}\big).
%(our $\tilde{f}$\, is $f$ in Ref.~\cite{PetrovLesHouches2010}).
        In order for the STM equation for $t_3$ [Eq.~(\ref{eq:STMt3_swap})]
        and the STM equation for $\tilde{f}$\,
to have the same kernel (for $E=1+i 0^+$),
        we identify
        $t_3(\qq , {\bf q'} ; 1+i 0^+)$ with $\sqrt{3q^2/4-1-i0^+} \ \, \tilde{f}\big(\sqrt{3}\,\qq/2 \big)$.
Using the above expression of
$\tilde{f}(\pp)$
with the change of variable $q=\frac{2}{\sqrt{3}} \, {\rm cosh} \, u_1$\,,
        we get
        $t_{3,l}(q , q' ; 1+i 0^+) \propto {\rm sinh}\, u_1 \, \int d\rho \,\rho\,J_s(\rho)\,j_l\big( \cosh(u_1) \, \rho \big)$.
      We multiply this expression by \,$q\propto\cosh\,u_1$ due to the change of function Eq.~(\ref{eq:defft3l}).
  Moreover, we take $s$ as an 
 imaginary parameter $i\omega$, in order to get an oscillating function of $\omega$, in agreement with
 App.~\ref{Appeigenvec}. Finally, we take a combination of $J_{i\omega}$ and $J_{-i\omega}$ to obtain a real even function.  
 For $l=0$, this yields
  \begin{equation}
    \sinh(u_1)\cosh(u_1)\int_0^{\infty}
    d\rho\,\rho\,j_0\big(\cosh(u_1)\,\rho\big)\ \frac{J_{i\,\omega}(\rho)+J_{-i\,\omega}(\rho)}{2} \ = \ \cosh\left(\frac{\pi}{2}\,\omega\right)\,\cos(\omega u_1)
    \nn
  \end{equation}
We will thus try \,$\cos(\omega u_1)$ as an eigenvector for $l=0$.
  For $l=1$, we get
  \bea
    \sinh(u_1)\cosh(u_1)\int_0^{\infty} 
    d \rho\,\rho\,j_1\big(\cosh(u_1)\,\rho\big)\ \, \frac{J_{i\,\omega}(\rho)-J_{-i\,\omega}(\rho)}{2\,i}
     \ &=& \ \frac{\sinh\left(\frac{\pi}{2}\,\omega\right)}{\omega}
    \nn \\ && \times
    \big[\omega\,\cos(\omega u_1)-\tanh(u_1)\,\sin(\omega u_1)\big]
    \nn
  \eea
We will thus try \,$\omega\,\cos(\omega u_1)-\tanh(u_1)\sin(\omega u_1)$ as an eigenvector for $l=1$.
  In the next two subsections, we will prove that these guesses for the eigenvectors are indeed correct.
 This constitutes exact non-trivial results.
  
  \subsection{Eigenvalues and eigenvectors for $l=0$}\label{eigenvr22l0}
  
  We first consider the $l=0$ case.
  We will show that the function
    $\cos(\omega u)$
is an eigenvector of $r_{2\,2}$\,,
  for an arbitrary real positive $\omega$.
  From Eq.~(\ref{eqdefr22})
   the action of $r_{2\,2}$ onto the function $\cos(\omega u)$ yields
  \be
   I_0(u_1,\omega) \ = \ \frac{4}{\pi\sqrt{3}} \ \int_0^{\infty}du\ \,
  Q_0\left(
  \frac{\frac{3}{4}-\cosh^2(u)-\cosh^2(u_1)}{\cosh(u)\cosh(u_1)}
  \right)\
  \cos(\omega u)
  \label{eqdefI0}
  \ee
       which we rewrite as
\,$I_0(u_1,\omega) \, = \, \int_{-\infty}^{\infty}du\ F_0(u,u_1,\omega)$\,
with
\be
F_0(u,u_1,\omega) \ = \ \frac{2}{\pi\sqrt{3}} \ \,
  Q_0\left(
  \frac{\frac{3}{4}-\cosh^2(u)-\cosh^2(u_1)}{\cosh(u)\cosh(u_1)}
  \right)\ e^{\,i\omega u}
  \label{eq:def_F0}
  \ee
Let us consider the rectangular contour $\Gamma$
connecting the points $-R$, $R$, $R+i\,\pi$ and $-R+i\,\pi$.
In the limit of infinite $R$,
we have
\bea
I_0(u_1,\omega) \ &=& \ \frac{1}{1+e^{-\pi\,\omega}} \ \, \int_{\Gamma}d u\ F_0(u,u_1,\omega)
\label{eq:I0=int_gamma}
  \eea
  where we used the fact that the contributions from the two vertical segments tend to zero for $R\to\infty$
  \big[because for $R\to \infty$ and fixed real $t$, we have $|\cosh(\pm R + i\,t)| \sim e^R / 2$,
      and thus using Eq.~(\ref{eq:Q0_z_inf}),
      we get $|F_0( \pm R + i\,t , u_1 , \omega)| = O(e^{-R})$\big]
  as well as the property
$F_0(u+i\pi,u_1,\omega)=-e^{-\pi\,\omega}\ F_0(u,u_1,\omega)$
\big[which follows from $\cosh(u+i\,\pi)=-\cosh(u)$ and $Q_0(-z)=-Q_0(z)$\big].
  Replacing $Q_0$ in Eq.~(\ref{eq:def_F0}) by its expression~(\ref{eq:defQl}), injecting into Eq.~(\ref{eq:I0=int_gamma}),
  and changing the order of integrations, we get
   \be
  I_0(u_1,\omega) \ = \ 
  \frac{  \cosh(u_1)}{\pi\sqrt{3}\,\big(1+e^{-\pi\,\omega}\big)}\ 
  \int_{-1}^{1}dv\ J_0(v,u_1,\omega)
  \label{eqI0}
  \ee
 and
  \be
  J_0(v,u_1,\omega) \ = \ - \,\int_{\Gamma}d u\
\, \frac{\cosh(u) \ e^{\,i\omega u}}
     {\cosh^2(u) + \cosh^2(u_1)
       + v\,\cosh(u_1)\cosh(u)
      - \frac{3}{4}}
  \label{eq:J0}
  \ee
 We can now perform the integration over $u$.
 The singularities inside the contour $\Gamma$
of the integrand in Eq.~(\ref{eq:J0})
are two simple poles at $u=u_\sigma$\,
with $\sigma = +$ or $-$\,,
which satisfy the equation
\be
\cosh^2(u_\sigma) \ + \ \cosh^2(u_1)
      \ + \ v\,\cosh(u_1)\cosh(u_\sigma)
     \ - \ \frac{3}{4} \ = \ 0
\label{eq:u_sig_impl}
\ee
More explicitly, $u_\sigma$ is defined as
the unique solution
such that $0 < \Im(u_\sigma) < \pi$\,
    of the equation
  \be
  \cosh(u_{\sigma}) \ = \ -\,\frac{v}{2}\ \cosh(u_1)\ + \ i\ \sigma
  \  \frac{\sqrt{-\Delta}}{2}
  \ \ \ \ \    {\rm with} \ \  \Delta  \ = \  3  \, +  \, (v^2-4) \ \cosh^2(u_1)
  \nn
  \ee
  (we have $\Delta < 0$ since we can discard the edge case $v^2 = 1$).
       The existence and uniqueness of $u_\sigma$ follows from the fact that the function $\cosh$ is a one-to-one mapping from the half-strip $\{ \, 0 < \Im(z) < \pi  \ \, {\rm and} \ \Re(z) > 0\, \}$ onto the upper half plane
       $\{ \, \Im z > 0 \, \}$.

Cauchy's theorem then yields
 \be
  J_0(v,u_1,\omega) \ = \ - \, 2\pi i \ \sum_{\sigma=\pm}\ e^{\,i\omega u_{\sigma}}\ \frac{\cosh(u_{\sigma})}{\sinh(u_{\sigma})\ i\sigma\sqrt{-\Delta}}
  \nn
 \ee
  Injecting this into Eq.~(\ref{eqI0}),
 we evaluate the integral over $v$ thanks to the change of variable $v \longrightarrow u_\sigma$\,: We have
 \be
 d v \ = \ -\,i\,\sigma\ \frac{\sqrt{-\Delta}\ \sinh(u_{\sigma})}{\cosh(u_1)\,\cosh(u_{\sigma})}\ d u_{\sigma}
\label{eqdv}
 \ee
 so that many factors cancel out and we obtain
 \be
    I_0(u_1,\omega) \ = \ 
  \frac{2\, i}{\sqrt{3}\,\big(1+e^{-\pi\,\omega}\big)}\ 
\   \sum_{\sigma}
\   \int_{\gamma_\sigma} d u_{\sigma}\ 
e^{\,i\omega u_{\sigma}}
\nn
  \ee
     where $\gamma_\sigma$ is a path going from
       $u_\sigma = u_\sigma^{(-)}$ to $u_\sigma = u_\sigma^{(+)}$ with
       \be
       u_\sigma^{(\pm)} \ = \ \sigma\,u_1 \ + \ i\ \left(\frac{\pi}{2} \, \pm \,  \frac{\pi}{6} \, \right)
       \label{eq:u_bornes}
       \ee
       This finally yields
  \be
    I_0(u_1,\omega)
    \ = \ - \, \frac{4\,\sinh\left(\frac{\pi}{6}\,\omega\right)}{\sqrt{3}\ \omega\,\cosh\left(\frac{\pi}{2}\,\omega\right)}\ \cos(\omega u_1)
\nn   \ee
We conclude that $I_0(u_1,\omega)=   \lambda(\omega)\, \cos(\omega u_1)$,
 which means that the function $\cos(\omega u)$ is an eigenvector of $r_{2\,2}$\,, with the eigenvalue
 \begin{equation}
  \lambda(\omega)\ = \ - \  
  \frac{4\ \sinh(\frac{\pi}{6}\,\omega)}{\sqrt{3}\ \omega\,\cosh(\frac{\pi}{2}\,\omega)} \qquad (\textrm{for}\quad l=0).
  \label{eq:lambda_l0}
\end{equation}
 
 \subsection{Eigenvalues and eigenvectors for $l=1$}\label{eigenvr22l1}
 We turn to the $l=1$ case.
We will see that the spectrum comprises a continuous part and one discrete eigenvector.
   We will first show that the function 
   \be
    \omega\,\cos(\omega u) \ - \ \tanh(u)\sin(\omega u)
   \label{eq:g_omega_l1}
   \ee
is an eigenvector of $r_{2\,2}$\,,
for an arbitrary real positive $\omega$. Then we will show that
\be
g_{\rm ds}(u) \ = \ \frac{1}{\cosh u}
\label{eq:gds}
\ee
is also an eigenvector of $r_{2\,2}$.

From Eq.~(\ref{eqdefr22}) the action of $r_{22}$ onto the function in Eq.~(\ref{eq:g_omega_l1}) yields
   \be
  \Ir_1(\cosh u_1,\omega) \ = \ 
\frac{4}{\pi\sqrt{3}}\int_0^{\infty}du\ \,  Q_1\left(  \frac{\frac{3}{4}-\cosh^2(u)-\cosh^2(u_1)}{\cosh(u)\cosh(u_1)}  \right)\,\big( \omega\cos(\omega u)-\tanh(u)\sin(\omega u)\big)\nn
   \ee
   (we make the dependence on \,$\cosh u_1$\, explicit, which will be useful in App.~\ref{AppImt3largek}). Since the integrand is an even function of $u$,
   \be
 \Ir_1(\cosh u_1,\omega)
  \ = \
  \int_{-\infty}^{\infty}du\ \,
         \Fr_1(u, \cosh u_1,\omega)
    \label{eq:I1=F1}
   \ee
where
   \be
     \Fr_1(u, \cosh u_1,\omega)
  \ = \
      \frac{2}{\pi\sqrt{3}} \ 
  Q_1\left(
  \frac{\frac{3}{4}-\cosh^2(u)-\cosh^2(u_1)}{\cosh(u)\cosh(u_1)}
  \right)\,e^{\,i\omega u}\ \big(\omega+i\,\tanh u\big)
    \label{eq:def_F1}
  \ee
    Since  $Q_1$ is an even function, we get the identity
     $\Fr_1(u + i\,\pi, \cosh u_1,\omega) = e^{-\pi\,\omega}\,   \Fr_1(u, \cosh u_1,\omega)$,
      which allows us to
      rewrite the integral over $u$ as an integral over the rectangular contour $\Gamma$\,
connecting the points $-R$, $R$, $R+i\,\pi$ and $-R+i\,\pi$, with $R\to\infty$.
Then, replacing $Q_1$ by its expression~(\ref{eq:defQl})
  and changing the order of integrations yields
  \be
    \Ir_1(\cosh u_1,\omega) \ = \ 
  \frac{\cosh(u_1)}{\pi\sqrt{3}\,\big(1-e^{-\pi\,\omega}\big)} \ \,
  \int_{-1}^{1}dv\ v \
        \Jr_1(v,\cosh u_1,\omega)
  \label{eq:I1=int}
  \ee
  where
  \be
  \Jr_1(v,\cosh u_1,\omega)
\ = \ -\,\int_{\Gamma}d u\ \exp(i\,\omega u)
\   \frac{
\omega\ \cosh(u) + i \, \sinh(u)}
  {\cosh^2(u_1) + \cosh^2(u) + v\,\cosh(u_1)\cosh(u) - \frac{3}{4}}
  \nn
  \ee  
  Inside the contour $\Gamma$, the integrand has poles at $u = u_\pm$\,, and thus by Cauchy's residue theorem
    \be
\Jr_1(v,\cosh u_1,\omega) \ = \ - \,
  2\pi i\sum_{\sigma=\pm}\exp(i\omega u_{\sigma})
\   \frac{
  \omega\ \cosh(u_{\sigma}) + i\,\sinh(u_{\sigma})}
  {\sinh(u_{\sigma}) \ i\sigma\sqrt{-\Delta}}
  \nn\ee
    Injecting this into Eq.~(\ref{eq:I1=int}),
    performing the change of variable $v \longrightarrow u_\sigma$\,
    and using Eq.~(\ref{eqdv}) yields
    \be
     \Ir_1(\cosh u_1,\omega) \ = \
    \frac{2\, i}{\sqrt{3} \,
\big(1-e^{-\pi\,\omega}\big)    }
  \ \sum_{\sigma=\pm}\ \,
  \int_{\gamma_\sigma} d u_{\sigma}
  \ \exp(i\omega u_{\sigma})
  \ \big[
    \omega+i\,\tanh(u_{\sigma}) \,
    \big]
  \  v(u_\sigma , u_1)
  \label{eqI1_int1}
  \ee
    where the origin $u_\sigma^{(-)}$ and the destination $u_\sigma^{(+)}$
of the integration path $\gamma_\sigma$ are given by Eq.~(\ref{eq:u_bornes}),
and where
\,$v(u , u_1) \, = \, [3/4-\cosh^2(u_1)-\cosh^2(u)] / [\cosh(u_1)\,\cosh(u)]$\,
due to Eq.~(\ref{eq:u_sig_impl}).
  Next, we use $\exp(i\omega u)
  \big[
  \omega+i\,\tanh(u) \, \big]= -i\,\cosh(u)\,\frac{d}{d u}\left(\frac{\exp(i\omega u)}{\cosh(u)}  \right)$
and integrate by parts, which yields
\be
\Ir_1(\cosh u_1,\omega) \ = \
\frac{2\, i}{\sqrt{3} \,
  \big(1-e^{-\pi\omega}\big)}
  \ \sum_{\sigma=\pm} \ \left\{
  \, \Big[ e^{\,i\,\omega u} \ v(u,u_1)\, \Big]_{u = u_\sigma^{(-)}}^{u = u_\sigma^{(+)}}
  \ \, + \ \,
  \frac{2}{\cosh(u_1)}\ \int_{\gamma_\sigma} du\ e^{\,i\,\omega u}\,\sinh(u) \,
  \right\}
  \label{eq:I1_sum-int}
  \ee
  To evaluate this expression, we use the identity
$v(u_\sigma^{(\pm)},u_1) = \pm 1$\,, and we notice that
  the $\sigma=-$ contribution in Eq.~(\ref{eq:I1_sum-int}) is obtained from the
  $\sigma=+$ contribution
  by changing $u_1$ into $-u_1$.
  The integral in Eq.~(\ref{eq:I1_sum-int}) is conveniently evaluated using
  \be
  \int_{\gamma_+} du\ e^{\,i\,\omega u}\ e^{\,\pm u} \ = \
  \left[ \frac{e^{(i\omega \pm 1)u}}{i\omega\pm1}\right]_{u = u_\sigma^{(-)}}^{u = u_\sigma^{(+)}}
  \ \, = \ \, \frac{e^{(i\omega \pm 1)\left(u_1+i\frac{\pi}{2}\right)}}{i\omega \pm 1}
  \ \left(
    e^{(i\omega \pm 1)\,i\frac{\pi}{6}}
   \, - \,
    e^{-(i\omega \pm 1)\,i\frac{\pi}{6}}
    \right)
 \nn \ee
  After rearranging terms we obtain
 \be
 \Ir_1(\cosh u_1,\omega)
 \ = \ \frac{4}{\sqrt{3}\,(\omega^2+1)\sinh(\frac{\pi}{2}\,\omega)}
\ \big[\omega\,\cosh(\tfrac{\pi}{6}\,\omega)-\sqrt{3}\,\sinh(\tfrac{\pi}{6}\,\omega)\big]
\ \big[\omega\,\cos(\omega u_1)-\tanh(u_1)\sin(\omega u_1)\big]
\nn
 \ee
    {\it i.e.},
    \be
       \Ir_1(\cosh u_1,\omega) \ = \  \lambda(\omega)
       \ \big[\omega\,\cos(\omega u_1)-\tanh(u_1)\sin(\omega u_1)\big]
       \label{eq:I1=}
    \ee
    with
  \begin{equation}
\lambda(\omega) \ = \ - \  \frac{4\,\left[\sqrt{3}\,\sinh(\frac{\pi}{6}\,\omega)-\omega\,\cosh(\frac{\pi}{6}\,\omega)\right]}{\sqrt{3}\,(\omega^2+1)\sinh(\frac{\pi}{2}\,\omega)} \ \ \ \ \ \ \quad (\textrm{for}\quad l=1).
\label{eq:lambda_l1}
\end{equation}
  We conclude that the function defined in Eq.~(\ref{eq:g_omega_l1})
  is an eigenvector of $r_{2\,2}$ with the eigenvalue
  given in Eq.~(\ref{eq:lambda_l1}). 
 
 We turn to $g_{\rm ds}$ defined by Eq.~(\ref{eq:gds}),
   and show that it is also an eigenvector of $r_{22}$. We have
  \be
   (r_{22}\,g_{\rm ds})(u_1) \ \equiv \ 
   I_{1,{\rm ds}}(u_1)\ = \ 
   \frac{4}{\pi\sqrt{3}}\int_0^{\infty}du\ \,  Q_1\left(  \frac{\frac{3}{4}-\cosh^2(u)-\cosh^2(u_1)}{\cosh(u)\cosh(u_1)}  \right)\,
   \frac{1}{\cosh(u)}   \nn\ee
   which we rewrite as
   \,$I_{1,{\rm ds}}(u_1)\, = \,
   \int_{-\infty}^{\infty}du\ F_{1,{\rm ds}}(u,u_1)$
   with
   \be
   F_{1,{\rm ds}}(u,u_1)\ = \
      \frac{2}{\pi\sqrt{3}} \ 
  Q_1\left(
  \frac{\frac{3}{4}-\cosh^2(u)-\cosh^2(u_1)}{\cosh(u)\cosh(u_1)}
  \right)\,\frac{1}{\cosh(u)}
   \nn\ee
$Q_1$ is an even function, which implies $F_{1,{\rm ds}}(u+i\,\pi,u_1)=-F_{1,{\rm ds}}(u,u_1)$.
Following the same approach as for $\Ir_1(\cosh u_1,\omega)$, we find
\be
   I_{1,{\rm ds}}(u_1)\ = \
\frac{\cosh(u_1)}{2\pi\sqrt{3}}\int_{-1}^{1}dv\ v \   J_{1,{\rm ds}}(v,u_1)
 \label{eq:I1bs=int}
\ee
where
\be
J_{1,{\rm ds}}(v,u_1) \ = \,-\int_{\Gamma}d u\ 
\   \frac{
1
}
  {\cosh^2(u_1) + \cosh^2(u) + v\,\cosh(u_1)\cosh(u) - \frac{3}{4}}
  \nn
  \ee  
and thus by Cauchy's residue theorem
    \be
  J_{1,{\rm ds}}(v,u_1) \ = \ - \,
  2\pi \sum_{\sigma=\pm}
\   \frac{
  1}
  {\sinh(u_{\sigma}) \sigma\sqrt{-\Delta}}
  \nn\ee
  Substituting this into Eq.~(\ref{eq:I1bs=int}),
and expressing
  the integral over $v$ as an integral over the $u_{\sigma}$'s, we get
\be
    I_{1,{\rm ds}}(u_1) \ = \
    \frac{i}{\sqrt{3}}
  \ \sum_{\sigma=\pm}\ \,
  \int_{\gamma_\sigma} d u_{\sigma}
  \  \frac{v(u_\sigma , u_1)}{\cosh(u_{\sigma})}
  \label{eqI1bs_int1}
  \ee
  The integrand is equal to $\big[3/4\ -\cosh(u_1)^2-\cosh(u)^2\big] \, / \, \big[\cosh(u_1)\cosh(u)^2\big]$\,, which yields
 \be
  I_{1,{\rm ds}}(u_1) \ = \ \frac{1}{\cosh(u_1)}\left\{\ 
  \sum_{\sigma=\pm}\ \left(\frac{3}{4} \, - \, \cosh(u_1)^2 \right)
  \ \left[\tanh(\sigma u_1+i\frac{2\pi}{3})-\tanh(\sigma u_1+i\frac{\pi}{3})
    \ \right] \,-\,i\,\frac{\pi}{3}
  \right\}
  \nn
 \ee
Using $\tanh(\sigma u_1+i\frac{2\pi}{3})-\tanh(\sigma u_1+i\frac{\pi}{3})=i\,\tfrac{\sqrt{3}}{2}\,/\,\big[\,\tfrac{3}{4}-\cosh(u_1)^2\big]$,
we obtain
\,$I_{1,{\rm ds}}(u_1) \ = \ \big[\,2\pi/\big(3\sqrt{3}\,\big)\,-\,1\big]\,/\,\cosh(u_1)$. We conclude
that $g_{\rm ds}(u)$ is an eigenvector of $r_{22}$\,,
with the eigenvalue 
 \be
 \lambda_{\rm ds} \ = \ \frac{2\pi}{3\sqrt{3}} \, - \, 1
\label{eq:lambda_ds}
 \ee

 \subsection{Closure relation}
 
In this subsection, we derive the following closure relation for the $l=1$ eigenvectors of $r_{22}$\,:
\be
\delta(u-u') \ = \ \int_0^{\infty}d\omega\ g_{\omega}(u)\,g_{\omega}(u') \ + \
 g_{\rm ds}(u) \, g_{\rm ds}(u')
\label{eq:closure}
\ee
 for any $u>0$ and $u'>0$,
     where $g_{\rm ds}$ is the discrete eigenvector given by Eq.~(\ref{eq:gds}),
     and
     \be
     g_\omega(u) \ = \ \sqrt{\frac{2}{\pi\,(\omega^2+1)}} \ \
     \big[
       \omega\,\cos(\omega u) \ - \ \tanh(u)\sin(\omega u)
       \,\big]
       \label{eq:g_omega_l1_norm}
     \ee
is the eigenfunction of $r_{22}$ introduced in Eq.~(\ref{eq:g_omega_l1}) times an appropriate normalization factor.
 It will prove convenient to rewrite
Eq.~(\ref{eq:g_omega_l1_norm}) as
      \be
  g_{\omega}(u) \ = \ \frac{\bar{g}_{\omega}(u) \, - \, \bar{g}_{-\omega}(u)}{\sqrt{2\pi(\omega^2+1)}}
  \label{eq:g_om_gbar}
  \ee
where
\be
\bar{g}_{\omega}(u) \ = \ (\omega+i\,\tanh u)\ e^{i \omega u}
\label{eq:def_gbar_omega}
\ee
We will also use the identities
\be
\bar{g}_{\pm i}(u) \ = \ \pm \,i\,g_{\rm ds}(u)
\label{eq:gbar_i}
\ee
and
\be
\lambda(\omega) \ \underset{\omega\to i}{\longrightarrow} \ \lambda_{\rm ds}
\label{eq:lambda_i}
\ee
   [Eq.~(\ref{eq:gbar_i})
     follows from Eq.~(\ref{eq:gds}), while
Eq.~(\ref{eq:lambda_i})
     follows from Eqs.~(\ref{eq:lambda_l1}) and~(\ref{eq:lambda_ds})].

To derive the closure relation Eq.~(\ref{eq:closure}), we start by introducing $R>0$ and defining
\be
I_R(u,u') \ = \ \int_0^R g_{\omega}(u)\,g_{\omega}(u') \ d\omega\nn
\ee
 Since $g_\omega(u)$ is an even function of $\omega$, we can replace $\int_0^R$\, by \,$\frac{1}{2}\int_{-R}^R$\,.
Using Eq.~(\ref{eq:g_om_gbar}) then yields
\be
I_R(u,u') \ = \ \frac{1}{2\pi}\ \int_{-R}^R \ d\omega \ \frac{\bar{g}_\omega(u) \ \big[ \, \bar{g}_\omega(u') \, - \, \bar{g}_{-\omega}(u') \big]}{\omega^2+1}
\nn\ee
We then consider the closed contour made of the interval $[-R,R]$ and a semi-circle $\Gamma_R$ of radius $R$ in the upper half plane.
Applying the residue theorem and using Eq.~(\ref{eq:gbar_i}), we find that the integral over the closed contour equals
 $-g_{\rm ds}(u)\,g_{\rm ds}(u')$, due to a pole at $\omega=+i$.
We then evaluate the integral over the semi-circle $\Gamma_R$ in the limit $R\to\infty$\,: Since $|\omega|=R$, we can replace $\omega^2+1$ by $\omega^2$ and
$g_\omega(u)$ by \,$\omega\,e^{i\omega u}$\,, which yields the integral 
$\int_{\Gamma_R}
e^{i\omega u} \, (e^{i\omega u'}+e^{-i\omega u'})
\ d\omega/(2\pi) \, = \, -\,(R/\pi)\ \big\{\, {\rm sinc}\big[R(u+u')\big]
\,+\, {\rm sinc}\big[R(u-u')\big]
\, \big\}$.
In the limit $R\to+\infty$, this tends to $-\delta(u+u')-\delta(u-u')$. For $u>0$ and $u'>0$, the first Dirac vanishes, so that
$I_{\infty}(u,u')-\delta(u-u') \, = \, -g_{\rm ds}(u)\,g_{\rm ds}(u')$\,,
which yields Eq.~(\ref{eq:closure}).

\section{ Diagonalization of \, $\tilde{r}_{22}$\,  at finite $a$}\label{Appeigenvec}

Let us denote by $g_{\omega}(u)$ and $\lambda(\omega)$ the eigenvalues and eigenvectors in the continuous spectrum of the symmetric kernel $\tilde{r}_{2\,2}$\,, with $\omega$ a parameter whose meaning will be clarified below.
  The eigenvalue equation reads
\begin{multline}
\int_0^{\infty}d u\ \sqrt{W(u_1)}\ 
A^{-1}\ Q_l\left(
\frac{\frac{3}{4}-\cosh^2(u_1)-\cosh^2(u)}{\cosh(u_1)\cosh(u)}
\right)
\ \sqrt{W(u)}\ g_{\omega}(u)
\ = \
    \lambda(\omega)\,g_{\omega}(u_1)
    \label{eq:eigeneq_l}
\end{multline}
where
we used the notations from Eqs.~(\ref{eq:def_A}) and (\ref{eq:def_w_alpha}).
We consider the limit $u_1\gg1$.
We introduce $\Lambda$ such that $1 \ll \Lambda \ll u_1$\,,
and we split the integral in Eq.~(\ref{eq:eigeneq_l}) in two parts: $I_{-}$ for $u<\Lambda$, and $I_{+}$ for $u>\Lambda$.

First, let us show that $I_{-}$ tends to zero.
For $u_1\to\infty$,
the argument of $Q_l$ is $\sim \cosh(u_1)/\cosh(u)\to\infty$,
and $A^{-1}\,Q_l(z) \sim a_l/z^{l+1}$ for $z\to \infty$.
Hence
$|I_{-}|\sim |a_l| / (\cosh u_1)^{l+1} \ \big|\int_0^\Lambda du\ (\cosh u)^{l+1}\, \sqrt{W(u)}\ g_{\omega}(u)\big|$. Assuming that $|g_{\omega}(u)|$ is bounded by a constant $M$ (which we verified numerically), the integral is bounded by $M\,\Lambda\,\cosh(\Lambda)^{l+1}$. As a consequence, $I_{-}$ is bounded by
$|a_l|\,M\,\Lambda\,(\cosh \Lambda/\cosh u_1)^{l+1}$.
This tends to zero for $u_1\to\infty$ if we chose, {\it e.g.}, $\Lambda = \sqrt{u_1}$\,.

Second, we evaluate the integral $I_{+}$. We can replace, at leading order, the argument of $Q_l$ by
$-(\exp(u-u_1)+\exp(u_1-u))=-2\cosh(u_1-u)$, since $u\gg 1$ and $u_1\gg 1$.
Therefore 
$I_{+}\simeq A^{-1}\int_{\Lambda}^{\infty}d u \ Q_l\big(-2\cosh(u_1-u)\big)\,g_{\omega}(u)$.
We now take as an ansatz 
\be
g_{\omega}(u)\sim\,\sqrt{\frac{2}{\pi}} \ \cos\big(\omega u+\Delta(\omega)\big)\ \ \ \ \ \  \textrm{for}\ u\gg1
\label{eq:gomegaasympt}
\ee
This defines $\omega$ which is a wave number for the variable $u$. The phase shift $\Delta(\omega)$ in general depends on $\omega$ and $k_0 a$, but we do not need to determine it.
The integral can be calculated in the limit $\Lambda\gg 1$:
\bea
\int_{\Lambda}^{\infty}d u\ Q_l\big(-2\,\cosh(u_1-u)\big)\,\cos\big(\omega u
&+&
\Delta(\omega)\big)
=
\int_{\Lambda-u_1}^{\infty}d u'\,Q_l(-2\,\cosh u'\,) \ \cos\big(\omega\,(u'+u_1)+\Delta(\omega)\big)\nn\\
&=&
\left(\,\int_{\Lambda-u_1}^{\infty}d u'\,Q_l(-2\,\cosh u'\,)\ {\rm cos}(\omega u'\,)\right) \ \cos\big(\omega u_1+\Delta(\omega)\big)
\nn\\&&
-\left(\,\int_{\Lambda-u_1}^{\infty}d u'\,Q_l(-2\,\cosh u'\,)\ {\rm sin}(\omega u'\,)\right) \ \sin\big(\omega u_1+\Delta(\omega)\big)
\nn
\eea
     Since  $u_1\gg\Lambda\gg 1$, the lower bounds of the integrals can be replaced with $-\infty$. The integral with a $\sin$ gives $0$ by parity, which yields
$I_{+}\simeq\,\big[A^{-1}\,\int_{\Lambda-u_1}^{\infty}d u'\,Q_l\big(-2\cosh(u')\big)\,{\rm cos}(\omega u')\big]\,\cos\big(\omega u_1+\Delta(\omega)\big)$. Therefore, there is an eigenvector $g_\omega(u)$ of $\tilde{r}_{2\,2}$
with the asymptotic behavior (\ref{eq:gomegaasympt}) and
 with the eigenvalue
\bea
\lambda(\omega)&=&A^{-1}\,\int_{-\infty}^{\infty}du\ \, Q_l(-2\,\cosh u)\,\cos(\omega u)
\label{eq:lambdal}
\eea
For $l=0$ and $l=1$,
we have checked numerically, by diagonalizing \,$\tilde{r}_{22}$\,, that the eigenvectors have an oscillating behavior in agreement with Eq.~(\ref{eq:gomegaasympt}).

Furthermore, for $l=1$, we find numerically that there is an additional isolated eigenvalue $\lambda_{\rm ds}$\,, located above the continuous spectrum $\{\lambda(\omega),\ \omega\geq0\}$,
  and approaching Eq.~(\ref{eq:lambda_ds}) in the unitary limit,
  with a corresponding eigenvector $g_{\rm ds}(u)$ which decays without oscillating at large $u$.

  Since $\tilde{r}_{22}$ is symmetric, its eigenvectors form an orthonormal basis.
  We conjecture that they satisfy the same closure relation Eq.~(\ref{eq:closure}) than at the unitary limit, because we chose the same prefactor for the large-distance behavior, {\it i.e.}, the unitary-limit eigenvectors (\ref{eq:g_omega_l1_norm}) satisfy Eq.~(\ref{eq:gomegaasympt}).

  \bibliographystyle{crunsrt}

  % hide bibliography from table of contents:
\let\oldaddcontentsline\addcontentsline% Store \addcontentsline
\renewcommand{\addcontentsline}[3]{}% Make \addcontentsline a no-op

\vspace*{80pt}
\bibliography{felix_copy}
\let\addcontentsline\oldaddcontentsline% Restore \addcontentsline

\end{document}